\documentclass{aa}  
\usepackage{graphicx,natbib,psfig,amssym} 
\newcommand{\ltsima} {$\; \buildrel < \over \sim \;$}  
\newcommand{\gtsima} {$\; \buildrel > \over \sim \;$}  
\newcommand{\lta} {\lower.5ex\hbox{\ltsima}}  
\newcommand{\gta} {\lower.5ex\hbox{\gtsima}}

\defcitealias{allen06}{Allen06}
\defcitealias{paper2}{Paper II}
\begin{document}  

\title{The accretion mechanism in low-power radio galaxies.}
\subtitle{}
  
\titlerunning{The accretion mechanism in radio galaxies}
\authorrunning{Balmaverde et al.}
  
\author{Barbara Balmaverde
\inst{1},   
Ranieri D. Baldi\inst{2} \and  
Alessandro Capetti\inst{1}}
\offprints{B. Balmaverde}  
     
\institute{INAF - Osservatorio Astronomico di Torino, Strada
  Osservatorio 20, I-10025 Pino Torinese, Italy\\
\email{balmaverde@oato.inaf.it}\\
\email{capetti@oato.inaf.it}\\
\and 
Universit\`a di Torino, Via P. Giuria 1, I-10125, Torino, Italy\\
\email{baldi@oato.inaf.it}}
\date{}  
   
\abstract{ We study a sample of 44 low-luminosity radio-loud AGN,
  which represent a range of nuclear radio-power spanning 
5 orders of magnitude, 
  to unveil the accretion mechanism in these galaxies. We estimate
  the accretion rate of gas associated with their hot coronae by analyzing
  archival Chandra data, to derive the deprojected density and temperature
  profiles in a spherical approximation. Measuring the jet power from
  the nuclear radio-luminosity, we find that the accretion power correlates
  linearly with the jet power, with an efficiency of conversion from rest mass
  into jet power of $\sim$ 0.012.  These results strengthen and extend the
  validity of the results obtained by Allen and collaborators 
for 9
  radio galaxies, indicating that hot gas accretion is the dominant process in FR~I
  radio galaxies across their full range of radio-luminosity.

We find that the different levels of nuclear activity are driven
by global differences in the structure of the galactic hot coronae. A
linear relation links the jet power with the host X-ray surface brightness.
This implies that a substantial change in the jet power must be accompanied by
a global change in its ISM properties, driven for example 
by a major merger.  This
correlation provides a simple widely applicable method to
estimate the jet-power of a given object 
by observing the intensity of its host X-ray emission.

To maintain the mass flow in the jet, the fraction of gas that crosses the
Bondi radius reaching the accretion disk must be $\gtrsim 0.002$.
This implies that the radiative efficiency of the disk must be $\eta \lesssim
0.005$, an indication that accretion in these objects occurs not only at a
lower rate, but also at lower efficiency than in standard accretion disks.

} \maketitle

\section{Introduction.}
\label{intro}

A fundamental problem in studying the physics of active galactic nuclei (AGN)
identifying the mechanism that controls the level of activity.  
More generally, we need to determine why nuclear activity occurs in active
galaxies but not in quiescent galaxies, and whether there is a discontinuity
in the physical properties of these two galaxy classes.
Addressing these
points is also important in light of the prominent role that AGN
feedback plays in the process of galactic evolution. Furthermore, AGN activity
is related to the growth, via accretion, of supermassive black holes (SMBH);
the clear evidence for a co-evolution of SMBH and host galaxies, represented
by the connections between galaxy physical 
properties and SMBH mass, implies that the
accretion process in AGN has a powerful influence on galaxy evolution.

Several comparisons between the properties of active and quiescent galaxies
have been performed with in attempts to isolate the quantities that correlate
most significantly with the level of nuclear activity.
We limit ourselves to radio-loud galaxies in this paper.
Since the research of for example \citet{auriemma77} it has been known 
that the probability of an object to have a radio luminosity above a certain
threshold is proportional to its optical luminosity. This was confirmed
by studies of the bivariate radio/optical luminosity function of
early-type galaxies \citep{best05,mauch07}. Objects of a given optical
luminosity can be associated with 
radio sources spanning a large range
of radio power.
Several authors explored the role of environment in setting the level of
radio-emission. The fraction of radio-loud AGN is largely
independent of the local galaxy density, but
a dependence is found on the larger-scale environment
\citep{best04}. A statistical trend links the amount of dust measured
from the absorption features, seen in their optical images, and radio-power
\citep[e.g.][]{deruiter02}.
\citet{valentijn83} found a positive correlation between the X-ray luminosity
of the hosting cluster and the radio-luminosity of the dominant cluster galaxy.
Similarly, \citet{burns90} showed that central, massive galaxies in clusters or
groups of galaxies with a cooling flow are more likely to
be radio-loud AGN than galaxies of similar mass that are not centered on
a cooling flow, suggesting that concentrated cooling atmospheres
can stimulate radio-loud AGN activity. 
These studies provide statistically important results, 
but it is still impossible to ascertain 
from any of these measurements,
on an object-by-object basis, whether a galaxy hosts an active nucleus. 
This implies that we cannot yet associate an observable quantity and,
consequently, physical process with the level of nuclear activity.

Substantial progress was reported by \citet{allen06}.  They studied 9 nearby,
low-power radio galaxies, selecting objects with well-defined cavities in
X-ray emitting gas, which were cospatial with the radio-lobes: these cavities
were most likely inflated by radio-jets. They were able to estimate the jet,
kinetic power required to inflate the cavities.  On the other hand, it was
possible, using the same X-ray images, to estimate the amount of hot gas
available for accretion onto the SMBH. This analysis revealed a significant
correlation between the accretion rate and jet power, which had a dispersion
of only 0.16 dex.  This represents not only an accurate and powerful method to
predict the AGN energy output, but implies that spherical hot-gas, accretion,
associated with the hot corona, is the main energy source powering the active
nucleus.

We extend the analysis performed by \citet{allen06} to two samples of
galaxies that host radio-loud nuclei of relatively low-power, which cover
almost five orders of magnitude in radio-power. The comparison of accretion
properties across the widest possible range of luminosity is clearly the next
crucial step in unveiling the connections (and diversities) between galaxies
showing different levels of nuclear activity.  We examine data for the sample
of 29 early-type ``core'' galaxies, selected by \citet{paper2}, and for the
sample of 33 FR~I radio galaxies, which were extracted from the 3C sample as
defined in \citet{chiaberge:ccc}; in the combined set of galaxies 
44 have observational data in the Chandra archive
(more details of the sample selection are given in Sect.  \ref{sample}).  With
respect to \citet{allen06} we estimated the jet power using a relation
between radio-core luminosity and jet kinetic power derived
by \citet{heinz07}. 
This relation provides a straightforward, robust measure of the jet power
(with an rms error of 0.4 dex), based only on the core power, and enables the
number of objects for which accretion and jet power can be compared, to be
increased. 
Furthermore, the range of radio-power that can be studied covers almost five
orders of magnitude; this corresponds to the full range of
radio-luminosity of FR~I, spanning from the faintest level of detectable
activity in radio galaxies up to include galaxies whose radio-emission exceeds
the luminosity marking the transition between FR~I and FR~II objects.

\section{Sample selection and data preparation}
\label{sample}

\begin{table*}
  \caption{Basic data and Chandra observations log.}
\label{basic}
\centering
\begin{tabular}{l | c c |c c c c c}
\hline \hline
Name &\multicolumn{2}{|c|} {Chandra summary} & \multicolumn{5}{c}{Basic Data}\\
     & Obs. Id & Exp. time & D     & Log $\nu$ L$_{core}$  & $\sigma$  & $M_{K}$  &  Log $(M_{BH}/M_{\odot})$  \\
\hline  
    3C~028          & 3233   & 50.38  &  666.7 & $<$38.86  & --  &  -25.71  &    8.99        \\
    3C~031          & 2147   & 44.98  &  67.15 &     39.40 & 278 &  -25.65  &    8.70         \\
    3C~066B         & 828    & 45.17  &  83.68 &     39.90 &  -- &  -26.42  &    9.32        \\
    3C~075          & 4181   & 21.78  &   91.0 &     39.30 & 301 &  -25.71  &    8.84        \\
    3C~078          & 3128   & 5.23   &  112.1 &    40.88  & 271 &  -26.18  &    8.61        \\
    3C~083.1        & 3237   & 95.14  &  98.65 &     39.10 & --  &  -26.84  &    9.51        \\
    3C~084          & 3404   & 5.86   &  69.31 &     42.10 & 259 &  -26.08  &    8.58         \\
    3C~189          & 858    & 8.26   &  165.8 &    40.54  & --  &  -26.22  &    9.23        \\
    3C~264          & 4916   & 38.33  &  85.46 &    39.91  & 271 &  -25.17  &    8.67        \\
    3C~270          & 834    & 35.18  &  29.69 &     39.22 & 309 &  -25.10  & 8.72$^{b}$      \\
    3C~272.1$^{*}$  & 803    & 28.85  &  14.11 &   38.57   & 282 &  -24.53  & 9.00$^{b}$      \\
    3C~274$^{*}$    & 1808   & 14.17  &  17.38 &    39.90  & 333 &  -25.39  & 9.53$^{b}$     \\
    3C~293          & 5712   & 3.10   &  174.1 &    40.29  & 185 &  -25.36  &    8.14        \\
    3C~296          & 3968   & 50.08  &  96.99 &    39.62  & 299 &  -26.17  &    8.80        \\
    3C~317          & 890    & 37.23  &  134.3 &    40.64  & 216 &  -26.09  &    8.26         \\
    3C~338$^{*}$    & 497    & 19.72  &  118.7 &    39.96  & 310 &  -26.50  &    8.89        \\
    3C~346          & 3129   & 46.69  &  569.3 &    41.74  & --  &  -26.03  &    9.14         \\
    3C~348          & 1625   & 15.00  &  544.9 &    40.36  & --  &  -26.13  &    9.19        \\
    3C~438          & 3967   & 47.9   &  907.7 &    41.14  & --  &  -26.47  &     9.34       \\
    3C~442          & 6392   & 33.13  &  103.1 &    38.12  & --  &  -24.39  &    8.40         \\
    3C~449          & 4057   & 29.56  &  67.46 &    39.06  & 253 &  -24.96  &     8.54       \\
    3C~465          & 4816   & 50.16  &  118.1 &    40.37  & 356 &  -26.64  &    9.14         \\
\hline		   										      
UGC~0968         & 6778 &  15.14 &  32.39  & 36.94   & 253 &   -25.39  & 8.54            \\
UGC~5902         & 1587 &  31.9  &  12.68  & 35.83   & 207 &   -24.25  & 8.00$^{b}$      \\
UGC~6297         & 2073 &  39.0  &  13.63  & 36.46   & 224 &   -23.68  & 8.33            \\
UGC~7203         & 3995 &  5.13  &  31.89  & 37.44   & 184 &   -24.08  & 7.98	       \\
UGC~7386         &  398 &  1.43  &  10.72  & 38.38   & 238 &   -22.97  & 8.43	          \\
UGC~7629$^{*}$   &  321 &  40.1  &  12.53  & 37.73   & 291 &   -25.09  & 8.78	         \\ 
UGC~7760$^{*}$   & 2072 &  55.14 &  5.227  & 37.30   & 253 &   -21.86  & 8.54	           \\
UGC~7797         & 6785 &  15.19 &  29.77  & 38.05   & 224 &   -24.61  & 8.33             \\
UGC~7878$^{*}$   &  323 &  53.05 &  14.83  & 36.90   & 203 &   -24.43  & 8.16	           \\
UGC~7898         &  785 &  37.35 &  16.41  & 37.46   & 336 &   -25.34  & 9.30$^{b}$       \\
UGC~8745         & 6787 &  15.18 &  27.95  & 37.81   & 232 &   -25.07  & 8.39	            \\
UGC~9706$^{*}$   & 4009 &  30.79 &  25.28  & 37.31   & 238 &   -25.06  & 8.43               \\
UGC~9723         & 2879 &  34.18 &  13.79  & 36.92   & 159 &   -23.82  & 7.73              \\
NGC~1316         & 2022 & 30.23  &  20.64  & 37.82   & 228 &   -25.99  & 8.36             \\
NGC~1399         &  319 & 56.66  &  16.23  & 37.20   & 344 &   -24.75  &  9.07            \\
NGC~3557         & 3217 &  371.99&  38.09  & 37.94   & 273 &   -25.70  &  8.67          \\
NGC~4696$^{*}$   & 1560 &  85.84 &  36.89  & 38.65   & 254 &   -25.69  &  8.55           \\
NGC~5128         & 463  & 19.6   &  5.2    & 39.05   & 120 &   -24.64  &  8.38$^{b}$    \\
NGC~5419         &4999  & 15     &  53.92  & 38.42   & 332 &   -26.14  &  9.02           \\
IC~1459          &  2196& 60.17  &  19.97  & 39.38   & 306 &   -24.70  &  9.18$^{b}$     \\
IC~4296          & 3394 & 25.4   &  48.17  & 39.46   & 336 &   -25.69  &  9.04          \\
\hline	
NGC~507$^{*}$    & 2882 & 44.21  &  65.02  & 36.45   & 310 &    8.688  &  8.89            \\
\hline \hline
\end{tabular}
	   			   
Basic data and Chandra observations log for the radio galaxies 
of the 3C/FR~I and CoreG sample (with the addition of 
    NGC~507). Objects from the
    Allen's sample (marked with *). Column description: (1) name, (2) Chandra
    observational identification number, (3) exposure time [ks], (4) distance
    [Mpc], (5) nuclear radio luminosity (5GHz) [erg s$^{-1}$], (6) stellar
    velocity dispersion [km s$^{-1}$] from the HyperLeda database,
    (7) total K band galaxy's absolute magnitude from 2MASS (except for 7
    sources, see Sec. \ref{The black-hole mass}), (8) logarithm of black-hole
    mass in solar unity (see text for details).
\end{table*}

In \citet{paper1}, we focused on a sample of luminous, nearby early-type
galaxies to study the link between their
host properties and the active nucleus. 
We used archival HST observations to analyze
their surface brightness profiles and to separate these early-type galaxies
into core and power-law galaxies on the basis of the slope of their nuclear
brightness profiles, following the Nuker scheme \citep{lauer95}. We obtained a
sub sample of 29 ``core'' galaxies (hereafter CoreG), that is galaxies 
for which the surface-brightness profiles were
fitted with a
double power-law of innermost slope $\gamma \leq 0.3$.  In \citet{paper2} we
demonstrated that the CoreG hosted a radio-loud nucleus and that they
could be considered to be miniature radio galaxies. In fact, they are drawn from
the same population of early-type galaxies as the FR~I hosts, according to
of the surface brightness profiles, and distributions of optical
luminosities and black-hole masses, they differ from ``classical'' FR~I
radio galaxies only in terms of their lower level of nuclear activity.

We combine these CoreG with a
sample of low-luminosity radio galaxies of morphology FR~I
extracted from the 3C catalogue of radio-sources \citep{bennett62}
in a way defined by \citet{chiaberge:ccc}. Three galaxies (namely 3C~270, 3C~272.1,
and 3C~274) are in common between the two samples\footnote{For simplicity, we include
these sources in Table \ref{basic} in the FR~I sample.}.
All 9 galaxies (with the exception of NGC~507) considered by \citet{allen06}
are in common with our combined FR~I/CoreG sample.

We searched for Chandra observations available in the public archive up to
August 2007 and found data for 22 of the 
33 objects belonging to the FR~I sample:
with respect to the sample considered by \citet{balmaverde06}, we were able
to add three sources observed, 3C~264, 3C~293, and 3C~442.
We found data for 24 of the 29 CoreG:
with respect to \citep{paper2} newly obtained observations
were found for UGC~0968, UGC~7797, UGC~8745. All together, we 
consider
44 galaxies that have a range of nuclear radio-power covering 
5 orders of magnitude.
The information on the Chandra observations and basic data for these
galaxies are presented in Table \ref{basic}.

We reduced all data using the Chandra Data analysis CIAO v3.4, with the CALDB
version 3.3.0. We reprocessed the data from level 1 to level 2 in the standard
way, that is using the CIAO tool acis\_process\_events, removing bad pixels and
applying pixel and PHA randomization. When possible, we improved the
rejection of cosmic events for VFAINT mode observations. We rejected time
intervals corresponding to high background levels and we filtered by selecting a standard set
of ASCA grades (02346).

We modeled the spectra obtained using XSPEC version 12.3.0. We used only events
filtered between 0.3 and 8 keV for our analysis and rebinned the spectrum to
a minimum of 20 counts per bin to ensure that a
$\chi^{2}$ statistics could be applied.

We adopt H$_o$=75 km s$^{-1}$ Mpc$^{-1}$ and q$_o$=0.5.

\begin{figure*}
\centerline{
\psfig{figure=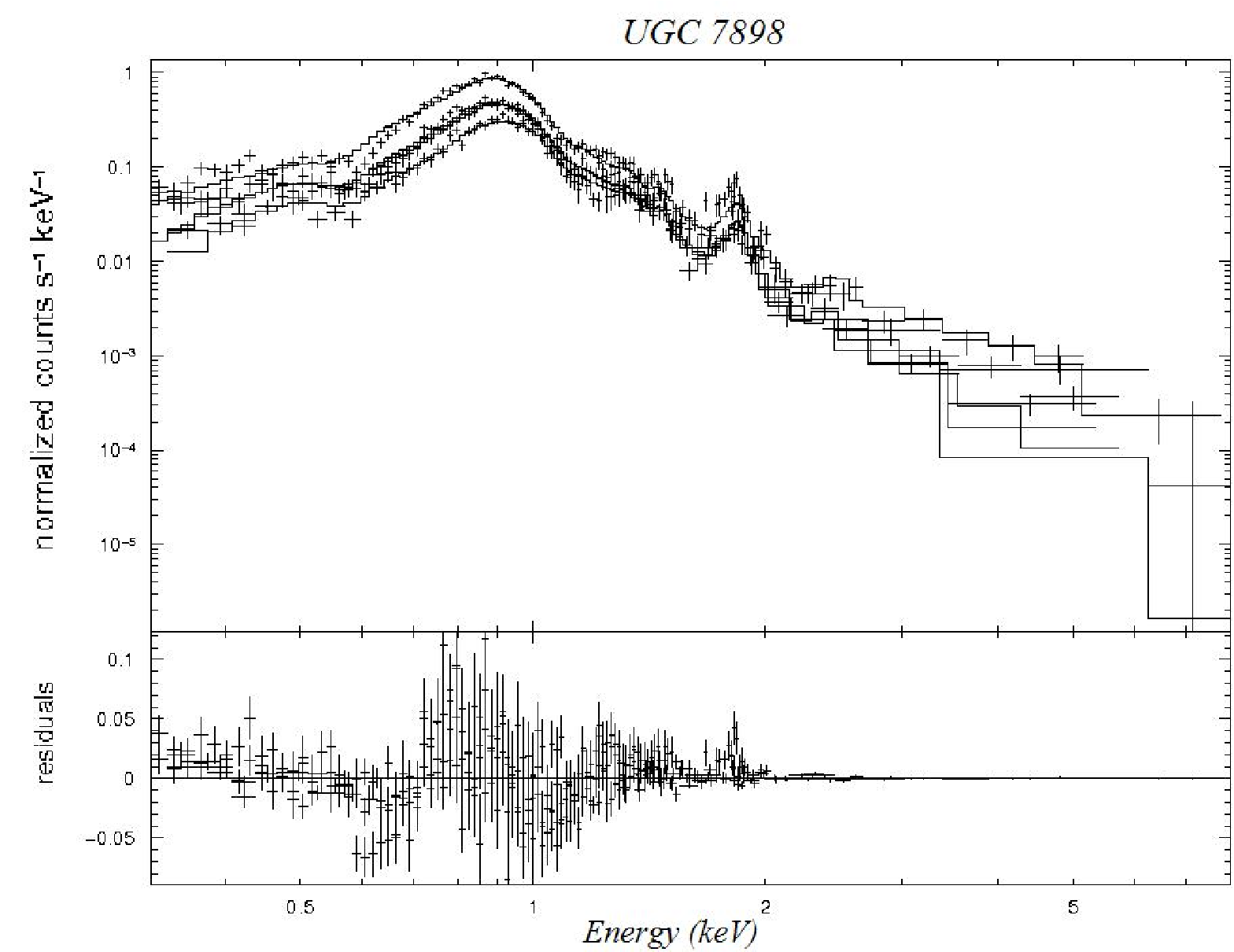,width=0.5\linewidth,rheight=0.5\linewidth,height=0.48\linewidth}
\psfig{figure=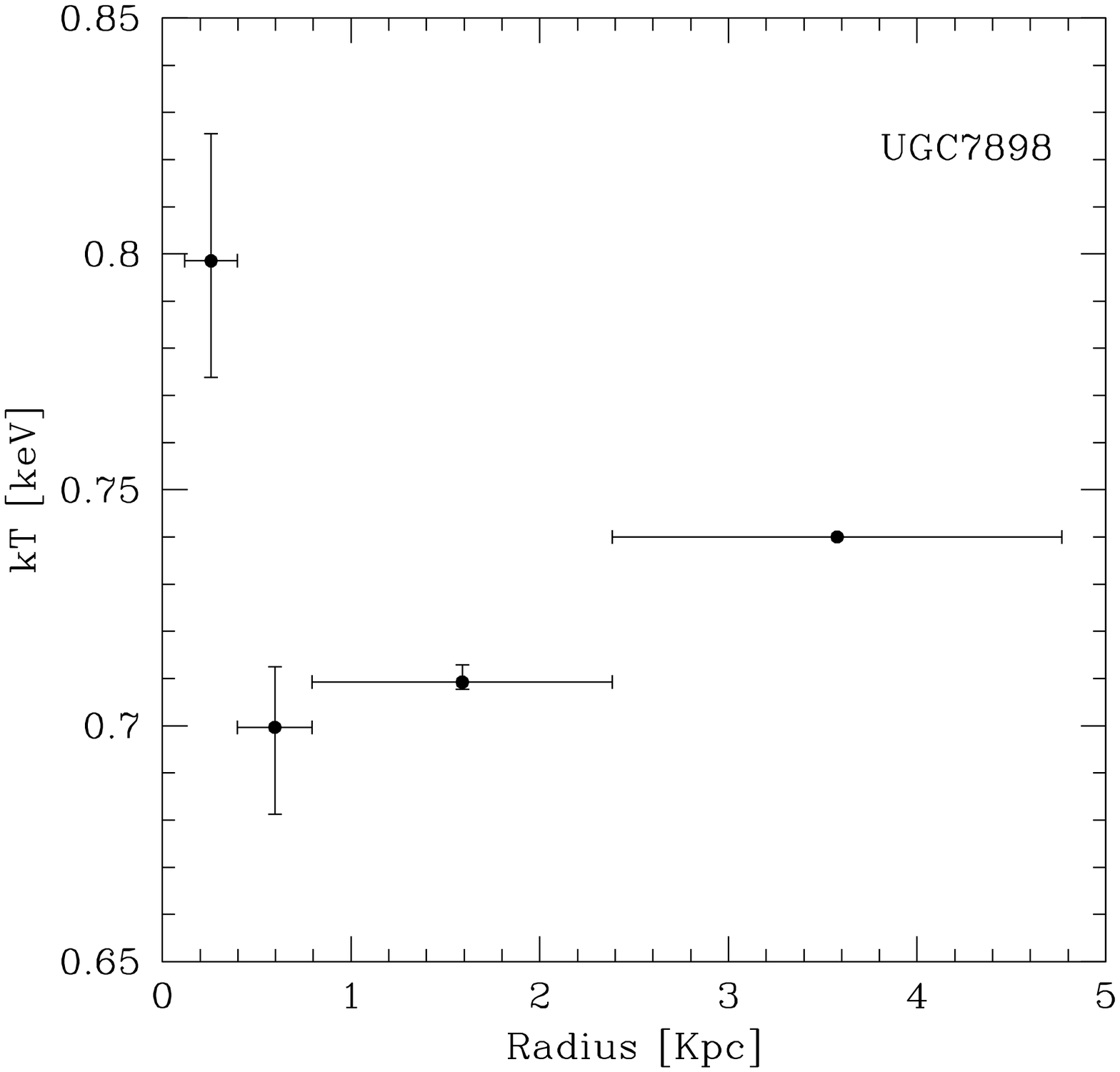,width=0.5\linewidth }}
\caption{Example of the procedure followed to determine the Bondi power for
  NGC~7898: in the left panel, we show the spectrum extracted in four annuli,
  centered on the nucleus, that was modeled with a PROJECT*PHA(MEKAL+POWERLAW)
  law; in the right panel, we show the corresponding deprojected temperature
  profile of the annuli.}
\label{es1}
\end{figure*}

\begin{figure*}
\centerline{
\psfig{figure=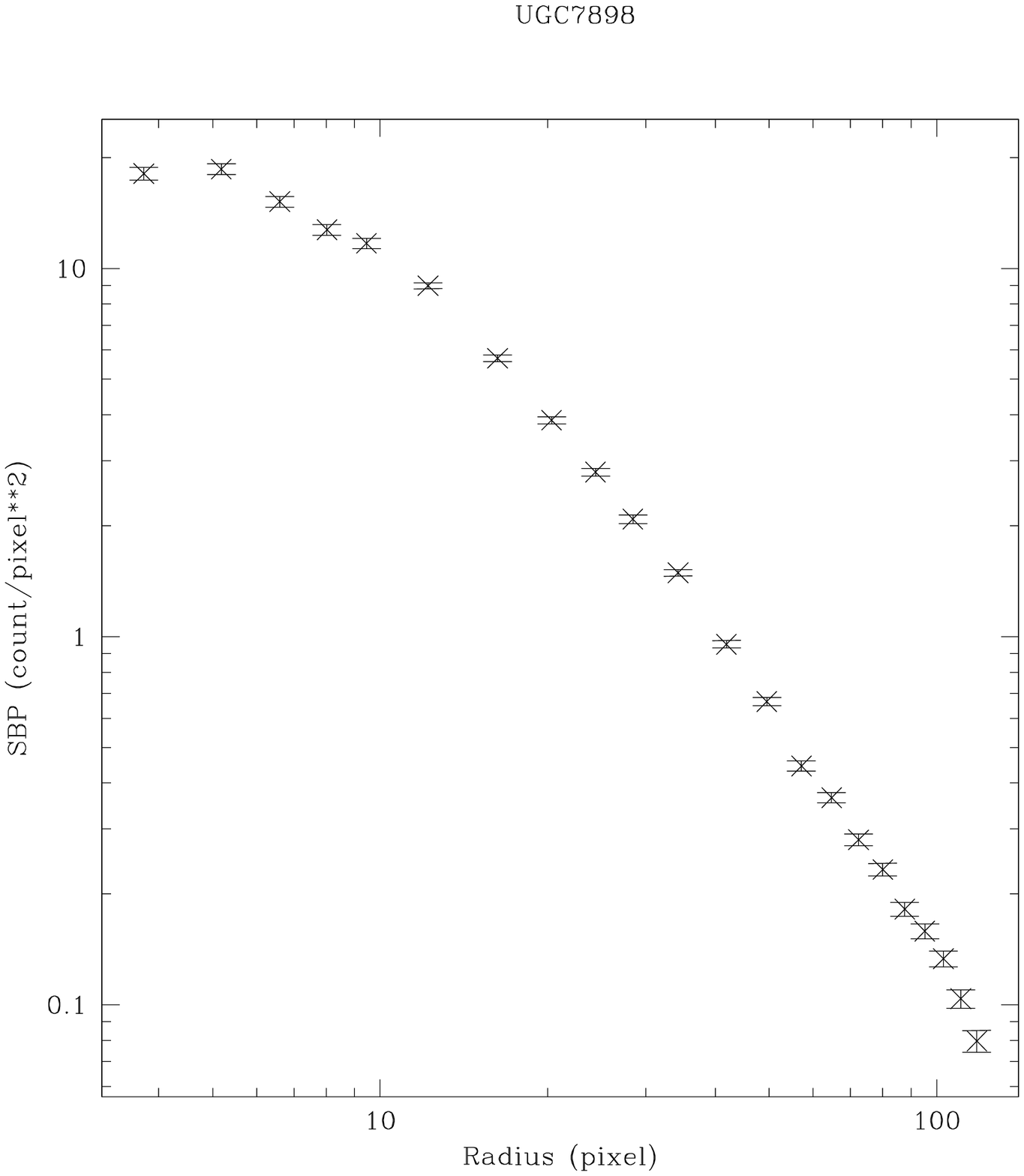,width=0.5\linewidth}
\psfig{figure=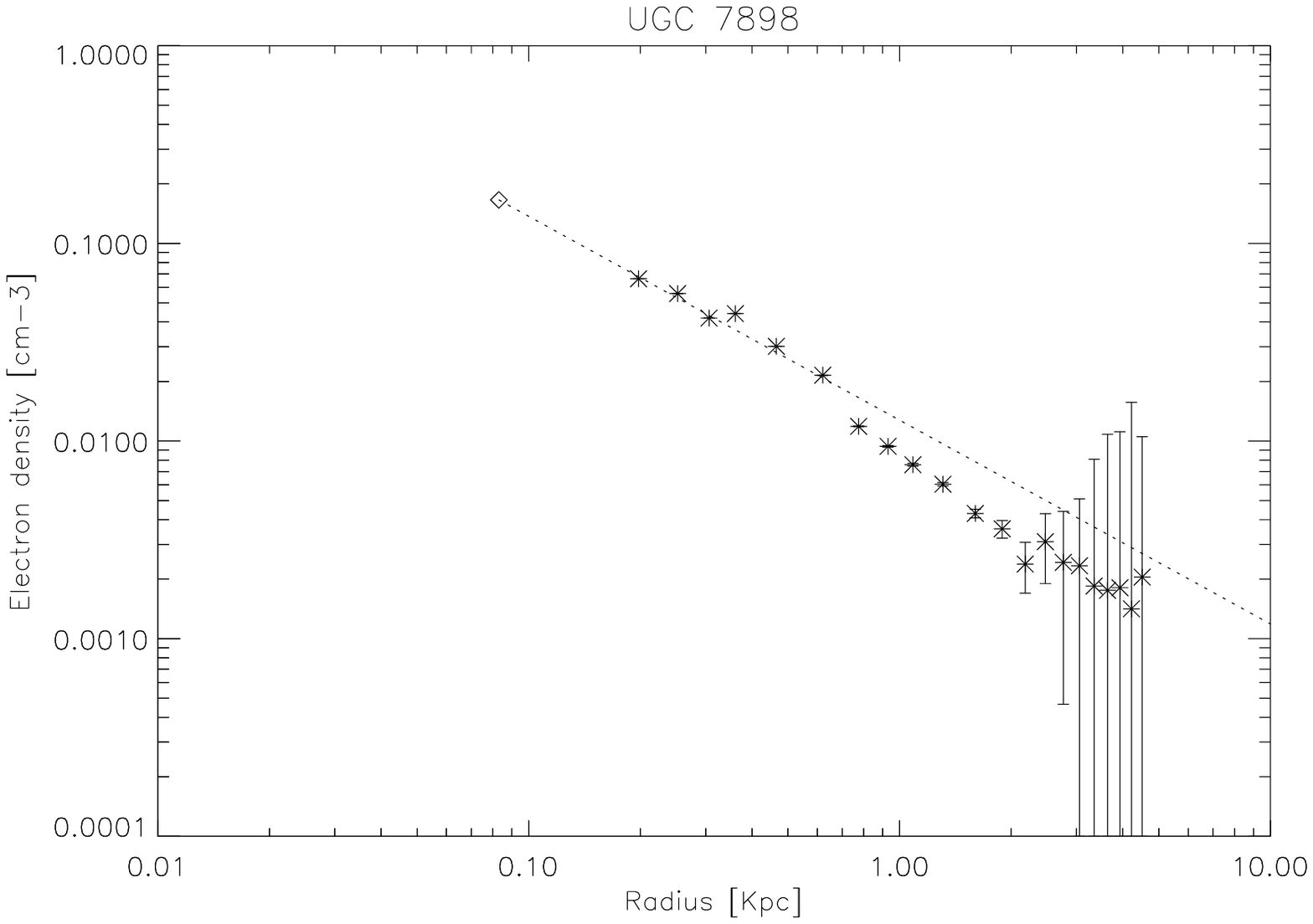,width=0.5\linewidth,rheight=0.5\linewidth,height=0.5\linewidth}}
\caption{Example of the procedure followed in this work to determine the Bondi
  power for NGC~7898: in the left panel, we present the X-ray surface
  brightness profile of the galaxy; in the right panel, we show the derived
  deprojected electron density profile and the power-law extrapolation used
  to estimate its value at the Bondi radius (marked with a diamond).}
\label{es2}
\end{figure*}

\section{Calculation of the Bondi accretion rates}
\label{pbondi}
Under the assumption of spherical symmetry and negligible angular momentum,
the accretion rate onto a SMBH can be estimated by the \citet{bondi52}
approximation as:
$$
\dot{M}_{B}=\pi \; \lambda \; c_s \; \rho_B \;{r_B}^2
$$ 

\noindent
where $\lambda$ takes the value of 0.25 for an adiabatic index $5/3$, $c_s$ is
the sound speed in the medium ($c_s=\sqrt{\Gamma\,k\,T\,/\mu\,m_p}$), $
r_B=\frac{2\,G\,M_{BH}}{{c_s}^2} $ is the Bondi accretion radius, and $\rho_B$
is the gas density at that radius, ($\rho=1.13 n_e m_p$), so that

$$\dot{M}_{B} = \frac{4.52 \pi \lambda G^{2} m_{p}}{(\frac{\Gamma}{\mu m_{p}})^{3/2}}   n_e \; {M_{BH}^2} \; (kT)^{-3/2}$$

The calculation of the Bondi accretion rate then requires the measurement of
three quantities: the electron gas density at the Bondi radius, the gas
temperature at the same radius, and the mass of the central black hole.

Unfortunately not even the Chandra telescope, which provides presently the
highest angular resolution in X-rays, is able to resolve the accretion radius
in most galaxies. Different strategies have been devised to estimate the gas
and temperature density at the accretion radius.  For example,
\citet{gliozzi03} fitted the electron density profile with a model of gas in
isothermal, hydrostatic equilibrium in spherical symmetry.
\citet{Hardcastle07} derived the Bondi power for radio galaxies from the 3CRR
catalogue, adopting density and temperature values comparable to that
typically measured ($\rho=5\;10^5\;m^{-3}$ and kT = 0.7 keV) in nearby FR~I.
However, the results of \citet{allen06} indicated strong departures from
equilibrium profiles and large variations in the central gas densities,
suggesting that the accretion rates derived with these assumptions can be
affected by large errors.

In this work, we determine the Bondi accretion rate following the same
technique as \citet{allen06} and \citet{birzan04}, which is described in
detail below. Briefly, we deproject the temperature and electron density
profile, assuming spherical symmetry. We then extrapolate the measured density
to the accretion radius and describe the density profile with a power-law,
assuming that the temperature is constant inside the innermost measurement
radius. Given the temperature, the electron density at the accretion radius,
and the black-hole mass, we calculate the Bondi accretion.

\subsection{The IGM temperature profile}

To characterize the thermal and emissivity properties of the
intergalactic hot gas, we extract the spectrum in four annuli (see Fig.
\ref{es1}, left panel) centered on the peak of X-ray emission. Since a X-ray
point source associated with the AGN is present in most galaxies, we extract
the spectrum of the IG medium avoiding the innermost region within a radius of
1$\farcs$5. The outer radius is set to where the diffuse emission reaches the
background level.

To measure the dependence of temperature on radius, we deproject the spectra
by assuming spherical symmetry and using the PROJECT model in XSPEC. This model
projects prolate ellipsoidal shells onto elliptical annuli, calculating the
projection matrix of each of the overlying shells onto the area of each annulus
and computing the total projected emission on that annulus. The fitting occurs
simultaneously for all annuli and, after completion, the model provides the
best-fitting temperature and abundance values in each shell (Fig. \ref{es1},
right panel).  We adopt a single-temperature plasma model (MEKAL) and
foreground absorption (PHABS), taking into account projection effects, that
is the model PROJECT*PHA(MEKAL). The hydrogen column density was tied between
annuli and allowed to vary. With this fitting procedure we determine the  deprojected profile of temperature. We then assume that the temperature measured
inside the innermost annulus (usually bound between 1$\farcs$5 and 5$\arcsec$
in radius) is a good estimate of the temperature at the Bondi radius (see Table
\ref{bondi}).

\subsection{The electron density profile}
\label{densityprof}

We determine the X-ray surface brightness profile (SBP)
extracting, the counts in a series of circular annuli centered on the central
AGN (see Fig. \ref{es2}, left panel, for an example; all SBP are shown in the
Appendix A (Fig. \ref{sbpprof}).
When present, we mask the X-ray jet and
other unrelated sources; we then subtract the background measured far from the
source of interest.
To avoid contamination from non-thermal emission due to the central
AGN, we exclude the inner 1.5 arcsec region.
When a point
source is present, we use the PSF libraries to create an image of the
PSF for the specific observation. We normalize it to the source flux in a
circle of 1 arcsec centered on the peak of the emission and subtract the
wings of the PSF brightness profile from the profile of the source.

As a general rule, we extract the X-ray counts from 4 annuli in the first
5 arcsec, then increase the size of the extraction region outward.
For weaker sources, we use a coarser sampling of the brightness profile
to maintain sufficient signal-to-noise ratio.  
The next step is to deproject the observed SBP,
to derive the number of counts emitted per unit volume as a function of
radius. Assuming spherical symmetry, the counts contribution 
provided by each spherical shell to the inner ones can be determined
by pure geometric considerations, following the calculation by
\citet{kriss83}. Briefly, we express the observed SBP as a matrix
product between the deprojected count rate in the spherical shell
and a weight matrix, defined as the portion of volumes of each 
overlying shell seen in projection on each annulus.

Our aim is to compare the observed counts per unit volume, with the
theoretical value expected for a thermal plasma of a specific
electron density and metal abundances. The XSPEC fitting model provides us
with the value of the thermal normalization and the model predicted count
rates (deprojected) in each of the annuli.  Comparing these values with
the counts per unit volume obtained by deprojecting the SBP, we adjust the
thermal normalization to recover the counts emitted in each of the
circular annuli used for the spatial analysis.  The density was then
calculated from the normalization of the thermal component, assuming
$n_e=1.2\,n_H$ (for a fully-ionized gas). The definition of the thermal
normalization is
$$
K=\frac{10^{-14}}{4\pi D_A^2(1+z)^2)}\int n_e n_p dV
$$ where $z$ is the redshift of the source, $D_A$ is the angular-diameter
distance at that redshift, $n_e$ and $n_p$ are the electron and proton
number-density, respectively, and the integration is performed over the volume
of the projected shells. Assuming a uniform density distribution along the
shell, we invert this equation to solve for the electron density.
Iterating this procedure for each spherical shell, we obtain the electron
density profile (see Fig. \ref{es2}, right panel, for an example; all profiles
are shown in the Appendix A, Fig. \ref{neprof}).

To estimate the gas density at the accretion radii,
we perform a fit on the innermost points of the density profile in a power-law form, $n_e(r)
\sim r^{- \alpha}$, following \citet{allen06} and as explained in more detail in
Sect. \ref{quality}.

\subsection{The black-hole mass}
\label{The black-hole mass}

The mass of the central black hole can be determined in various ways:
when available, we prefer to use the BH mass estimates obtained from direct gas
or stellar dynamical measurements listed in \citet{marconi03}. 

Otherwise, we derive the SMBH  masses using the stellar velocity
dispersion (from the HyperLeda database database).  We apply the relation
derived by \citet{tremaine02}:

\centerline{
log (M$_{BH}$/M$_{\odot}$)=8.13 + 4.02 $\,$ log ($\sigma$ / 200 km s$^{-1})$ }

\noindent
which has an intrinsic dispersion of about 0.23 dex.  Alternatively, for the 7
objects for which velocity dispersion measurement is unavailable, we
estimate the SMBH mass using the K-band absolute total magnitude (from 2MASS
data archive\footnote{For 3C~066B, 3C~075, 3C~083.1, 3C~338, 3C~449, and
  3C~465 the K band luminosity is converted from H-band \citep{donzelli07},
  assuming H - K = 0.21 and for 3C~442 we convert the V-band luminosity from
  \citet{smith89}, assuming V - K = 3.30}) by using the relation derived by
\citet{marconi03}:

\centerline{
log (M$_{BH}$/M$_{\odot}$) = 8.21 + 1.13 $\,$ (log L$_K$ -10.9).} 

\noindent
The observed
scatter in this case is $\sim$ 0.5 dex.

\subsection{Error budget}
\label{Error}

The uncertainties in the results of the deprojection were estimated
using a Monte Carlo simulations. We perturbed randomly each point of the SBP
with a Gaussian distribution of the perturbations, which had an 
amplitude determined by the 
error bar. We obtained 1000 different realizations of the SBP. 
Each of them was deprojected, and we evaluated, in each shell, the mean value
of the counts rate and the scatter. Then we propagate the errors in the
calculation of the electron density, taking into account the errors in the
counts rate and in the temperature measurement at the accretion radius.

\begin{figure}
\centerline{
\psfig{figure=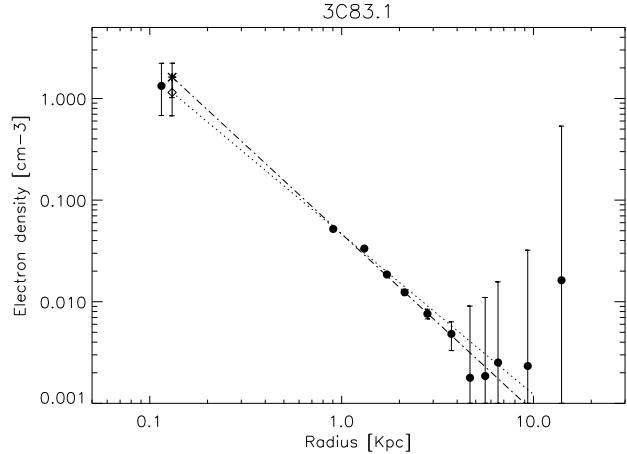,width=1.0\linewidth}}
\caption{Example of the procedure followed to determine the density at the
  Bondi radius in the case of 3C~83.1.  The dotted (dashed-dotted) line
  represents the best-fit linear relation to the density profile obtained by using the
  innermost 3 (6) density measurements, while the diamond (star) is its
  extrapolation to the Bondi radius. The adopted value for n$_e$ is the filled
  circle at r$_B$.}
\label{density}
\end{figure}

Another source of uncertainty is associated with the extrapolation of the
electron density to r$_{B}$. This is related to the uncertainties in the
parameters of the power-law fit describing the density behavior but also to
the choice of the range of radii to be included.  We evaluated the differences
in n$_e$, repeating the extrapolation using from 3 to 6 points of the
electron density profile. The adopted value for n$_e$ is the average value
of these measurements and its uncertainty spans the full range given by the
overlap of each individual error bar. This method is graphically explained
in Fig. \ref{density}.

Concerning uncertainties in the black-hole mass
determination, we assume conservatively an error of a factor 3 (0.5 dex), 
regardless of the method used. 

The final error in $P_{B}$ is estimated by propagating the errors of all
relevant quantities.  We note that since the Bondi accretion power
is tied to the electron density, related in turn to radius via the value of
$\alpha$, the effective dependence of $P_{B}$ on the black-hole mass
is not simply from
M$_{BH}^2$, but from M$_{BH}^{2 - \alpha}$. Given that the mean $\alpha$ value
estimated (see Table \ref{bondi}) is 1, the propagated uncertainty is
essentially linear with the black-hole mass error.

\begin{table*}
  \caption{Measured Bondi accretion rates and jet power.}
\label{bondi}
\centering
\begin{tabular}{l | c c c c c c c c c c}
\hline \hline
Name &  Flag & r$_{B}$ & r$_{min}$/r$_{B}$ & $\alpha$  & KT & $n_e$ &$\dot{M}_{B}$& $\dot{P}_{B}$ & P$_{j}$ & ${\cal S}_{1kpc}$ \\
            &                    &\tiny{[pc]}          &  & & \tiny{[keV]}       &  \tiny{[cm$^{-3}$]}  & \tiny{[10$^{-3}$M$_\odot yr^{-1}$]} & \multicolumn{2}{c}{\tiny{$\,\,\,\,\,\,$[10$^{43}$ erg s$^{-1}$]}}        &  \\
\hline \hline
3C~028   & R &  19  & 101 &                      & 1.70$\pm$0.06 &                        &         &                      & $<$ 3.3  & -3.26\\
3C~031   & E &  21  &  36 &                      & 0.82$\pm$0.08 &                        &         &                      &    7.9   & -3.53 \\
3C~066B  &   &  69  &  11 &  1.2$_{-0.3}^{+0.1}$ & 0.42$\pm$0.08 &0.35$_{-0.16}^{+0.26}$& 350 & 390($\pm$ 31\%)    &   17.8   & -3.72\\
3C~075   & W &  41  &  22 &                      & 0.57$\pm$0.07 &                        &         &                      &    6.7   & -4.19\\ 
3C~078   & W &  11  &  96 &                     & 1.19$\pm$0.08  &                        &         &                      &   87.7   & -2.74\\ 
3C~083.1 &   & 131  &   7 &  1.6$_{-0.1}^{+0.4}$ & 0.66$\pm$0.06 & 0.92$_{-0.48}^{+0.57}$& 2100& 1300($\pm$ 26\%)     &    4.9   & -3.79\\
3C~084   & C &  --  &  -- &                      &               &                        &         &                      &  637.0   & -2.53 \\
3C~189   & W &  91  &  17 &                      & 0.62$\pm$0.09 &                        &         &                      &   50.5   & -3.01\\
3C~264   & E &  29  &  26 &                      & 0.52$\pm$0.08 &                        &         &                      &   18.1   & -3.32\\
3C~270   &   &  32  &   9 & 1.2$_{-0.2}^{+0.3}$  & 0.55$\pm$0.02 & 5.4$_{-3.1}^{+3.3}$  &  460    & 260($\pm$ 31\%)      &    5.9   & -3.72 \\
3C~272.1 & * &  30  &  -- & 0.55$_{-0.18}^{+0.19}$&0.71$\pm$0.05&0.92$_{-0.47}^{+0.99}$   &   85  &49($_{-29 \%}^{+30 \%}$)& 1.5$\pm$0.46& -4.05 \\
3C~274   & * & 117  &  -- & 0.00$_{-0.10}^{+0.10}$&0.80$\pm$0.01&0.166$_{-0.032}^{+0.036}$&   240   &145($_{-40 \%}^{+28 \%}$) & 3.4$\pm$1.67 & -2.95\\
3C~293   & W &  --  &  -- &                      &               &                        &         &                      &   33.6   & -3.45\\
3C~296   &   &  32  &  28 & 1.7$_{-0.2}^{+0.3}$  & 0.66$\pm$0.03 & 42$_{-29}^{+42}$ &4000     &2280($\pm$ 37\%)      &   11.3   & -3.21\\
3C~317   & C &  --  &  -- &                      &               &                        &         &                      &    59.4  & -3.40 \\
3C~338   & * &  20  &  -- &0.50$_{-0.06}^{+0.07}$&1.3$\pm$0.7    &0.96$_{-0.29}^{+0.38}$  & 55    & 30.9($_{-26\%}^{+34\%}$)& 1.6$\pm$0.47& -2.84\\
3C~346   & E &  55  &  94 &                      & 0.84$\pm$0.05 &                        &         &                      &  354.9   & -2.71 \\
3C~348   & R &  34  & 339 &                      & 1.5$\pm$0.1   &                        &         &                      &   37.7   & -2.11\\ 
3C~438   & R &  13  & 815 &                      & 1.41$\pm$0.09 &                        &         &                      &  133.8   & -2.40\\
3C~442   & W &   6  &  17 &                      & 1.42$\pm$0.11 &                        &         &                      &    1.0   & -4.15 \\
3C~449   &   &  18  &  32 & 1.4$_{-0.2}^{+0.4}$  & 0.64$\pm$0.08 &18$_{-14}^{+15}$& 530    &302($\pm$ 38\%)     &    4.6   & -4.04\\
3C~465   &   &  42  &  27 & 1.4$_{-0.3}^{+0.1}$  & 1.11$\pm$0.02 & 12$_{-4}^{+8}$&2500     &1400($\pm$ 25\%)      &   38.3   & -3.14 \\
\hline	                  								 
UGC~0968 & E &  21  &  10 &                      & 0.56$\pm$0.04 &                        &         &                      &    0.15  & -4.44\\
UGC~5902 & E &   8  &   7 &                      &0.44$\pm$0.05  &                        &         &                      &    0.024 & -5.39\\
UGC~6297 &   &   7  &   9 &  0.9$_{-0.7}^{+0.1}$ &1.08$\pm$0.36  &4.0$_{-1.5}^{+7.0}$  & 21    & 4.8($\pm$ 51\%)      &    0.067 & -5.48\\
UGC~7203 & W &  --  &  -- &                      &               &                        &         &                      &     0.33 & --    \\
UGC~7386 &   &  23  &   4 &  1.8$_{-0.3}^{+0.5}$ &0.39$\pm$0.10  &4.2$_{-3.2}^{+5.0}$  & 160     & 89($\pm$ 43\%)     &     1.5  & -5.35\\
UGC~7629 & * &  36  &  -- & 0.36$_{-0.12}^{+0.12}$&0.70$\pm$0.16 &0.82$_{-0.29}^{+0.52}$  &110      & 61.7($_{-23 \%}^{+25 \%}$)&0.81$\pm$0.23& -4.10 \\
UGC~7760 & * &  17  &  -- & 0.83$_{-0.08}^{+0.08}$&0.67$\pm$0.09 &1.26$_{-0.81}^{+3.48}$  &44     &23.4($_{-21 \%}^{+22 \%}$) &0.16$\pm$0.04& -5.18\\ 
UGC~7797 & E &  10  &  29 &                      & 0.71$\pm$0.08 &                        & 	    &     	           &    0.88  & -4.84 \\
UGC~7878 & * &  10  &  -- & 0.31$_{-0.09}^{+0.09}$&0.54$\pm$0.11 & 0.42$_{-0.15}^{+0.26}$ & 47    &1.95($\pm$ 24\%)      &    0.030$\pm$0.008& -4.02\\
UGC~7898 &   &  83  & 2.4 & 1.0$_{-0.2}^{+0.1}$  & 0.80$\pm$0.02 &0.16$_{-0.04}^{+0.05}$  & 110     &  65($\pm$ 24\%)    &    0.34  & -3.61\\
UGC~8745 &   &  15  & 22  & 1.7$_{-0.5}^{+0.2}$  &0.54$\pm$0.11  & 39$_{-25}^{+56}$ & 750     &  420($\pm$ 45\%)     &    0.60  & -4.75\\
UGC~9706 & * &  19  &  -- & 0.35$_{-0.21}^{+0.26}$&0.67$\pm$0.09 &0.33$_{-0.20}^{+0.56}$  & 13    &7.08($_{-40 \%}^{+43 \%}$)& 0.074$\pm$0.026& -3.50\\
UGC~9723 & R & 2.5  & 101 &                      & 0.73$\pm$0.07 &                        &         &  	                   &   0.14   & -5.19\\
NGC~1316 & E &  14  &  13 &                      & 0.54$\pm$0.06 &                        & 	    &                	   &    0.61  & -3.94\\
NGC~1399 &   &  44  &   4 & 1.0$_{-0.2}^{+0.1}$  &0.88$\pm$0.01  & 2.3$_{-0.8}^{+1.1}$ & 490     & 280($\pm$ 28\%)      &    0.22  & -3.96\\
NGC~3557 &   &  52  &   3 & 1.1$_{-0.5}^{+0.5}$  &0.30$\pm$0.09  &  1.1$_{-0.7}^{+1.2}$& 190     & 105($\pm$ 41\%)      &    0.74  & -4.94\\
NGC~4696 & * &  16  &  -- & 0.63$_{-0.30}^{+0.30}$&0.81$\pm$0.05 &1.58$_{-1.13}^{+4.27}$  &3.4     &25.1($_{-55 \%}^{+56 \%}$)& 0.79$\pm$0.30& -3.97\\
NGC~5128 & C &  --  &  -- &                      &               &                        &         &                      &     4.5  & -4.61\\
NGC~5419 & E &  45  &  11 &                      & 0.78$\pm$0.11 &                        &         &  	                   &     1.6  & -3.76\\
IC~1459  &   &  80  & 2.4 & 1.5$_{-0.2}^{+0.4}$  &0.63$\pm$0.03  &0.91$_{-0.44}^{+0.48}$  &53      & 300($\pm$ 24\%)      &     7.7  & -4.44\\
IC~4296  &   &  59  &   8 & 1.6$_{-0.2}^{+0.1}$  &0.62$\pm$0.02  &9.7$_{-3.5}^{+4.5}$  &3000     &1700($\pm$ 20\%)      &     8.7  & -3.45\\
\hline      														   
NGC~0507 & * &  37  &  -- & 1.1$_{-0.04}^{+0.04}$&0.74$\pm0.04$  & 3.43$_{-1.70}^{+2.89}$ & 450     & 257($\pm$ 9\%)       & 10.2$\pm$3.37& -3.77\\
\hline \hline	   	   			       							 
\end{tabular}	
	   										
Summary of the measured Bondi accretion rates and power of the FR~I,
CoreG, and Allen sample (marked with *). Column description: (1) name, (2) quality flag,
(3) Bondi accretion radius in parsec, (4) ratio between the minimum
radius reached by CHANDRA resolution r$_{\rm{min}}$ and r$_{B}$, (5) the logarithmic slope
of the density profile, (6) temperature at the accretion radius [keV], (7) electron number density at the accretion radius [cm$^{-3}$], (8)
Bondi accretion rate [$M_\odot yr^{-1}$], (9) predicted Bondi
accretion power [$10^{43}$ erg s$^{-1}$], (10) radio jet power 
[10$^{43}$ erg s$^{-1}$] (the error on $P_J$ is assumed to be 0.4 dex, the
dispersion of the \citet{heinz07} relation linking it to the core power
except for the Allen et al. objects for which we used their published values), 
(11) logarithm of the surface brightness at 1 kpc,
${\cal S}_{1kpc}$ [counts s$^{-1}$ arcsec$^{-2}$].\end{table*}

\subsection{Accretion rate estimates and data quality}
\label{quality}

The methods presented in the previous sections were applied to every
source of our sample to estimate the accretion power.
However, the quality of the data is clearly not always sufficient to obtain 
a measure of $P_B$ at a level of accuracy adequate for our purposes. 
In the following, we describe the quality criteria that we adopted to decide
the inclusion or exclusion of a given galaxy from the analysis.

\subsection*{Quality criterion W: weak sources}

The first criterion is based on the detection of a sufficient number of
photons to obtain a reliable estimate of temperature and to perform the density
deprojection.  We require at least 100 counts in the region used for the
temperature estimate and the presence of at least 5 bins in the surface
brightness profile. For this reason, we discard 3C~075, 3C~78, 3C~189,
3C~293, 3C~442, and UGC~7203. The counts in the X-ray image of UGC~7203, only
marginally in excess of the background level, are insufficient to measure even
its brightness profile.  These 6 sources are marked with a W (weak) quality
flag in Table \ref{bondi}.

\subsection*{Quality criterion C: complex morphologies}

In some cases, the X-ray galaxy morphology is quite complex, far away from the
assumption of spherical symmetry on which we based the deprojection
technique. This is the case for three galaxies, namely 3C~084 (Perseus A),
3C~317, and NGC~5128 (Cen A) as it can be seen from their respective Chandra
images published in the literature (\citealt{fabian00};
\citealt{blanton03}; \citealt{kraft00}).  These sources are marked with a C
(Complex) quality flag in Table \ref{bondi}.

\subsection*{Quality criterion R: large r$_{min}$/r$_{B}$ ratio} 

The Bondi radius of most galaxies in our sample is typically between 10 - 100
pc, which is always unresolved by Chandra data, with the only exception of
UGC~7629 (aka NGC~4472). However, in some cases, the ratio between the minimum
radius at which we are able to measure the electron density, r$_{min}$, and
r$_{B}$ is quite large, preventing a reasonable extrapolation of the density
profile.  For example, in 3C~348, the density at a Bondi radius of 34 pc
should be obtained from a measurement at $\sim$ 10 kpc.  All galaxies (namely
3C~028, 3C~346, 3C~348, 3C~438, as well as the core galaxy UGC~9723) with
r$_{min}$/r$_{B}$ $>$ 100 are excluded from the analysis at this level, and
marked with a R flag.

\subsection*{Quality criterion E: large extrapolation errors}

We required that the error in the extrapolation of the
logarithm of the electron density to the accretion radius, 
as described in Sect. \ref{Error}, 
must be smaller 1. We discarded for this reason 3C~31,
3C~264, and 3C~346 in the FR~I sample and UGC~968, UGC~5902, UGC~7797,
NGC~1316, and NGC~5419 among the CoreG (all marked with an E).

We note that the errors in $n_e$ of the remaining sources are substantially
smaller than the adopted threshold and its precise value is therefore not
important. Similarly, we also verify that all sources excluded by the R
criterion would also fall into this flag category, since the extrapolation over
a factor of 100 in radius always produces a large error in P$_B$,
which indicate that the limit chosen for r$_{min}$/r$_{B}$ is not important.

\medskip
To summarize, we have 6 FR~I sources and 8 CoreG for
with an accurate estimate of the accretion power at the Bondi radius,
reported in Table \ref{bondi}, to be added to the 9 sources of the
\citeauthor{allen06} sample, for a total of 23 objects.

\begin{table*}
\caption{Correlations summary}
\centering
\begin{tabular}{c c c c c c c c c } \hline
Var. A    & Var. B  & N. objects    &    $r$        &   $\rho$    &  P($\rho$)      &    Slope        &  q    &   rms  \\
\hline                                         
P$_{B}$      & P$_{j}$     & 23 & 0.84   &  0.76  &   3$\times10^{-5}$   &  1.10$\pm$0.11  & -1.91$\pm$0.20  & 0.40  \\
P$_{B}^{*}$  & P$_{j}^{*}$ &  9 & 0.94   &  0.93  &   0.0003             &  1.33$\pm$0.11  & -2.18          &  0.20 \\  
kT           & P$_{j}$   & 23 & -0.04    & -0.08  &   0.73               &  -1.25$\pm$0.66 & -0.06           &  0.82 \\
n$_{e}$      & P$_{j}$   & 23 & 0.34     &  0.29  &   0.18         &  1.13$\pm$0.21  & -0.17           &  0.77 \\
M$_{BH}$     & P$_{j}$   & 23 & 0.59     &  0.60  &   0.002        &  2.01$\pm$0.23  & -17.6          &  0.66 \\
M$_{BH}$     & P$_{j}$   & 44 & 0.46     &  0.48  &   0.001        &  2.05$\pm$0.20  & -17.4          &  0.94 \\
${\cal S}_{1kpc}$& P$_{j}$& 43& 0.73     &  0.72  &   8$\times10^{-8}$   &   1.17$\pm$0.11 &  4.89           &  0.67  \\
 \hline
\label{tabcorr}
\end{tabular}

Col. (1) and (2) the correlated variables; Col. (3) the number of objects used;
Col. (4) the linear correlation coefficient $r$; Col. (5) the Spearman's rank
correlation $\rho$ and (6) the probability of no correlation between the
variables; Col. (7) the slope of the correlation; Col. (8) the intercept q;
Col. (9) the rms scatter from the correlation. * : P$_{B}$-P$_{j}$ relation
using only data from the 9 objects in \citet{allen06}.
\end{table*}

\section{The jet kinetic power}
\label{jetpower}
Several estimators of jet kinetic power in radio galaxies 
have been devised. A widely used
method is based on the link between extended radio emission and jet power
\citep{rawlings91,willott99} that however suffers from substantial
uncertainties (e.g. variations in environment, long-term time variability
effects), which are particularly severe when dealing with individual objects.
Furthermore, the connection between extended radio emission and $P_{j}$ is
calibrated only for FR~II radio galaxies and cannot be simply extrapolated to
FR~I.

Results from X-ray studies with CHANDRA and XMM-Newton demonstrated the ability
of a jet associated with an active galactic nucleus to blow cavities or
`bubbles' in the surrounding X-ray emitting-gas (e.g. \citealt{boehringer93};
\citealt{fabian00}; \citealt{forman05}). These surface brightness depressions
associated with radio sources provide us with the most robust and precise
estimation of the jet kinetic power (e.g \citealt{churazov02}), assuming that
the jet energy inflates these bubbles at the sound speed.

Measuring the pressure and density of the gas inside the X-ray bubbles, we
can calculate the work done  by the jet on the surrounding medium to inflate
the cavities ($W\propto pV$). There are different sources of uncertainty in
the measurements of cavity energy: the projected bubble size and the shape and
composition of the contained plasma. The energy output is probably
underestimated for adiabatic losses, cavity disruption, undetected cavities,
and the omission of shock energy (e.g. \citealt{nusser06};
\citealt{binney07}). All these factors provide an uncertainty in the jet power
of at least a factor of a few. 
Unfortunately, only a limited number of
objects clearly shows surface brightness depressions associated with a radio
source.

\citet{heinz07} derived a relation between the radio core luminosity $L_\nu$
and the jet kinetic power $P_{j}$ using a sample of 15 galaxies that showed
jet-driven X-ray cavities. The slope of the relation is determined by
theoretical assumption about the jet emission, modeled by a power-law
spectrum, while the normalization is obtained empirically. The possible
drawbacks of this approach are related to nuclear variability (as opposed to
the time-averaged value obtained from the X-ray cavities) and Doppler
boosting. Nonetheless, they showed that a significant relation between these
two quantities exists in the form:
$$
P_{j}=P_0 \left (\frac{L_\nu}{L_0}\right)^{\frac{12}{17}}
$$
where $P_0$ is a constant which has a value $1.0^{+1.3}_{-0.6}\;10^{44}\,$ erg
\,s$^{-1}$, when $L_0$ is fixed at $L_0=7\;10^{29} $ erg\,Hz$^{-1}$ s$^{-1}$.
This relation then provides us with a simple and rather robust (with a scatter
of 0.4 dex) estimate of the total jet power based only on the measure of the
core power, a quantity available for all sources of our sample, with the only
exception of 3C~028.  

We note that since the research by \citet{heinz07}, two further measurements
of $P_{j}$ based on X-ray cavities have been published \citep{shurkin07}.
These two objects (namely NGC~1399 and NGC~4649) closely follow the $P_{j}$ -
L$_{core}$ relation, providing further support of its validity.  Given the
good agreement we do not need to re-estimate the value of $P_0$.
 
\section{Results}

\subsection{The connection between jet and accretion power}
\label{thecorrelations}

\begin{figure}
\centerline{
\psfig{figure=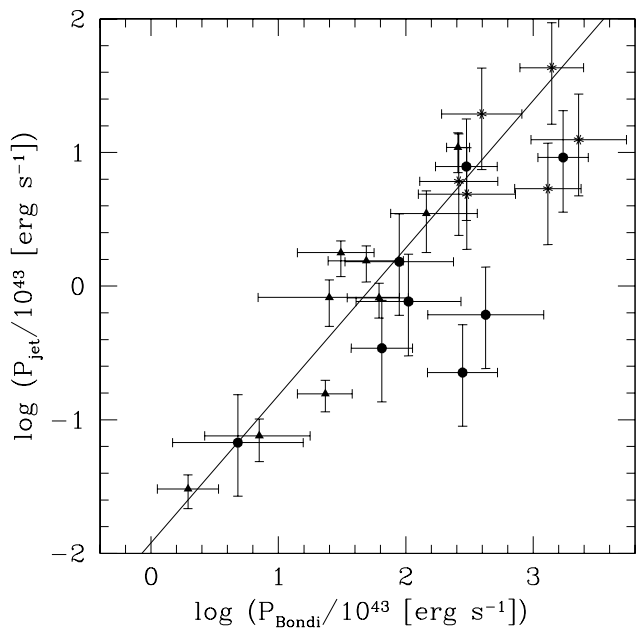,width=1.0\linewidth}}
\caption{Bondi accretion power versus jet power. The stars represent FR~I
  objects, the circles are the CoreG and the triangles represent the objects
  from \citeauthor{allen06} sample. We also plot the best-fit linear
  relation(solid line).}
\label{PbPj}
\end{figure}

The estimated accretion rates at the Bondi radius in the galaxies of our sample are in
the range from $\sim 3 \,\times\, 10^{-4} - 4\,\times\, 10^{-1}$ M$_{\odot}$ year$^{-1}$,
corresponding to an available accretion power of $\sim 2\,\times\,10^{42} - 2\,\times\,10^{45} $erg s$^{-1}$ 

In Fig. \ref{PbPj}, we compare the estimated jet
power and Bondi accretion power for the 23 objects for which
we were able to measure this quantity\footnote{With respect to \citeauthor{allen06} we exchanged the variables
on the two axis and used P$_{B}$ = $\dot{M}_{B}$c$^{2}$  directly without any
assumption about the efficiency of conversion of rest mass into energy.}. The
Bondi accretion power and the jet power are strongly correlated,
with a Spearman's rank coefficient of $\rho = 0.76$. The probability that this
is produced by from a random distribution is only 3$\times$10$^{-5}$. The linear
correlation coefficient is instead $r$ = 0.84 (see Table \ref{tabcorr} for a
summary of the statistical analysis).  The best-fit linear relations 
were derived to be
the bisectrix of the fits using the two quantities as independent variables;
this followed the suggestion of \citet{isobe90} and is the preferable for
problems that require the symmetrical treatment of the variables.
The best-fit linear relation, weighted for the errors of both the quantities,
is 

\medskip
\centerline{log(P$_{j}$) = (1.10$\pm$0.11)$\times$log(P$_{B}$) - 1.91$\pm$0.20}
\medskip

\noindent
in 10$^{43}$ erg s$^{-1}$ units. The rms dispersion of the data from the fit
is 0.4 dex. Therefore, the jet power is, within the uncertainties, linearly
related to the Bondi rate over 3 orders of magnitude with a scatter of less
than a factor of 3.  Unfortunately, all but 6 FR~I had to be excluded from the
analysis of the density profile (due partly to the presence of bright point
sources and to a larger than average distance) and this prevents being able
explore the behavior of the P$_{B}$ vs P$_{j}$ relation at the high end of
radio luminosities of the sample.  The efficiency of the conversion between
accretion and jet power is 1.2$^{+0.7}_{-0.4}$ \%. We note that the slope found
by \citeauthor{allen06} corresponds, in our notation, to 1.30$\pm$0.34 with a
normalization of 2.2$^{+1.0}_{-0.7}$ \% for P$_{j}$ = 10$^{43}$ erg s$^{-1}$,
consistent within 1 sigma confidence level with our values.

Since the Bondi accretion rate depends on the black-hole mass, the central gas
density and temperature, it is important to assess whether the relation
between P$_{j}$ and P$_{B}$, discussed above, is driven mostly by only
one of these quantities. To estimate these dependences, we
compare the jet power with each of these variables separately.

\begin{figure*}
\centerline{
\psfig{figure=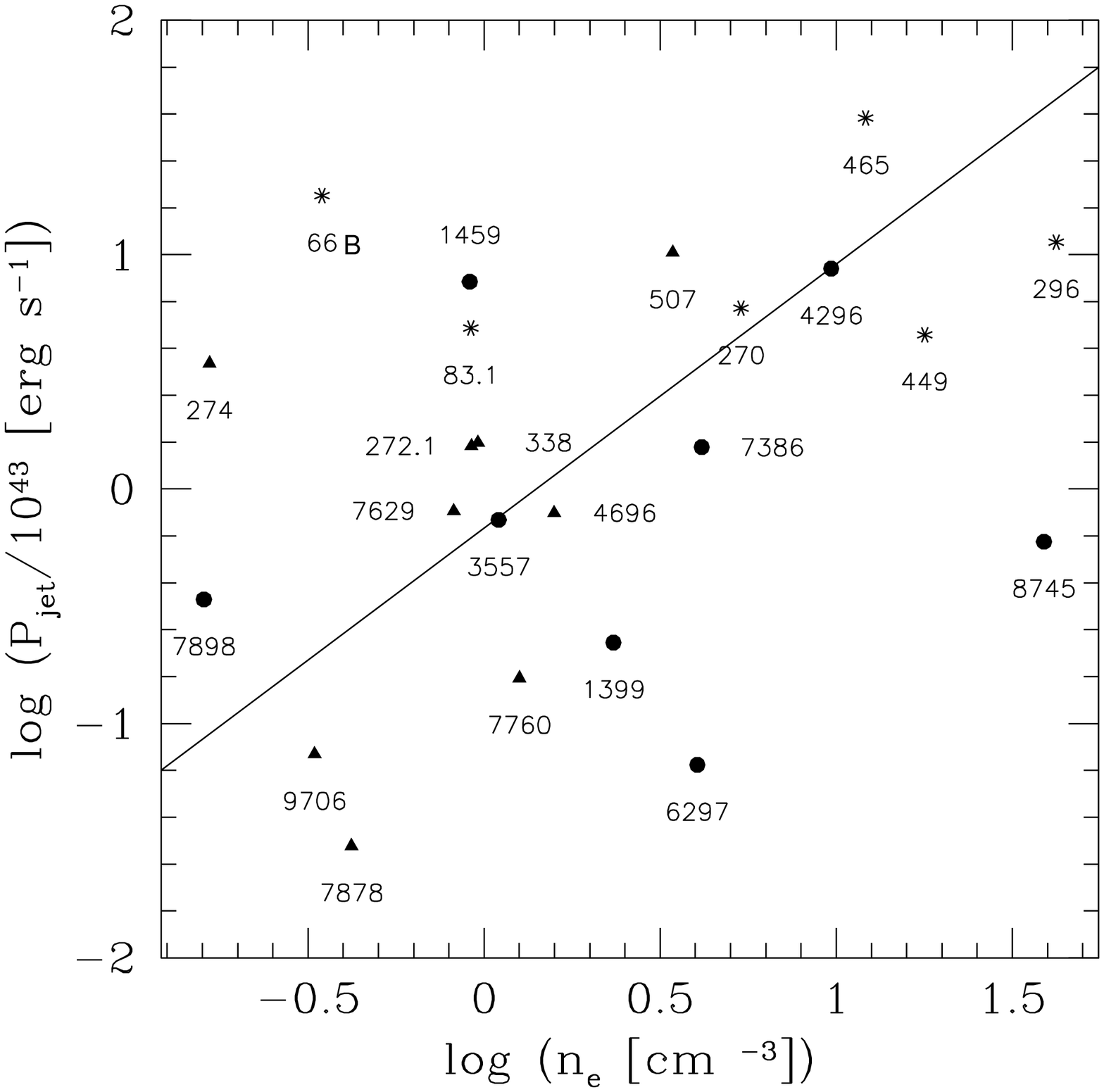,width=0.33\linewidth}
\psfig{figure=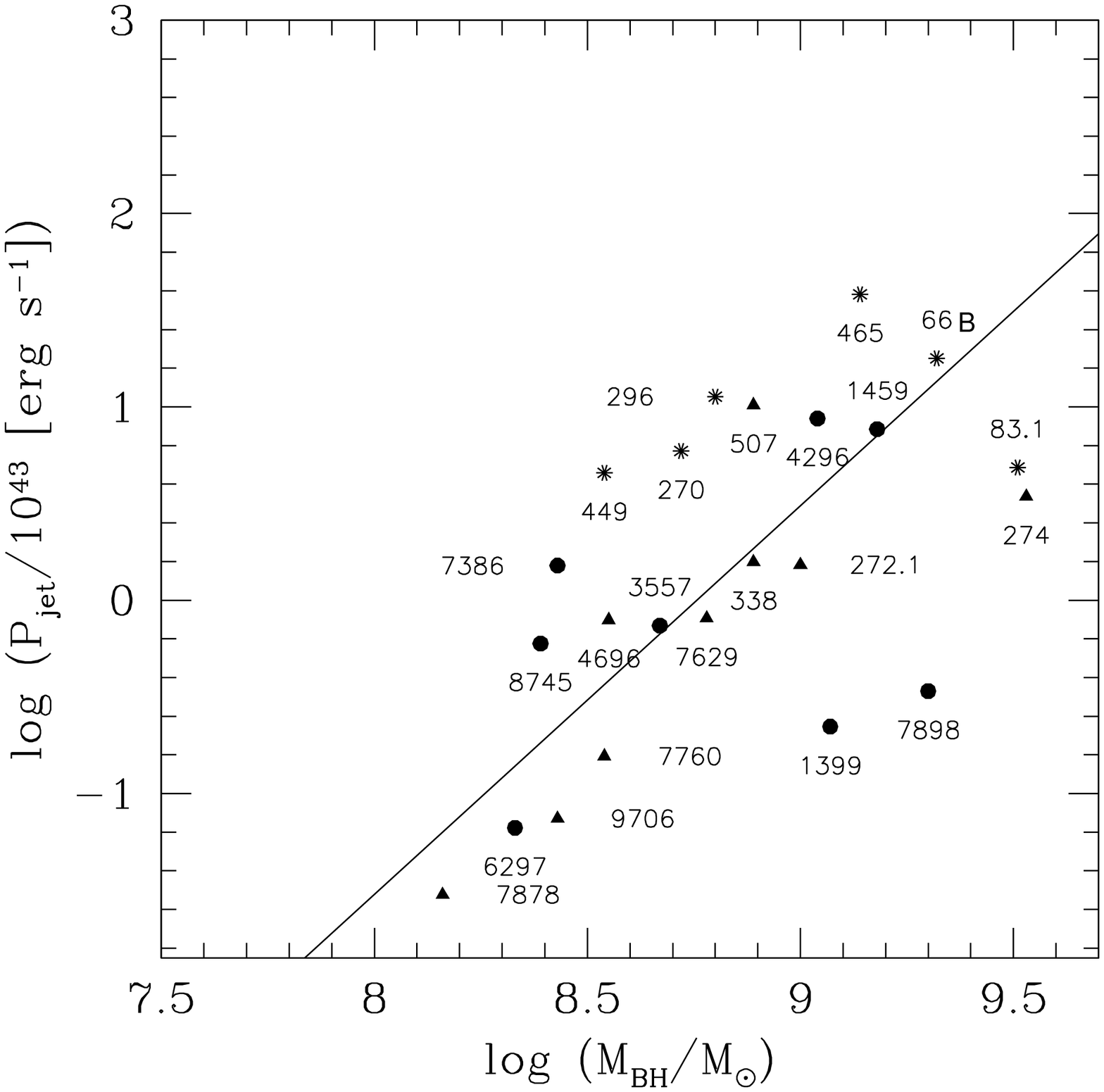,width=0.33\linewidth}
\psfig{figure=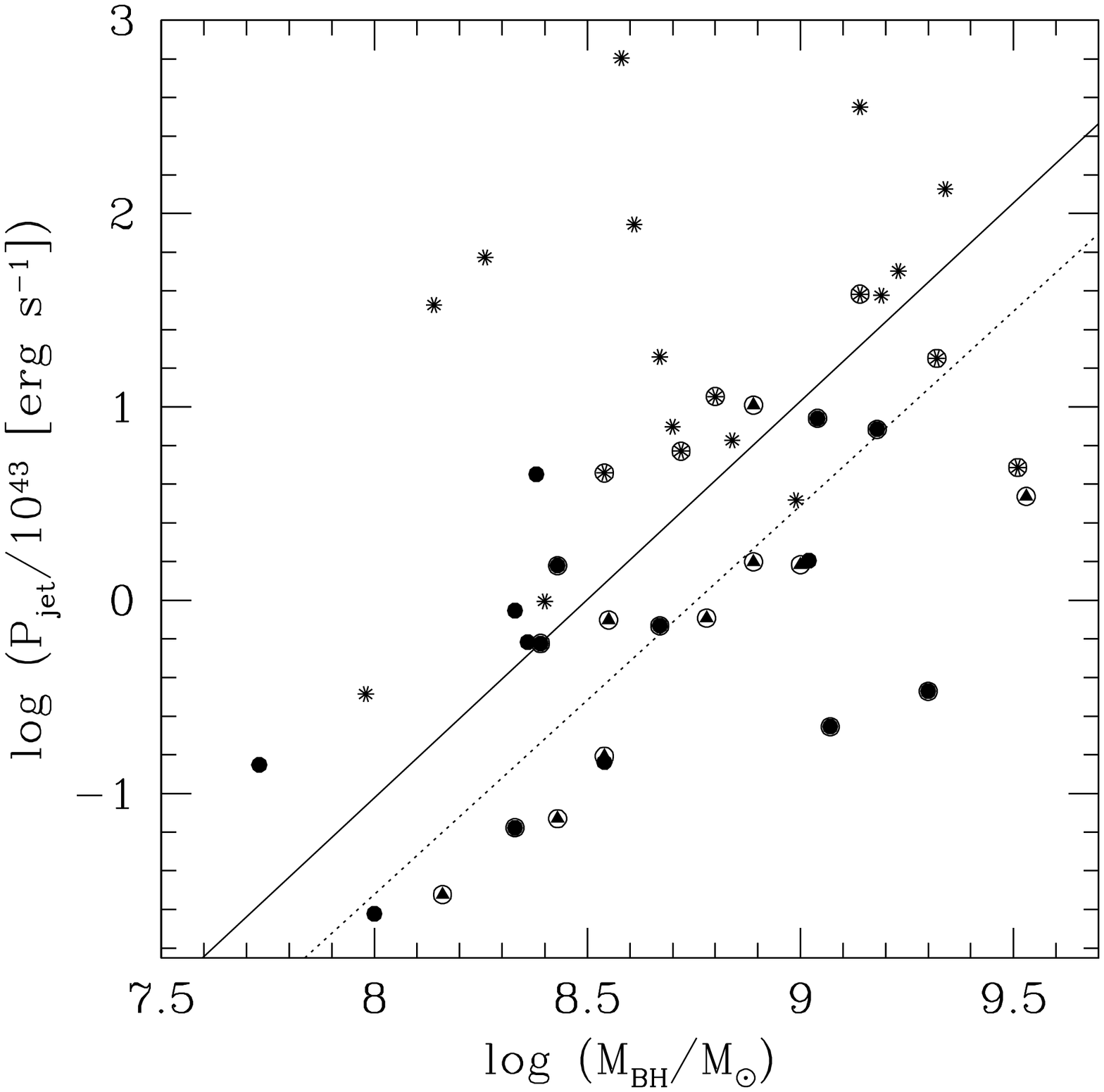,width=0.33\linewidth}}
\caption{The electron density at $r_B$ versus the jet power (left panel) for
  the sub sample of 23 objects for which we are able to measure the accretion
  power, with the linear best-fit linear relation. Middle panel) black-hole
  masses versus the jet power for this same sub sample. In the right panel, we
  compare the linear correlation (dashed line) obtained from the 23 objects
  (surrounded by a large circle) with that obtained from the entire sample
  studied (44 objects).}
\label{neMbhPj}
\end{figure*}

First, we found that there is no correlation between the jet
power and the temperature at r$_{B}$. 
Secondly, we considered how the electron density n$_{e}$ at
r$_{B}$ affects the jet power. Fig. \ref{neMbhPj} (left panel) shows a
weak correlation between n$_{e}$ and the jet power with a Spearman's rank
coefficient of $\rho = 0.29$ (with an associated probability of no correlation
of 18 \%) and a linear correlation coefficient of r = 0.34.  The derived
relation has a slope of 1.13$\pm$0.21.

Thirdly, we correlated P$_{j}$ with the black-hole masses of the 23 objects
for which we were able to estimate the accretion rate: Fig. \ref{neMbhPj}
(middle panel) shows a correlation with a slope $\sim$2 and a rms scatter of
0.66 dex. However, we can consider the
measurements of mass for all 44 galaxies in the sample, regardless of
the possibility to estimate P$_{B}$.  In particular, this enables us to include
the strongest radio sources, radically changing the picture.  
The strength of the correlation between M$_{BH}$ and P$_{j}$ is strongly
reduced (see Table \ref{tabcorr} and Fig.
\ref{neMbhPj}, right panel)
using the entire sample. Most importantly, we find objects 
at a given black-hole mass with P$_{j}$ differing by 3 orders
of magnitude and indeed the rms scatter of the residual is almost of 1 order
of magnitude. This result was presented in \citet{paper1} where we
discussed this issue using the same sample but without the further requirement
of the availability of Chandra data: while CoreG define a correlation
between M$_{BH}$ and L$_{core}$ (and consequently with P$_{j}$), FR~I
radio galaxies have much larger radio core luminosities at the similar value of
M$_{BH}$ which produces a large scatter in the P$_{j}$-M$_{BH}$ plane.

We conclude that the jet power shows a connection to the Bondi accretion
rate, which is stronger than those to M$_{BH}$, n$_{e}$, or T separately. This indicates that
P$_{j}$ vs P$_{B}$ is the primary relation that does not
descend from any of these quantities separately.

\subsection{Jet power and surface brightness profiles}

The analysis presented in the previous Section revealed the presence of a
strong correlation between the jet power and the central properties of the
galaxy, in particular with the gas density and black-hole mass.  However,
radio galaxies of different power differ also ion terms of 
the global properties of their X-ray emitting hot coronae.

\begin{figure*}
\centerline{\psfig{figure=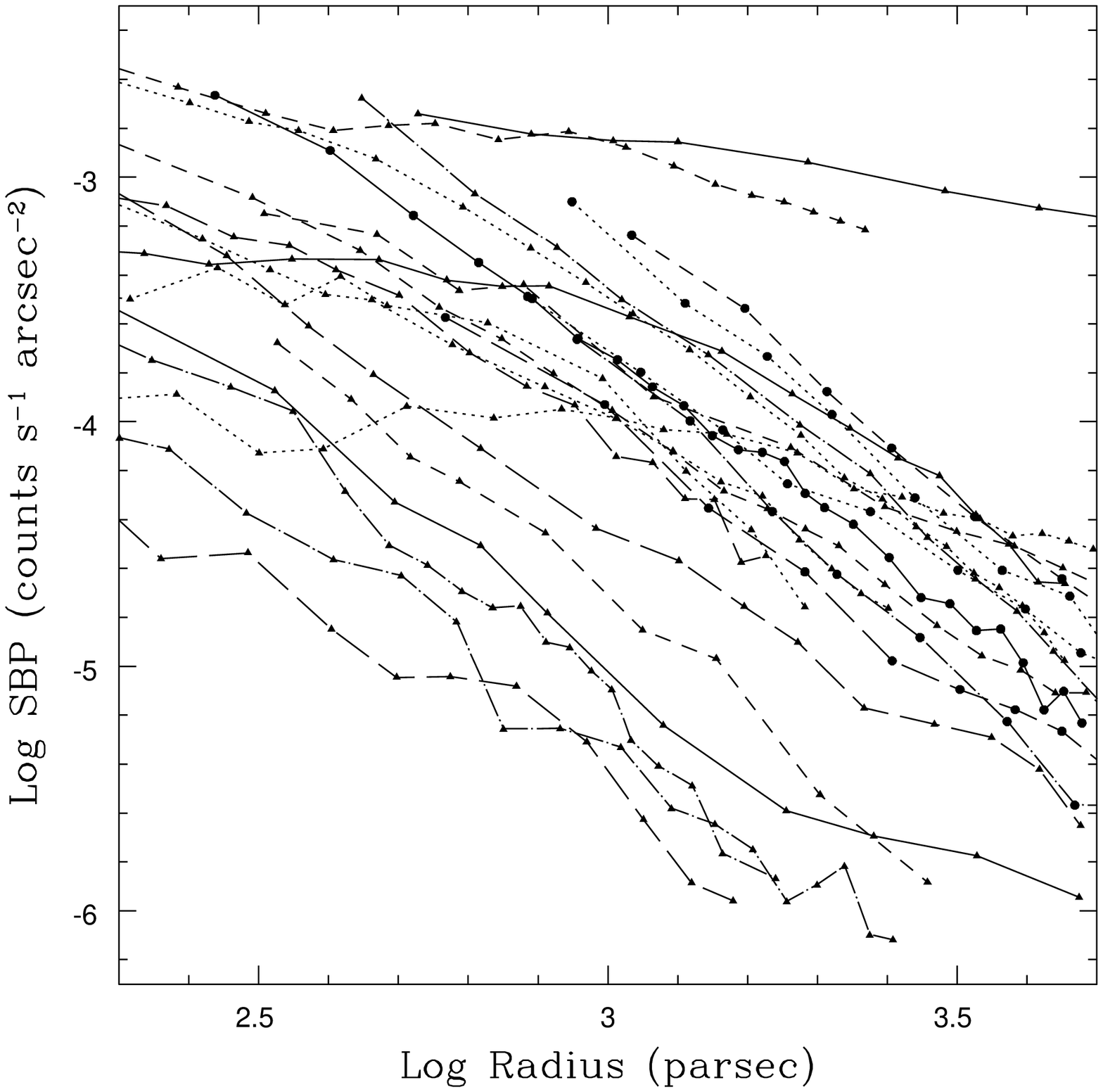,width=0.5\linewidth}
\psfig{figure=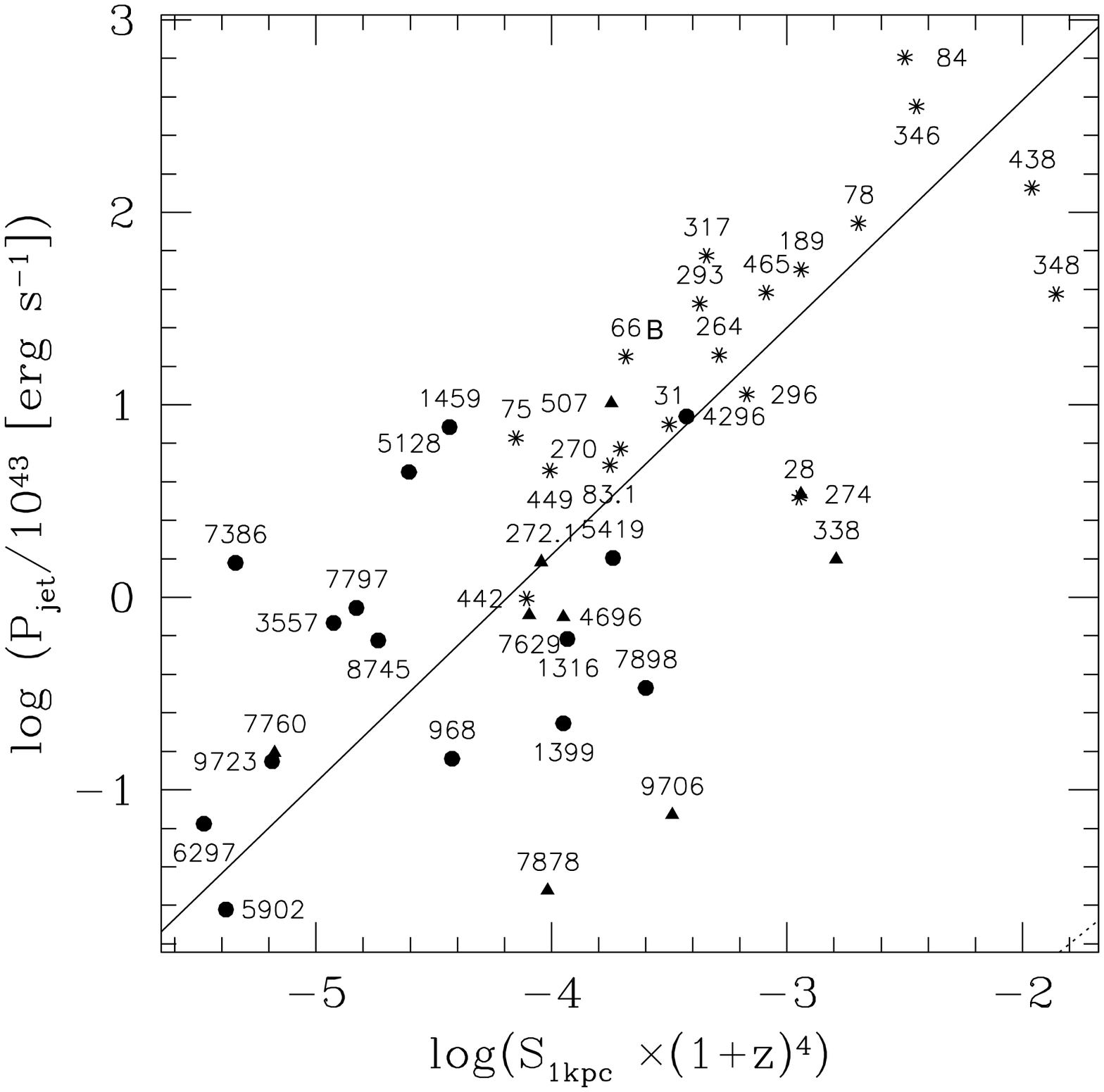,width=0.5\linewidth}}
\caption{Left panel: surface brightness profiles of the 23 objects, zooming
into the region around a radius of 1 kpc. 
They are described well by power laws, with a relatively small range of
slopes, while they differ
substantially in normalization. 
Right panel: Surface brightness at 1 kpc (${\cal S}_{1kpc}$) in units of 
counts s$^{-1}$ arcsec$^{-2}$, versus the jet power for
the extended sample of 43 galaxies. 
The solid line represents the best fit linear relation.}
\label{prof}
\end{figure*}

These differences are clearly evident in a comparison of their surface brightness profiles (shown
in Fig. \ref{prof} for the sub sample of 23 galaxies with estimated $P_B$).
Their SBP are described well by power-laws, with a relatively small range of
slopes ($\textless$$\Delta$log${\cal S}$/$\Delta$log$r$$\textgreater$ =
-1.7$\pm$0.6)\footnote{The average slope corresponds to the large scale
behavior of an isothermal $\beta$ model with $\beta=0.45$} while they differ
substantially in normalization.  The remaining sources of the sample show a
similar behavior with an even increased range of intensity\footnote{Only the
galaxy UGC~7203, showing extended emission only
marginally in excess of the background level, is excluded from this analysis}. 
To quantify these
differences, we estimated the value of the surface brightness at 1 kpc,
(hereafter ${\cal S}_{1kpc}$) a radius accessible for almost all galaxies, by
interpolating locally the profile with a power-law.  The galaxies of our
sample cover 3.5 orders of magnitude in ${\cal S}_{1kpc}$ (see Table
\ref{bondi}).
 
The intensity of the brightness profile at 1 kpc turns out to be strongly
correlated with the radio-jet power (see Fig. \ref{prof}, right panel).  The
correlation coefficient of this relation is 0.73, the rms is 0.67, and the
Spearman rank coefficient is 0.72. The probability, associated with this rank
value, that this is produced by a random distribution is 8$\times10^{-8}$.
The slope of the correlation is 1.17$\pm$0.11.

A significant advantage of this approach, based simply on the surface
brightness profile, is the possibility of including in the analysis also
galaxies for which the data were of insufficient quality to estimate the
electron density at the Bondi radius. In particular, this enables us to extend
the range of jet power probed to 4.5 orders of magnitudes. This is due to two
reasons.  As already noted, on the one hand, the most luminous radio
galaxies did not, in general, have reliable density profiles due to the
contamination of the nuclear source and to their average larger distance (e.g.
3C 84 and 3C 346). On the other hand, we find that the objects with the lowest
radio luminosity correspond to hot coronae of lower surface brightness, which 
were then excluded from the analysis due to insufficient count rates in
their X-ray images (e.g. UGC 5902 and UGC 9723).

Therefore, the jet power of a given object can be predicted with
an rms uncertainty of only a factor of 5 by looking at the 
intensity of the X-ray emission produced by the galactic hot corona. 

Since the brightness profiles are directly related to the gas electron
density, the large range of brightnesses observed reflects the different
distributions of hot gas in the centers of these galaxies.
A link between brightness profile and jet power is expected because the
electron density is one of the critical parameters that, with the BH mass,
determine the accretion flow.
It must however be noted that, in the case of Bremsstrahlung
emission, the expected dependence is in the form P$_{j}$ $\propto$ ${\cal
  S}_{1kpc}$$^{0.5}$.  \footnote{ We tested that the inclusion of the effects
  on temperature of the count rates produces a marginal improvement in the
  correlation, but leaves the slope unchanged.} 

This suggests that our assumption of simple homology in the hot coronae, 
described only by the normalization of the brightness profile, is probably
an over-simplification.  Other variables can influence the relation between
P$_{j}$ and ${\cal S}_{1kpc}$, for example, a dependence of the SBP shape
(e.g. its core radius or outer slope) on luminosity. There is an
indication of the existence of a fundamental plane for the X-ray emitting
gas, linking the temperature with the surface brightness and half-light (in
X-ray) radius \citep{diehl05}. It is likely that a more robust 
parametrization of
the hot corona properties, not simply its intensity at a fixed radius, might
reveal an even stronger link with the AGN properties, but this is beyond the
scope of this paper.

An alternative explanation for the steeper than expected dependence between
brightness profile and jet power originates in a result described by
\citet{best05}.  They found that the total mass of gas cooling within an
isothermal hot corona is $\dot M_{\rm cool} \propto L_X/T$ and that this has
the same dependence on black-hole mass as the radio-loud AGN fraction $\dot
M_{\rm cool} \propto M_{\rm BH}^{1.5}$. This implies that a linear relation
exists between the cooling rate, radio-power, and the X-ray luminosity (and
hence with the surface brightness) of the hot corona.

Despite our inability to interpret in
detail the correlation between P$_{j}$ and ${\cal S}_{1kpc}$, the
phenomenological connection between the brightness profile and jet power
has several important ramifications. First of all, considering the galaxies
with the highest radio luminosity, we already noted in Fig. \ref{neMbhPj}
(right panel) that they show a substantial upward scatter in the plane
comparing radio luminosity and black-hole mass.  They effectively break the
link between L$_{j}$ and SMBH mass defined by the less luminous radio-sources.
However, Fig. \ref{prof} (right panel) shows that these bright radio sources
are associated with the brightest coronae.  Their higher radio luminosity, with
respect to other galaxies of similar black-hole mass, is driven by the higher
density of their hot ISM, producing a higher accretion rate and a more luminous
extended X-ray emission.

Furthermore, the results presented in Sect. \ref{thecorrelations} connect the
jet power with the gas properties (and black-hole mass) at the Bondi radius.
Conversely, Fig. \ref{prof} (right panel) sets a link between the global
properties of the hot coronae and the level of AGN activity. The implication
of this result is that a substantial variation in
the jet power in a given galaxy 
must be accompanied by a global change in its ISM properties.
This can only occur when a dramatic event, such as a major merger, takes
place. This is an indication that the accretion flow is stable over
significant timescales and that,
most likely, the long-term level of jet power remains 
stable over the interval between major mergers.

Finally, the correlation between P$_{j}$ and ${\cal S}_{1kpc}$ provides us
with a simple and rather robust method to estimate the jet power, with an
accuracy of better than a factor of $\sim$ 5, based only on the measurement of
the brightness of the hot gas at a fixed radius, 1 kpc for the sample
considered here.  Clearly, this is applicable to X-ray data of relatively poor
quality and/or spatial resolution that would not enable a full spectral
fitting and deprojection down to a spatial scale comparable to the Bondi
radius. Indeed, this already found a useful application in the context of this
work, as it enabled us to understand the properties of the bright FR~I.  We
expect to be able to use this method also, for example, for more distant (and
hence more powerful) radio galaxies.

\section{Constraints on the accretion process 
and on its radiative efficiency}
\label{adaf}

In \citet{paper2}, we noticed that the X-ray nuclei of CoreG have
luminosities of typically $L/L_{\rm {Edd}} \sim 10^{-6} - 10^{-9}$ expressed
in fraction of the Eddington luminosity. 
Furthermore, the tight correlations
between radio, optical, and X-ray nuclear emission 
argue in favor of a jet origin for
the nuclear emission. In FR~I radio galaxies, the dominant contribution
from jet emission is confirmed directly by the high polarization
of their optical nuclei \citep{cccpol}.
This implies that the observed nuclear emission
does not originate in the accretion process and the values reported above
should be considered as upper limits.

These results add to the already vast literature reporting emission
corresponding to a very low fraction of the Eddington luminosity associated
with accretion onto supermassive black holes. This prompted the idea that in
these objects accretion occurs not only at a low rate but also at a low
radiative efficiency, such as in the Advection Dominated Accretion Flows
\citep[ADAF,][]{narayan95}. The ADAF models have been rather successful in
modeling the observed nuclear spectrum in several galaxies, such as e.g. the
Galactic Center and NGC 4258 \citep{narayan95,lasota96}. However, even ADAF
models over-predict the observed emission in the nuclei of
nearby bright elliptical galaxies
\citep{dimatteo00,loewenstein01,pellegrini05}.

This suggested the possibility that a substantial fraction of the mass
included within the Bondi's accretion radius might not actually reach the
central object, thus further reducing the radiative emission from the
accretion process. This may be the case in the presence of an outflow
\citep[Advection Dominated Inflow/Outflow Solutions, or ADIOS,][]{blandford99}
or strong convection \citep[Convection Dominated Accretion Flows, or
CDAF,][]{quataert00} in which most gas circulates in convection eddies rather
than accreting onto the black hole.  These arguments apply also to the
galaxies considered here, since the available accretion powers determined in
the Sect. \ref{pbondi} exceed by several orders of magnitude the observed
nuclear X-ray luminosities L$_X$.  More quantitatively, the comparison of the
values of L$_X$ reported in \citet{paper2} and \citet{balmaverde06} and P$_B$
correspond to typical ratios of L$_X$ / P$_B$ $\sim 10^{-4} - 10^{-5}$.

In the case of the radio-loud objects considered here, we have further
information that can be used to constrain the properties of the accretion
process, related to the presence of jets. In fact the mass flowing
along the jet, $\dot{m}_{\rm j}$, cannot exceed the mass rate at the
radius of accretion disk, $\dot{m}_{\rm disk}$.
The jet power, for a matter-dominated jet, can be expressed as
$ P_{\rm j} \sim \Gamma \dot{m}_{\rm j} c^2$
where $\Gamma$ is the jet Lorentz factor. 
If we denote with $f$ the fraction of gas crossing the Bondi radius
that actually reaches the accretion disk 
(that is $\dot{m}_{\rm disk} = f \dot{m}_{\rm B}$)
and since the correlation between jet and Bondi power gives us
$$ P_{\rm j} \sim 0.012 \,\, \dot{m}_{\rm B} c^2$$ 
the condition $\dot{m}_{\rm j} < \dot{m}_{\rm disk}$ translates into a lower limit
to $f$ as 
$$ f > 0.012 \,\, \Gamma^{-1}$$ 
Assuming the estimate of jet Lorentz for FR~I radio galaxies,
$\Gamma \sim 5$, derived from \citet{giovannini01} this limit becomes  
$$ f \gtrsim 0.0024 $$
Numerical simulations performed by \citet{zanni07} suggest that plasma within
the accretion disk can be channeled into the jet with a rather high
efficiency, $\dot{m}_{\rm j} \sim 0.2 - 0.55 \,\,
\dot{m}_{\rm disk}$ from which $ f \sim 0.01$.

On the other hand, 
the correlation between the nuclear X-ray luminosities and the radio-core 
luminosities derived in \citep{paper2} sets a limit to the disk emission as
$$ log(L_X) < 1.40 + log (\nu L_R)$$
The X-ray luminosity can be expressed as 
$$ L_X = \eta_x \dot{m}_{\rm disk} c^2 = \eta_x f \dot{m_B} c^2 = 0.012^{-1} \eta_x f P_j
$$ where $\eta_x$ is the efficiency of conversion of rest mass into energy 
radiated within the 2-10 keV band. 
Taking advantage of the link between radio-core luminosity and
jet power discussed in Sect. \ref{jetpower}, we obtain:
$$ log L_X = 24.84 + log (\eta_x f) + 12/17 \,\, log L_R $$ 
leading to
$$ log (\eta_x f) <  5/17 \,\, log L_R -13.74  $$ 
The limit on $\eta_x f$ is most stringent for the objects of lower
radio luminosity. For example, for $L_R = 10^{27}$ erg/s/Hz
\footnote{corresponding to log $\nu L_R = 36.7$}
this corresponds to $ \eta_x f < 1.6 \,\times\, 10^{-6}$. 
When combined to the limit of $f$ obtained above, this implies
$$ \eta_x < 1.3 \times 10^{-4} \,\, \Gamma \sim 6.5 \times 10^{-4} $$

To derive the total disk radiative efficiency we have to estimate the
bolometric correction corresponding to the 2-10 keV band. This is a complex
task, since little is known about the spectral energy distribution of our
sources and, in addition, their broad-band nuclear spectrum is dominated by
jet emission. \citet{marconi04} estimated bolometric corrections at different
levels of luminosity by using suitably-built AGN templates and found that the
2-10 keV bolometric correction decreases with decreasing luminosity, down to
$\sim 8$ for a bolometric luminosity of $\sim 10^{42}$ erg s$^{-1}$.  By
adopting this value, we derive a bolometric disk radiative efficiency of $\eta
\lesssim 0.005 $.  This limit is substantially smaller than the `canonical'
value of 0.1 and of the estimates derived from matching the mass function
of local SMBH with the AGN relics, $\eta \sim 0.08 $
\citep{marconi04,shankar07}.  This is an indication that accretion in these
objects occurs not only at a smaller rate, but also at lower efficiency than
in standard accretion disks.

\section{Summary and conclusions}

We have presented our
 results concerning the connection between the accretion rate
of hot gas, estimated in the Bondi's spherical approximation, and the jet
power in a sample of low-luminosity radio galaxies.  The sample was formed by
combining radio galaxies with FR~I morphology extracted from the 3C catalogue
and early-type galaxies with a `core' optical brightness profile.  Galaxies of
this latter sub sample have been shown to represent ``miniature''
radio galaxies, being drawn from the same population of early-type galaxies as
the FR~I hosts and differing from ``classical'' FR~I radio galaxies only in a
lower level of nuclear activity.  This combined FR~I/CoreG sample covers the
full range of radio-luminosity of FR~I, spanning five orders of magnitude.

In the Chandra public archive, there were observations available for 44 objects
of the sample. These data were used to derive profiles of X-ray brightness
and of gas temperature that, once deprojected, provided a measurement of the
electron gas density at the Bondi radius. When combined with the mass of the
central black hole, these values provided us with an estimate of the accretion
rate of hot gas, $P_B$. The quality and/or spatial resolution of the data was
sufficient to obtain a measure of $P_B$ at a level of accuracy
adequate for our purposes for a sub sample of 23
galaxies.

To estimate the jet power, we took advantage of an empirical
relation linking the jet kinetic power, $P_{j}$, to the
radio-core luminosity. This method freed us from the need of detecting
X-ray cavities inflated by a radio jet to measure its power and enabled us to
estimate $P_{j}$ with an rms error of 0.4 dex for all galaxies in our sample.

We found that a tight, linear relation links the accretion with the jet
power whose normalization sets the efficiency of the conversion
between accretion and jet power to $\sim$ 1.2\%.  These results
strengthen and extend the validity of the results obtained by \citet{allen06}
based on the analysis of 9 low-luminosity radio galaxies.

In addition, we found that the jet power is closely connected also with the
surface brightness of the X-ray emission, measured at the fixed radius of 1
kpc.  This is a simple quantity to measure; it was therefore possible to
consider in the analysis all galaxies, including those for which the data were
of insufficient quality to estimate the Bondi accretion rate. This method
could be extended to objects of the highest radio luminosity, which did not
provide useful estimates of P$_{B}$, showing that they are associated with the
brightest coronae.  This implies that the different levels of radio luminosity
are not only associated with nuclear quantities (such as the gas density and
black-hole mass) but also to global differences in the structure of the
galactic hot coronae.  A substantial change in the jet power must be
accompanied by a global change in its ISM properties, driven e.g. by a major
merger.

At the estimated accretion rates at the Bondi radius, $M_B$,
even ADAF models substantially
over-predict the observed X-ray nuclear emission
suggesting that a substantial fraction of the mass crossing the Bondi 
radius does not reach the central object. On the other hand, the mass rate at the
outer radius of the accretion disk must be sufficient to maintain the mass
flow along the jet. This requires that at least a mass 
fraction of $ \gtrsim 0.0024 M_B$ 
reaches the accretion disk.  The X-ray nuclear luminosity corresponding
to this accretion rate does not exceed the observed luminosities of the
nuclear X-ray sources (having assumed a bolometric correction of 8 for the
2-10 keV band) only when the radiative disk efficiency is $\eta \lesssim 0.005
$, substantially smaller than the `canonical' value of 0.1. This is an
indication that accretion in these objects occurs not only at a smaller rate,
but also at lower efficiency than in standard accretion disks.

The picture which emerges is that the accretion of hot gas associated with the
host galaxy corona represents the dominant process of fueling for FR~I
radio galaxies across their full range of radio luminosity, spanning from the
faintest level of detectable activity to radio-luminosities 
marking the transition between FR~I and FR~II objects.  The
differences of AGN power across this wide range of luminosity are determined
by the available supply of gas, which we found to be closely related to the
global properties of the hot corona. These results
provide a strong evidence for the presence of a feedback process linking the
host galaxy with the active nucleus since the same hot gas that provides the
supply to the supermassive black hole is also the repository of the kinetic
energy liberated into the ISM by the jets.
Clearly, it will be of great interest to
explore how the hot gas content is connected to the
evolution of the host galaxy and to its merger history. 

Another interesting issue is the fate of the gas within the Bondi
accretion radius. While on the one handm the spherical accretion provides, a
posteriori, a good description of gas inflow in the central regions of these
active galaxies, only a small fraction of the gas is allowed to reach
the accretion disk. 

At small radii, the spherical inflow must therefore break and the bulk of the
accreting gas must be removed from its large-scale motion toward the central
black hole. The
(almost) linear relation between jet power and accretion rate indicates that
the efficiency of this mechanism must have only a very weak dependence on the
amount of accreting gas. Apparently, also the fraction of
gas ultimately launched into the relativistic jet
is essentially independent of the jet power. 
Accounting for this complex phenomenology represents an important
challenge if we want to preceed in our understanding of the process of 
accretion and jet formation in active nuclei.

A crucial question that must also be addressed is which manifestation of
an active nucleus can be supported by hot accretion.  Although the galaxies
considered span a large range of radio power, it is important
to establish whether hot accretion is sufficient to power
FR~II radio galaxies of even higher luminosity or, alternatively, above some
threshold an additional source of fuel, possibly associated with an ISM phase of
lower temperature. The results presented here refer only to radio-loud AGN
that channel a significant fraction of the accretion power into jet kinetic
energy. However, hot accretion might also be important in powering radio-quiet
AGN. We will explore this possibility in a forthcoming paper.

\acknowledgements
We acknowledge the usage of the HyperLeda database (http://leda.univ-lyon1.fr).

\appendix
\section{}
\label{appendix}

\begin{figure*}
\centerline{
\psfig{figure=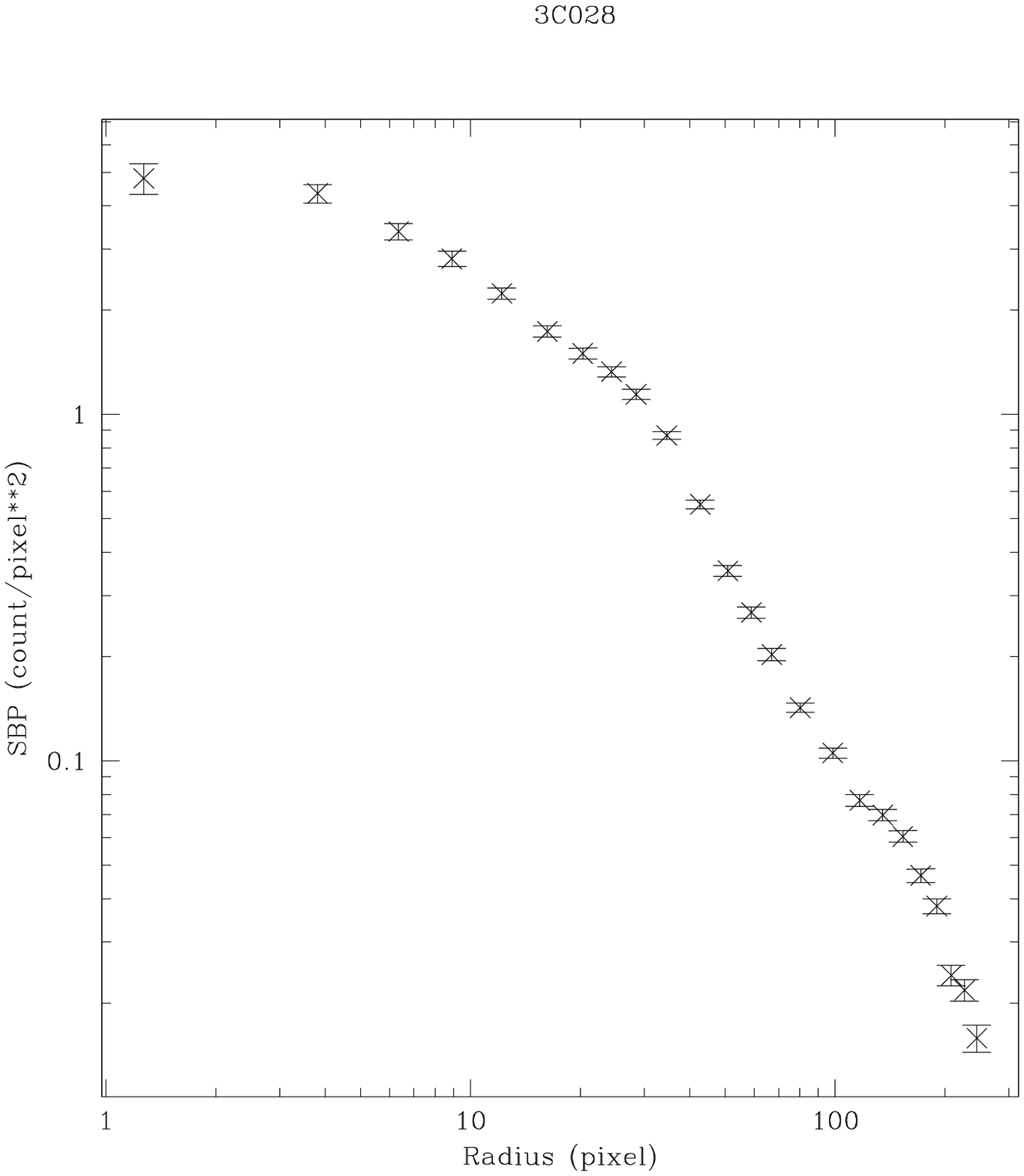,width=0.25\linewidth}
\psfig{figure=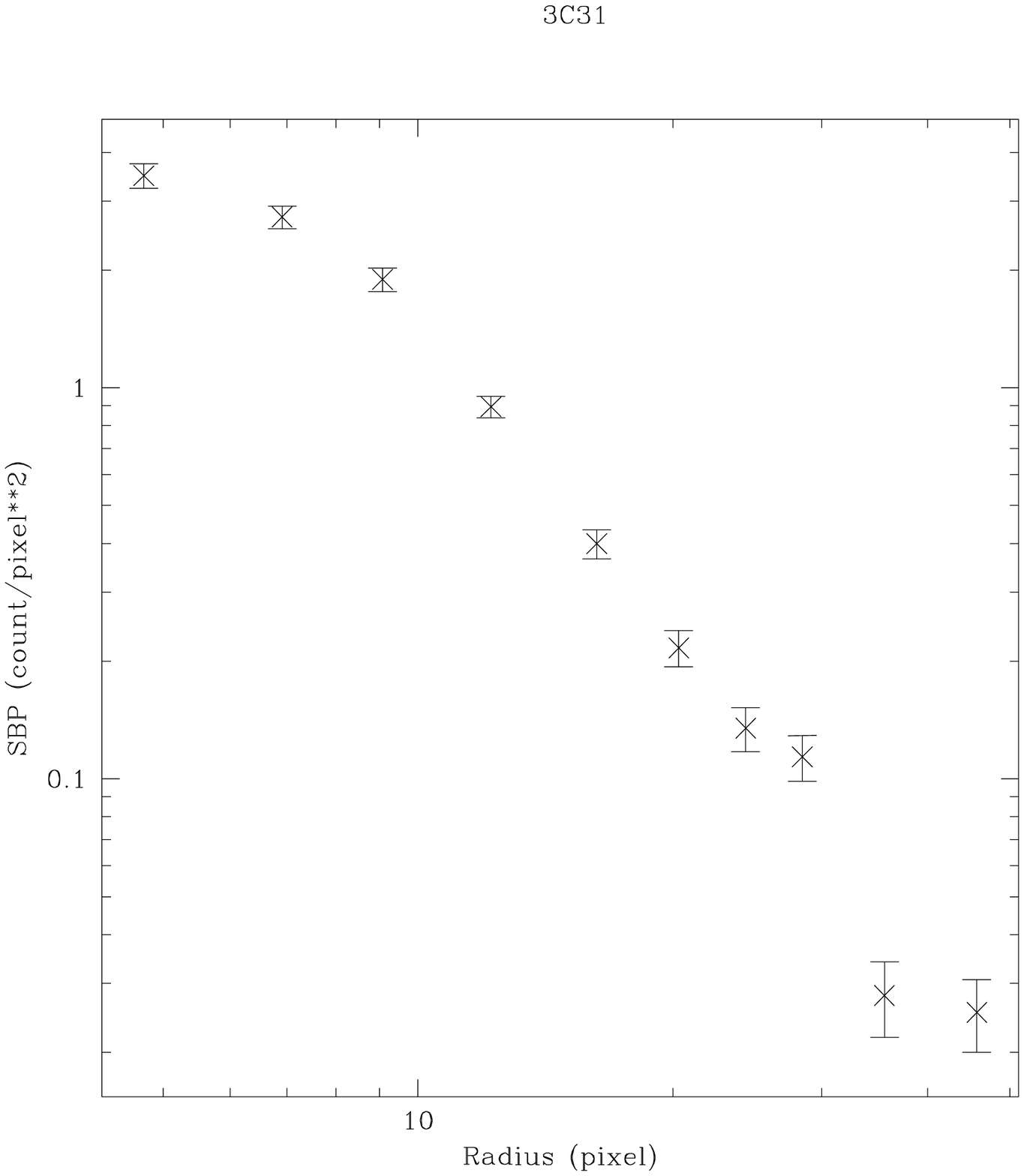,width=0.25\linewidth}
\psfig{figure=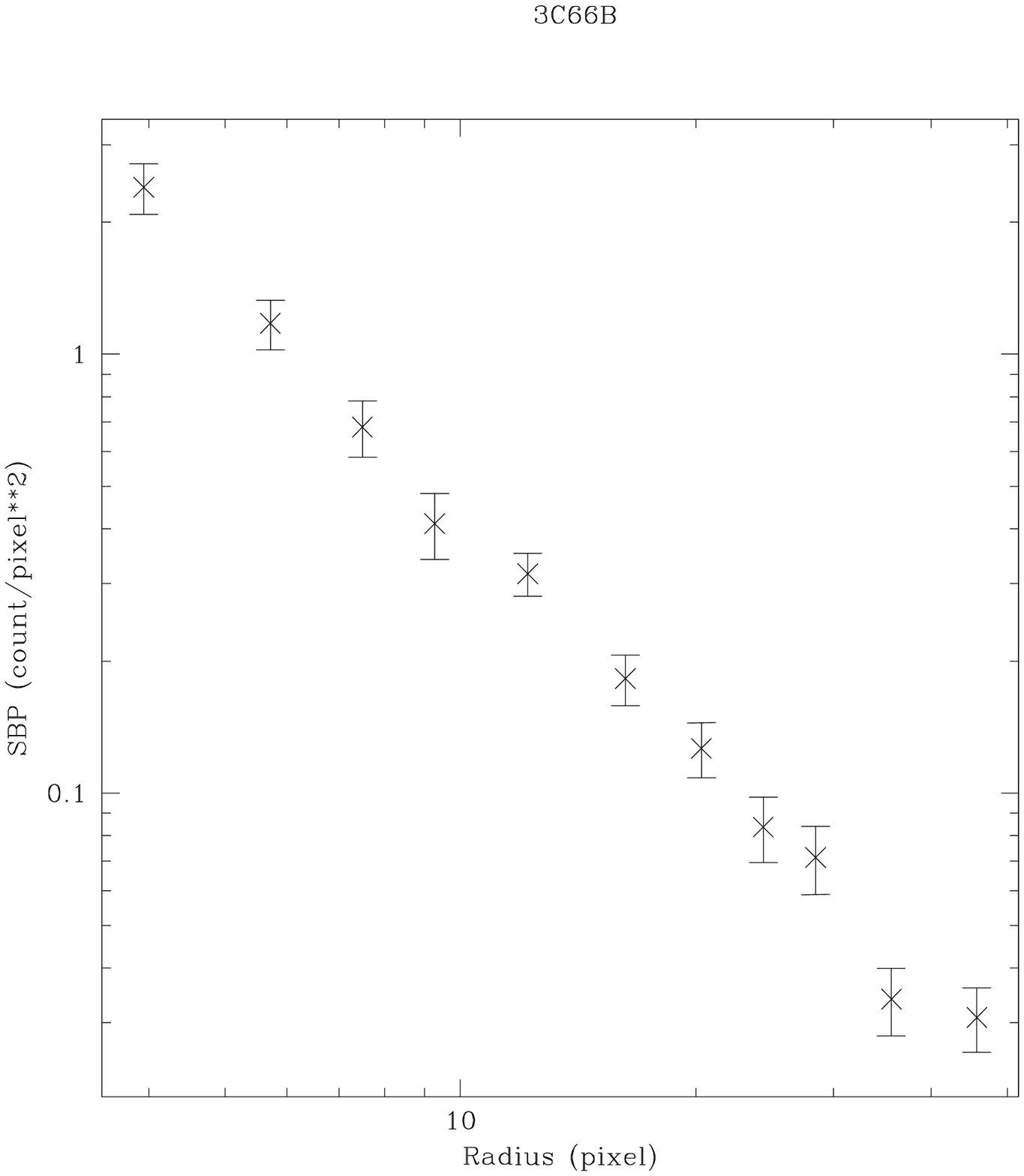,width=0.25\linewidth}
\psfig{figure=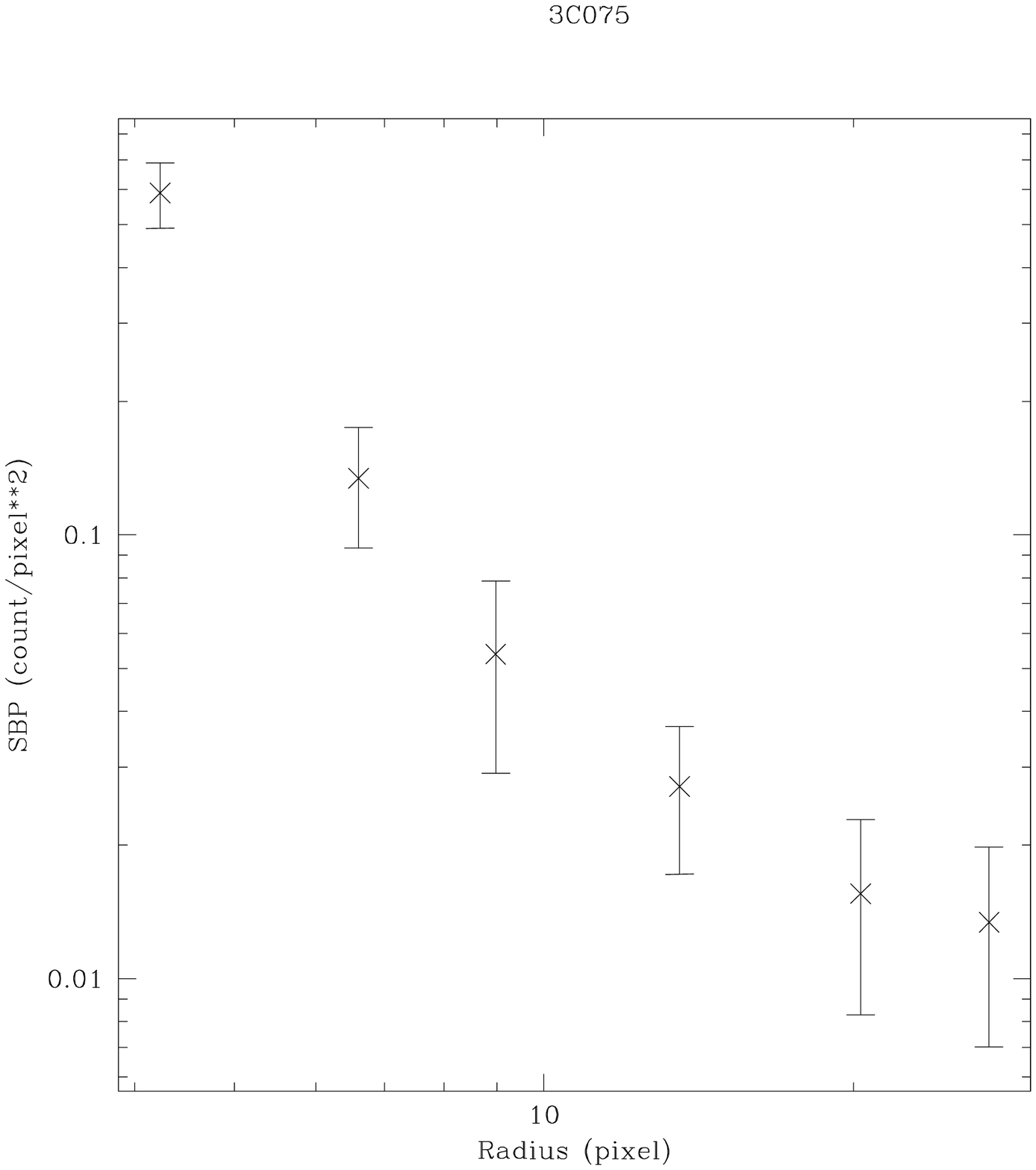,width=0.25\linewidth}}
\centerline{  			
\psfig{figure=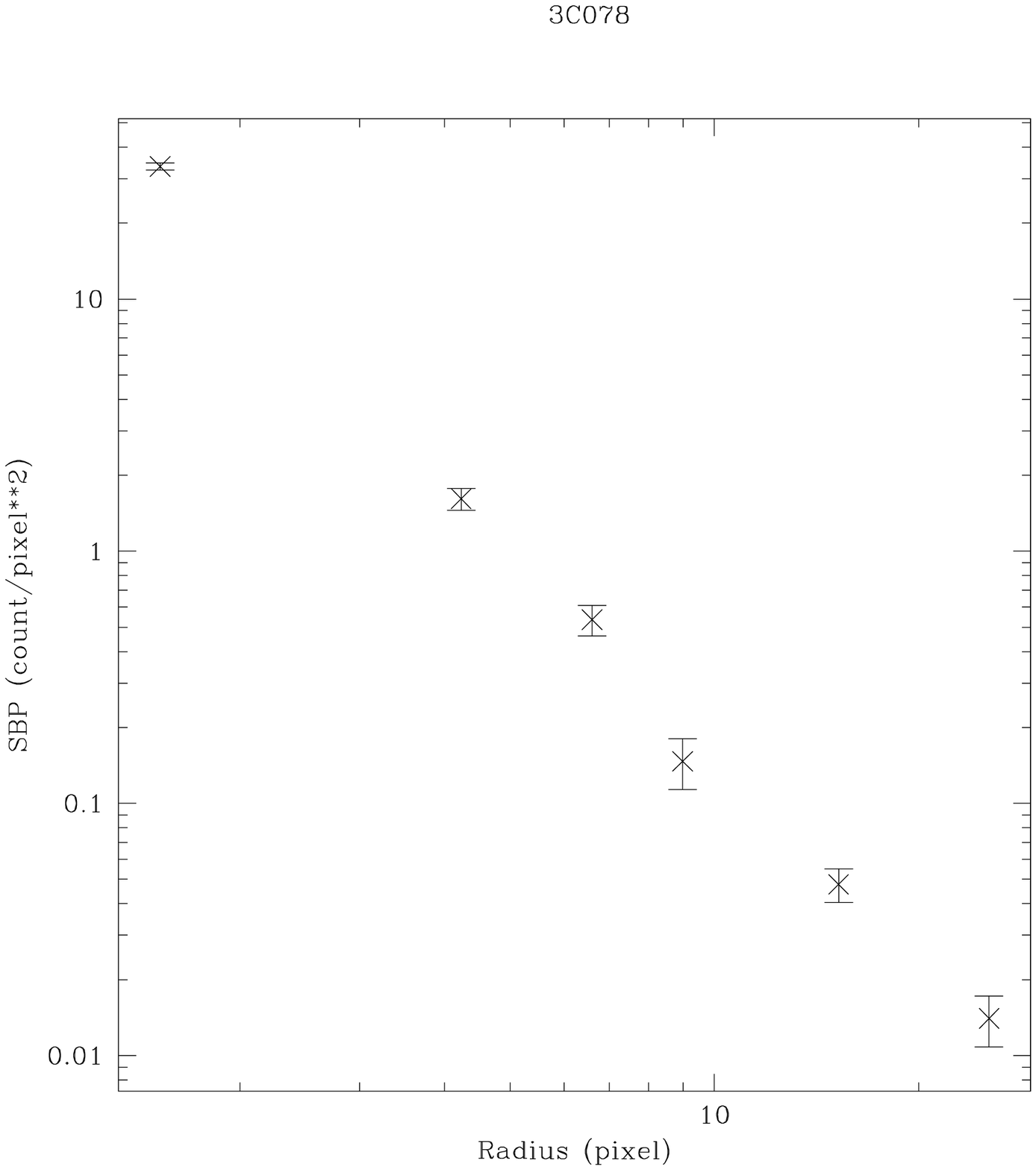,width=0.25\linewidth}
\psfig{figure=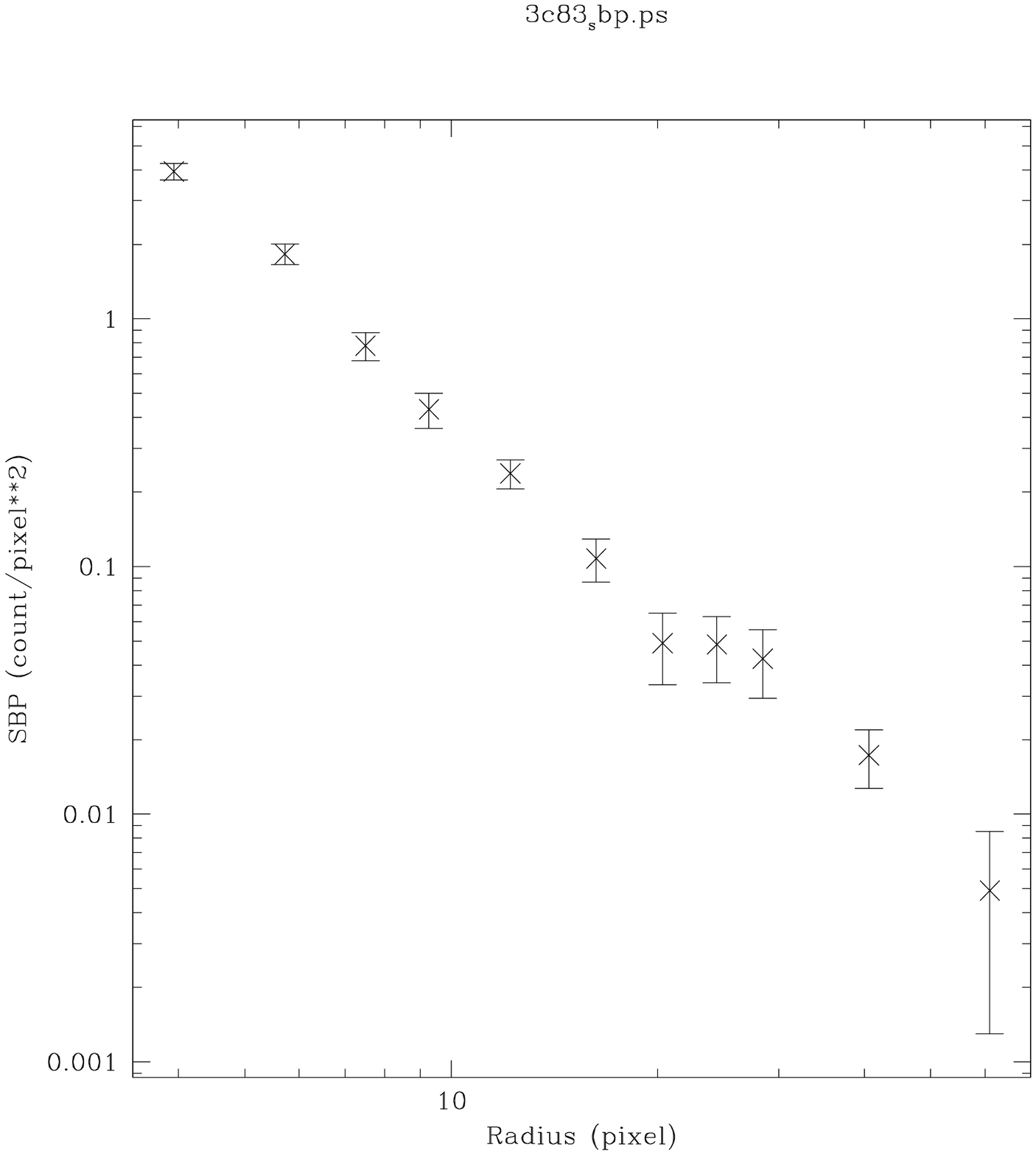,width=0.25\linewidth}
\psfig{figure=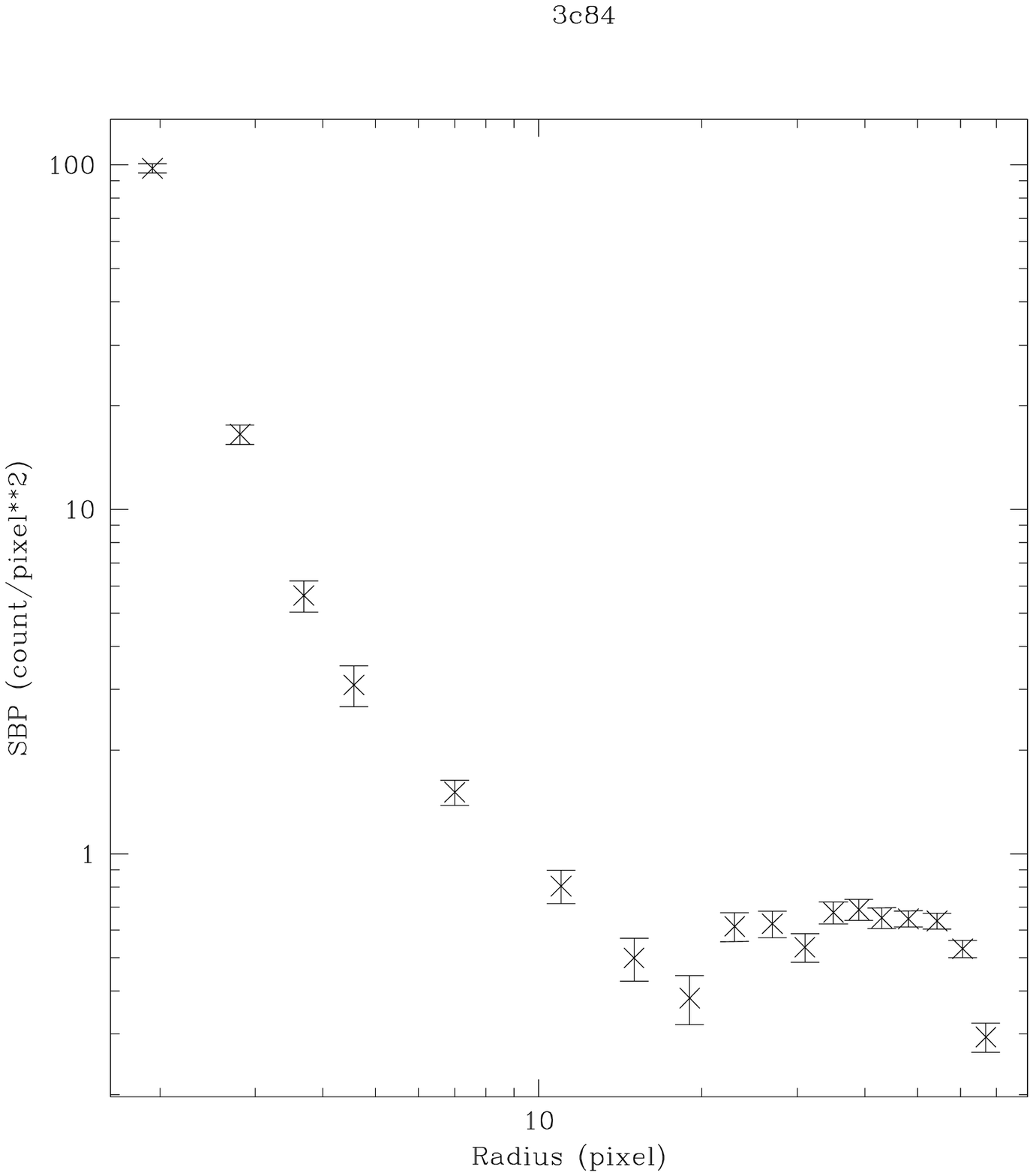,width=0.25\linewidth}
\psfig{figure=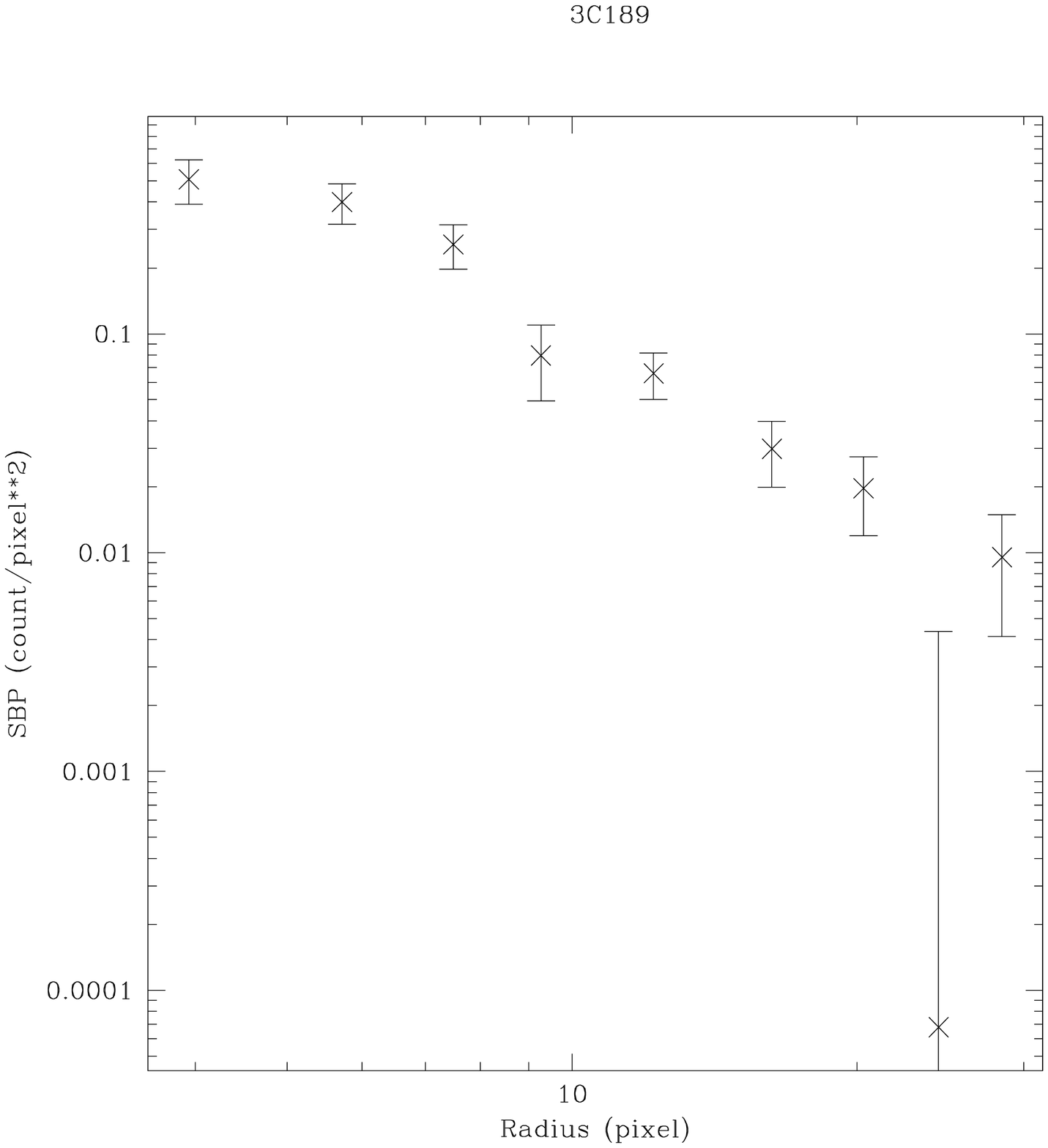,width=0.25\linewidth}}
\centerline{  			
\psfig{figure=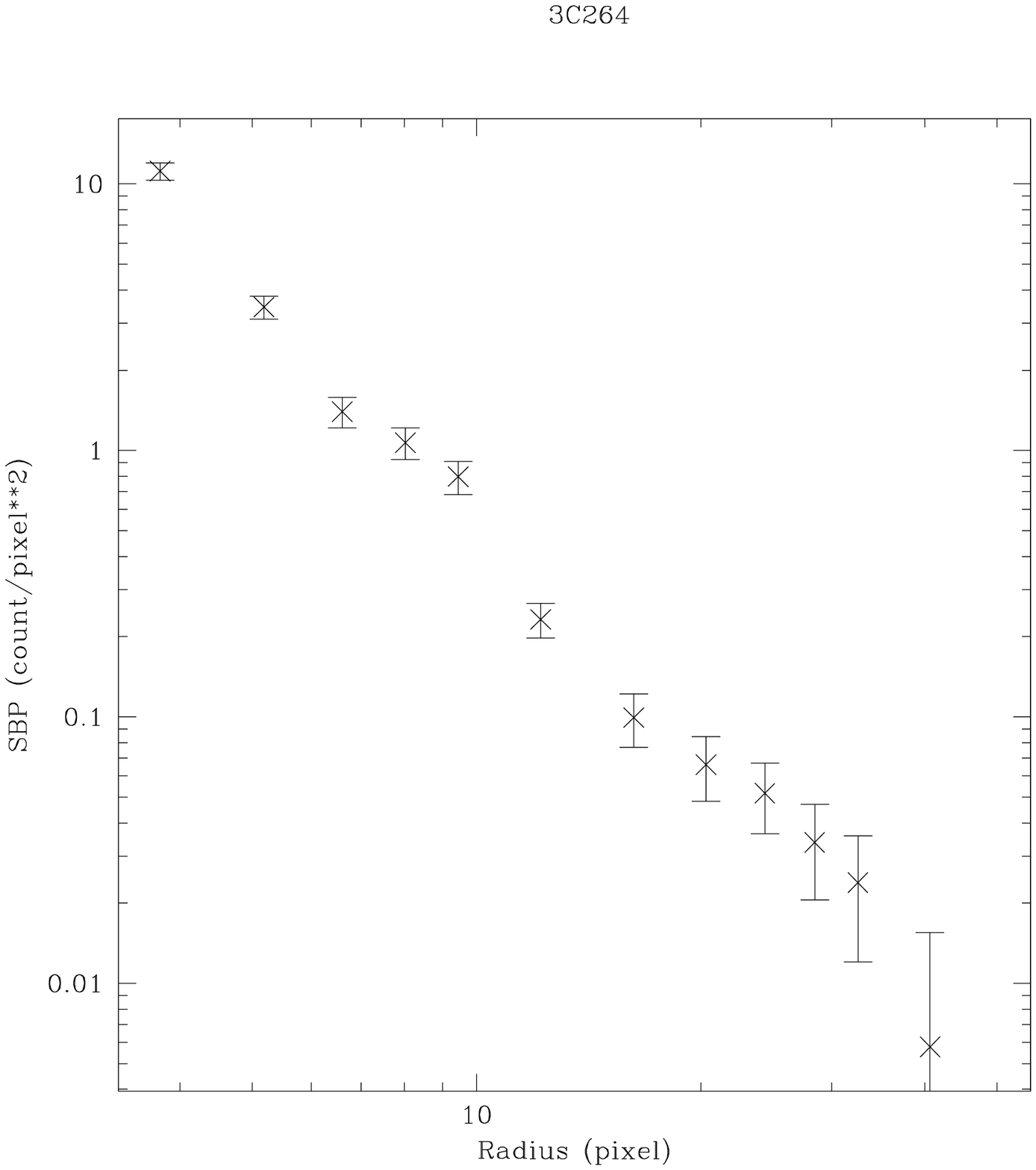,width=0.25\linewidth}
\psfig{figure=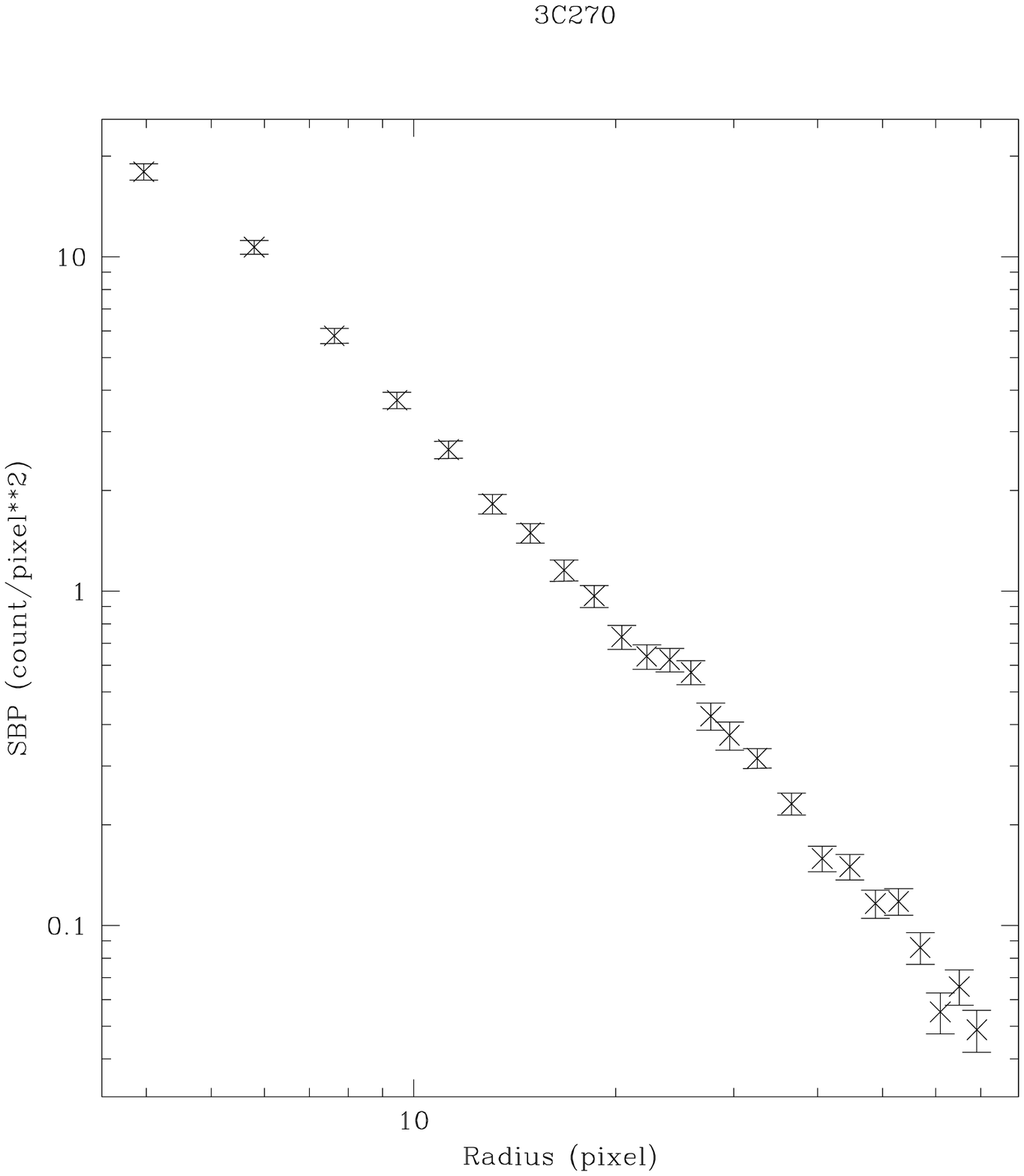,width=0.25\linewidth}
\psfig{figure=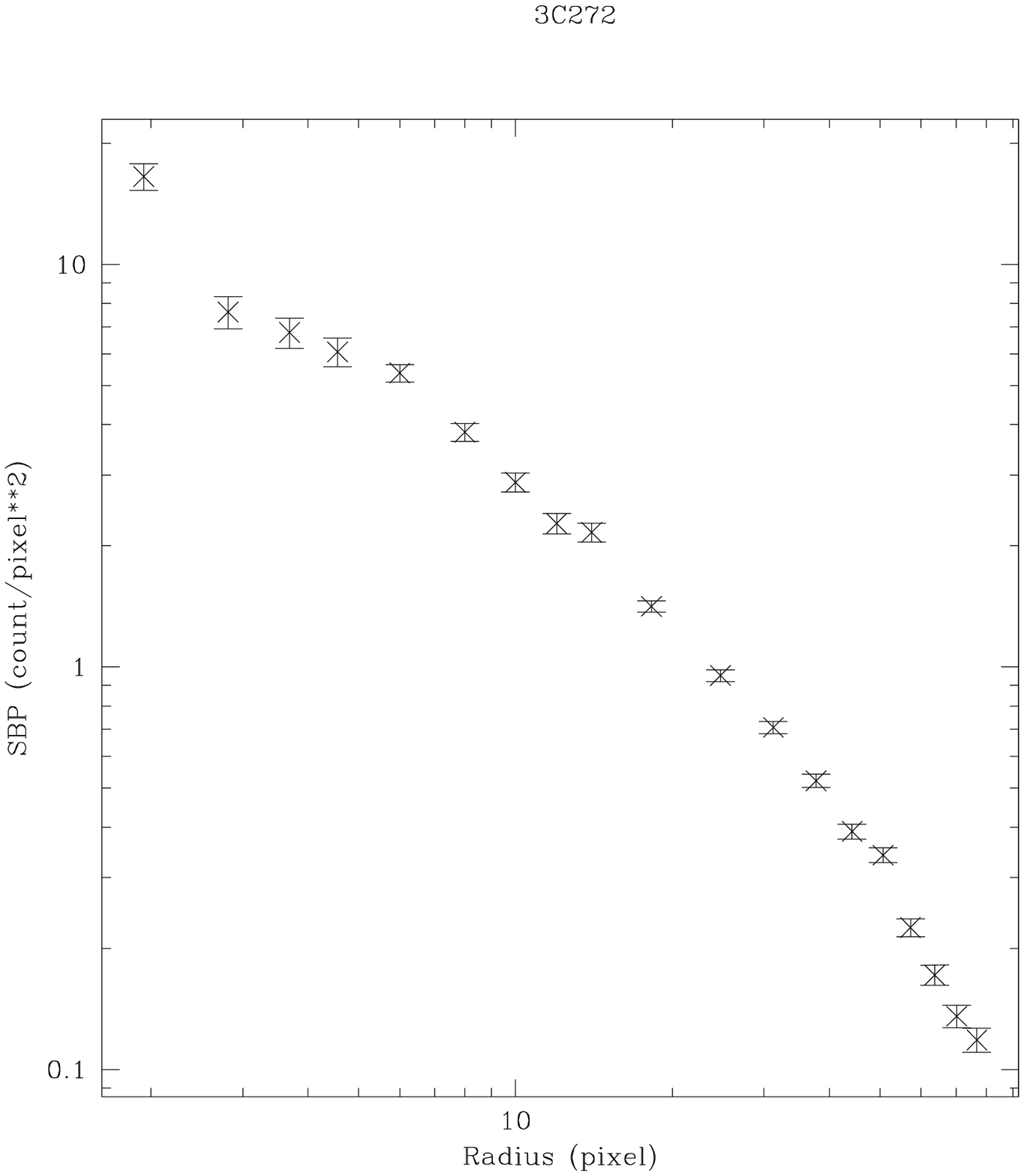,width=0.25\linewidth}
\psfig{figure=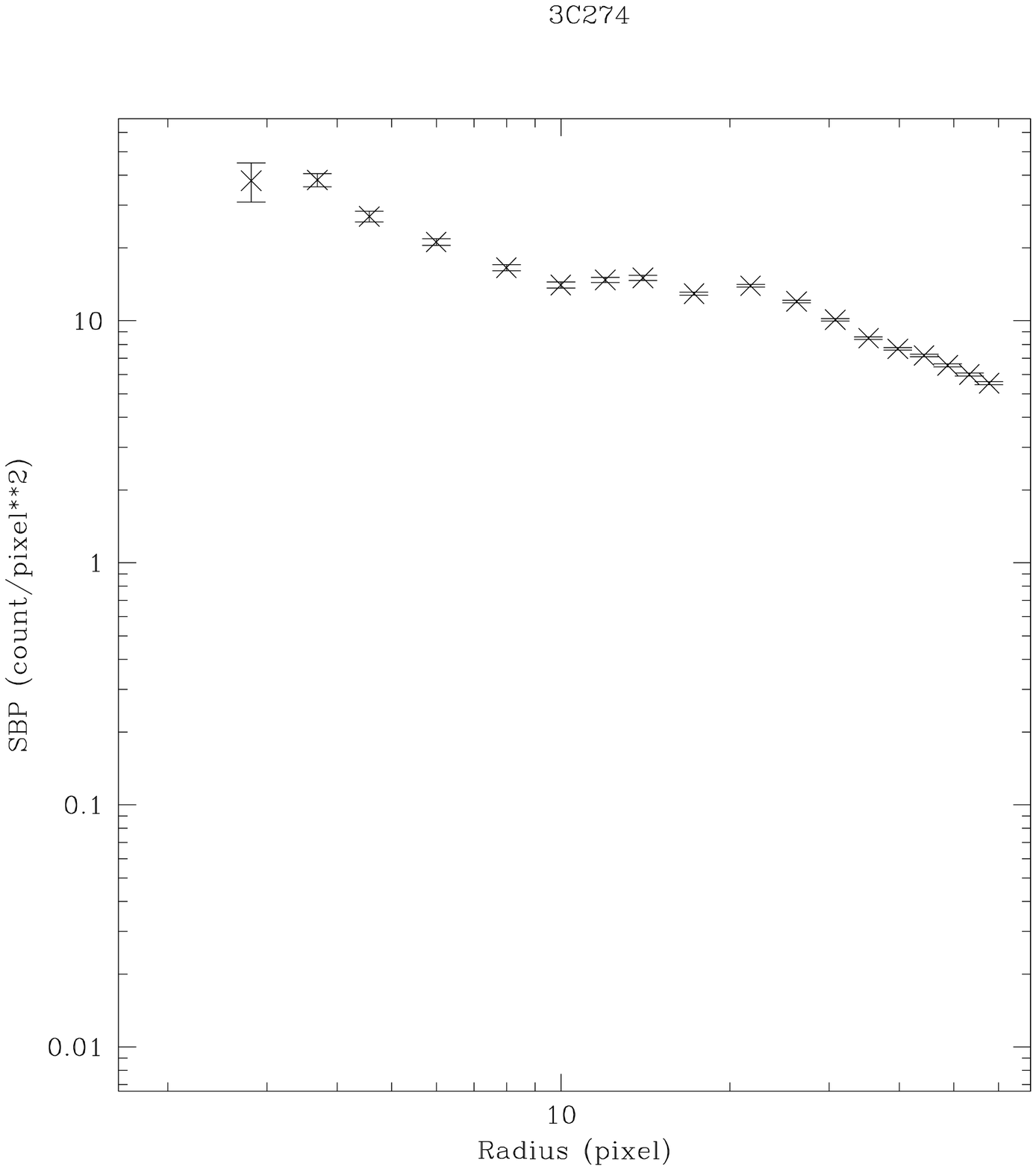,width=0.25\linewidth}}
\centerline{  			
\psfig{figure=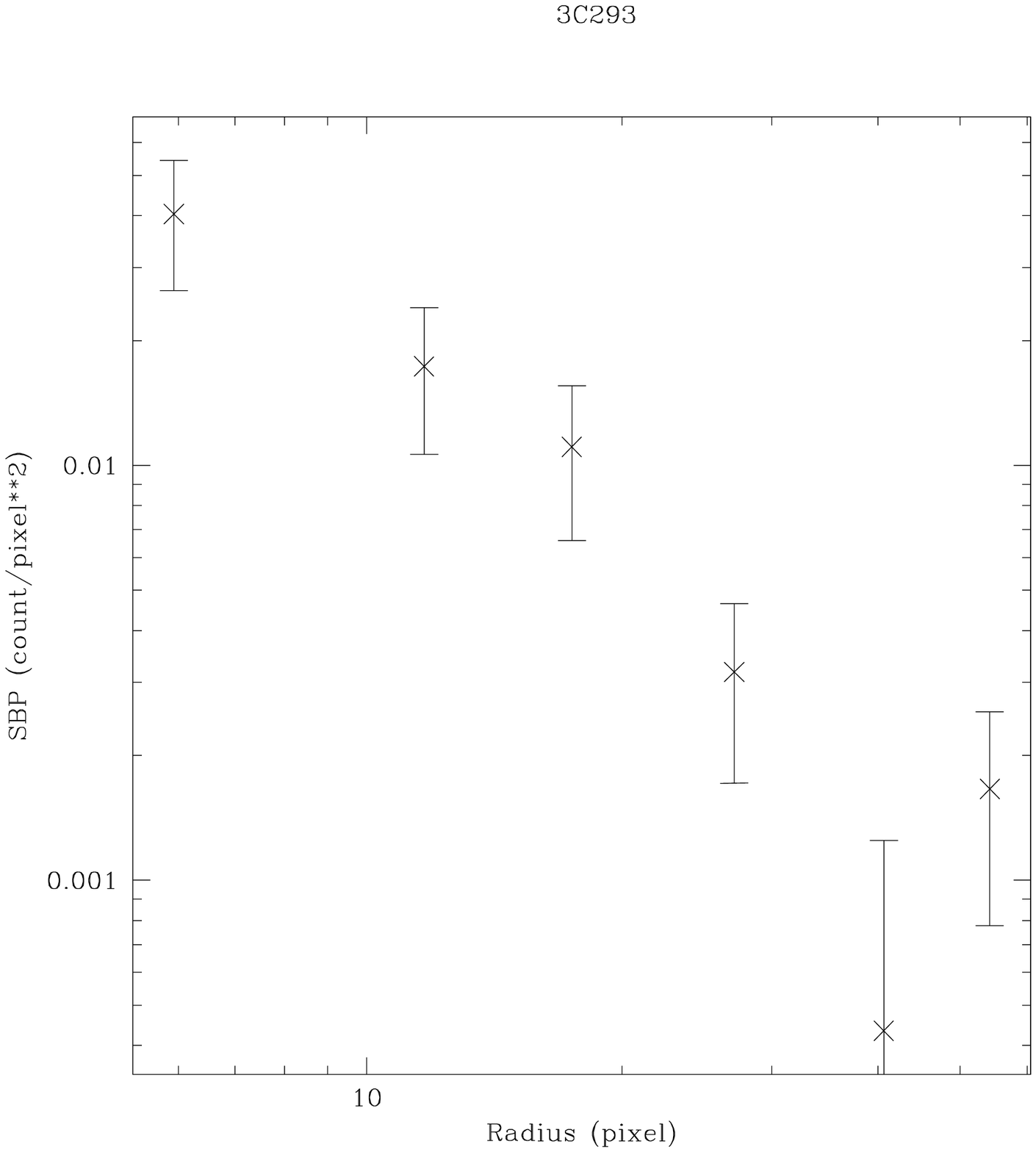,width=0.25\linewidth}
\psfig{figure=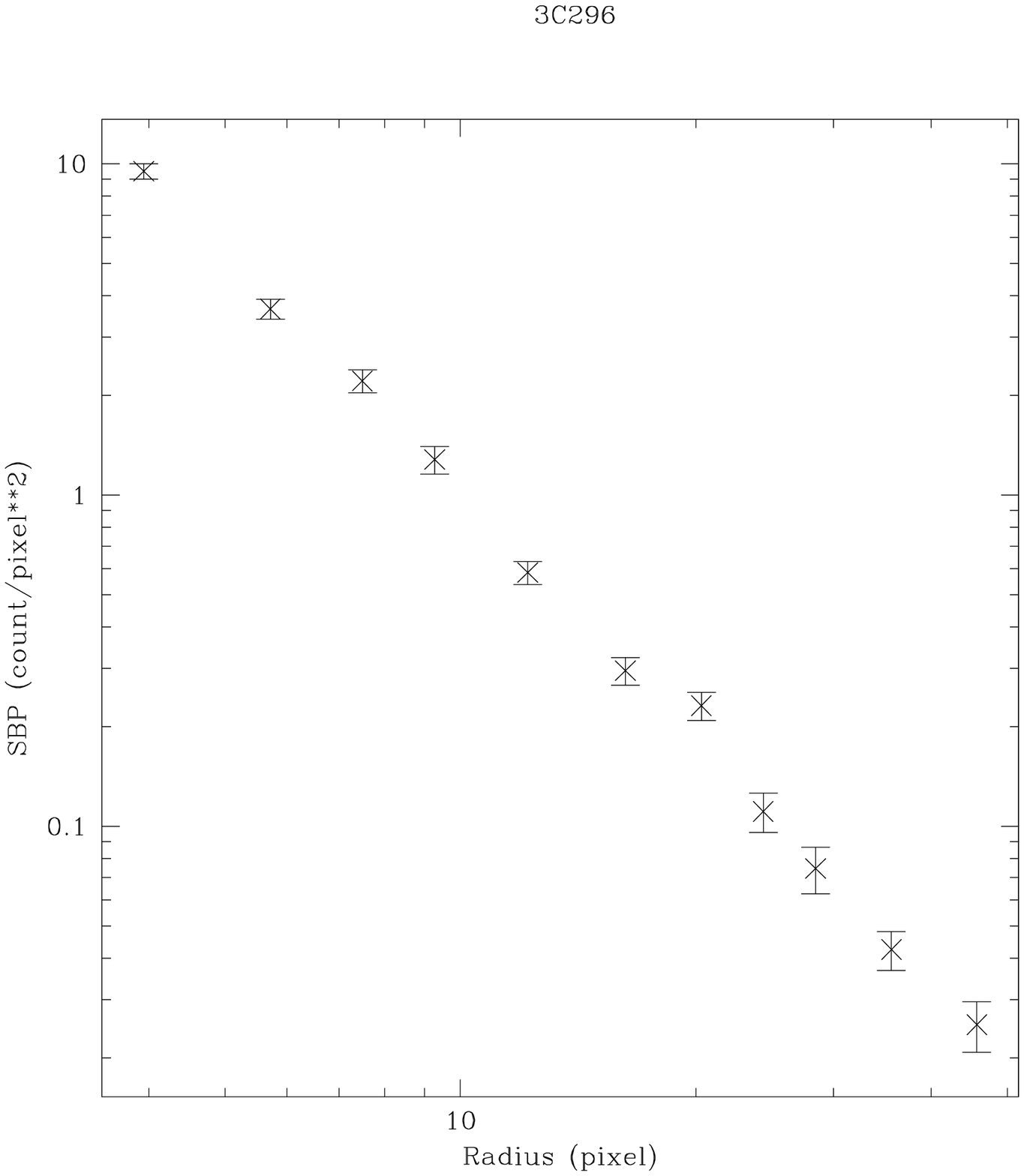,width=0.25\linewidth}
\psfig{figure=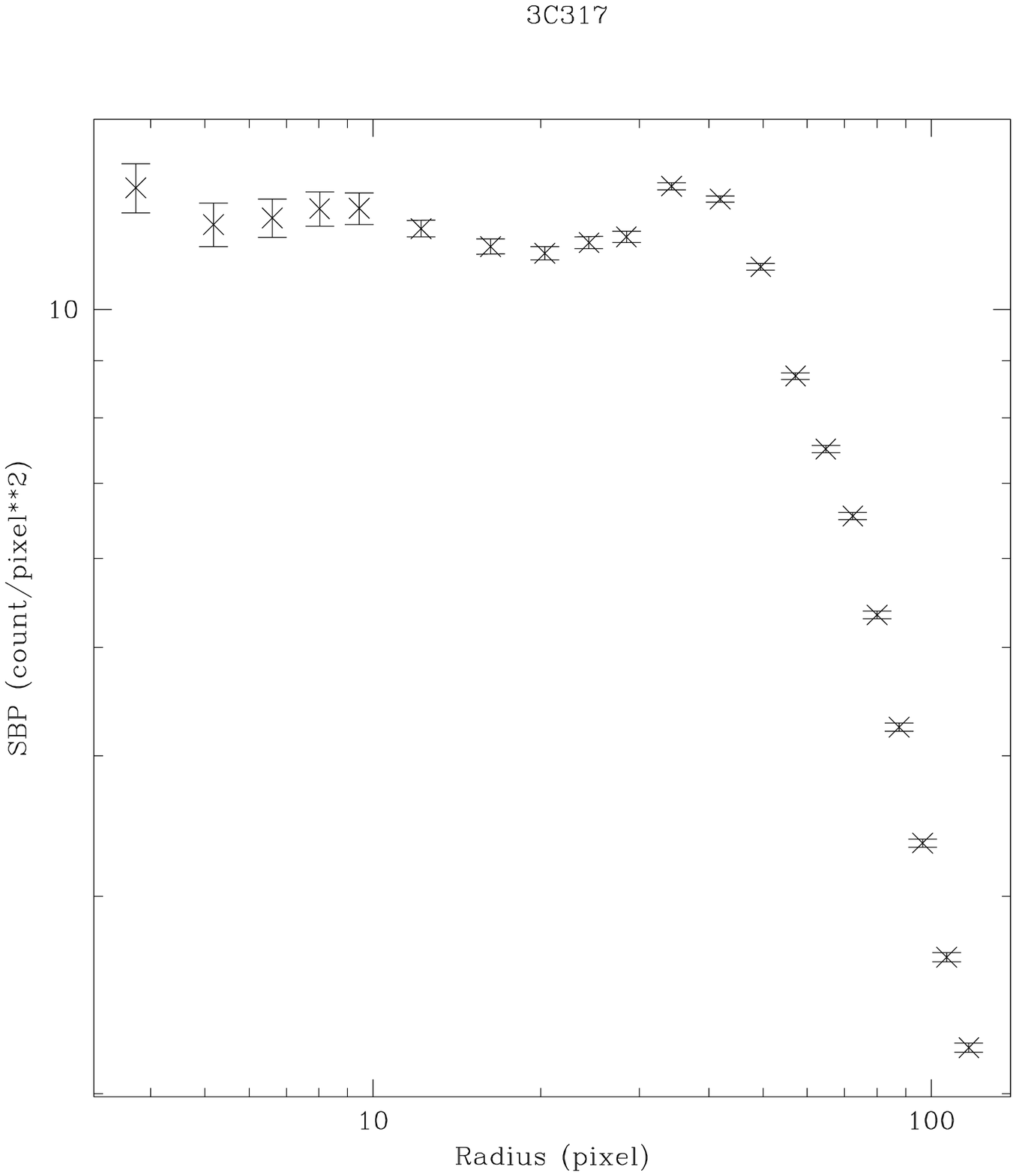,width=0.25\linewidth}
\psfig{figure=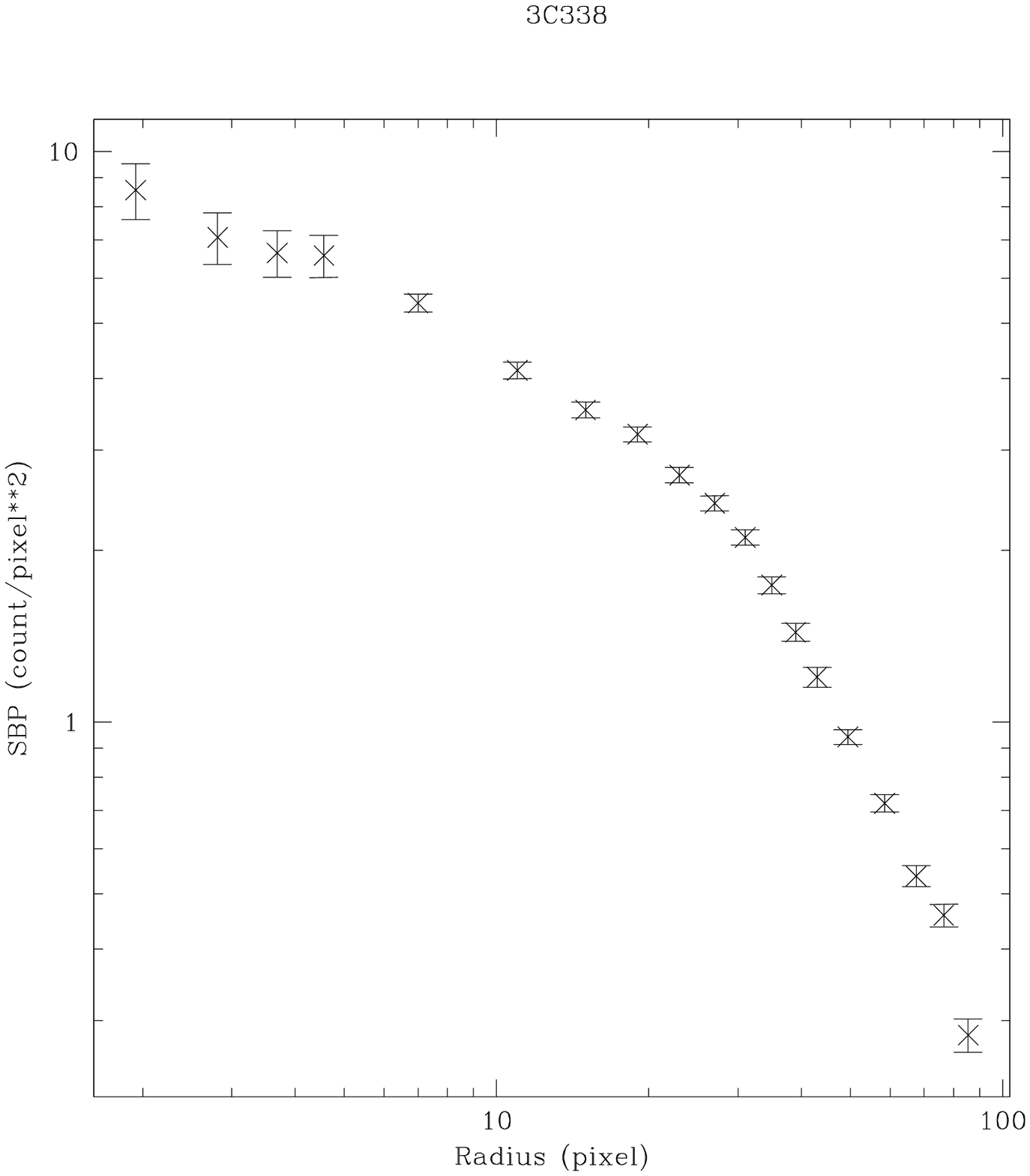,width=0.25\linewidth}}
\centerline{  			
\psfig{figure=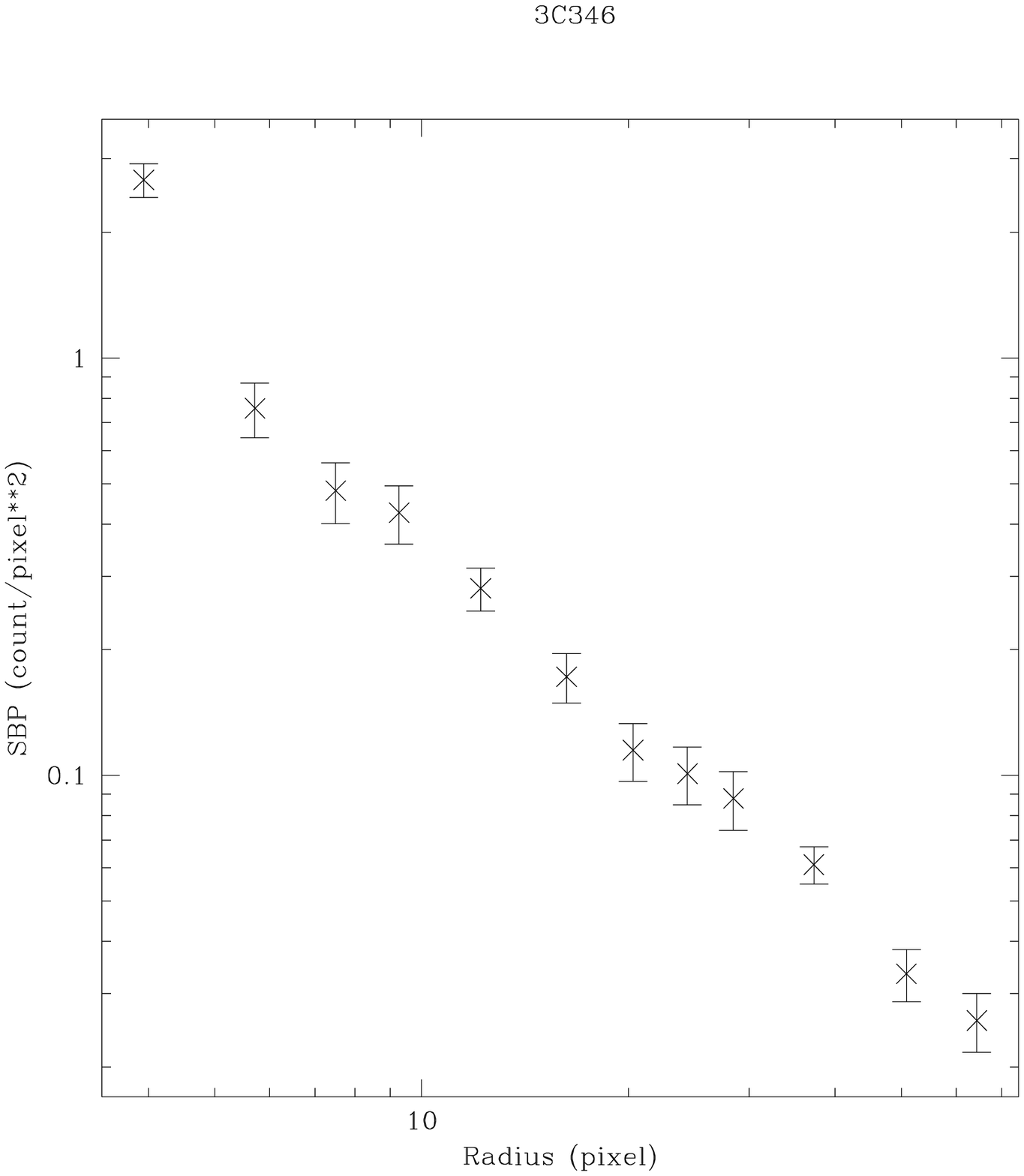,width=0.25\linewidth}
\psfig{figure=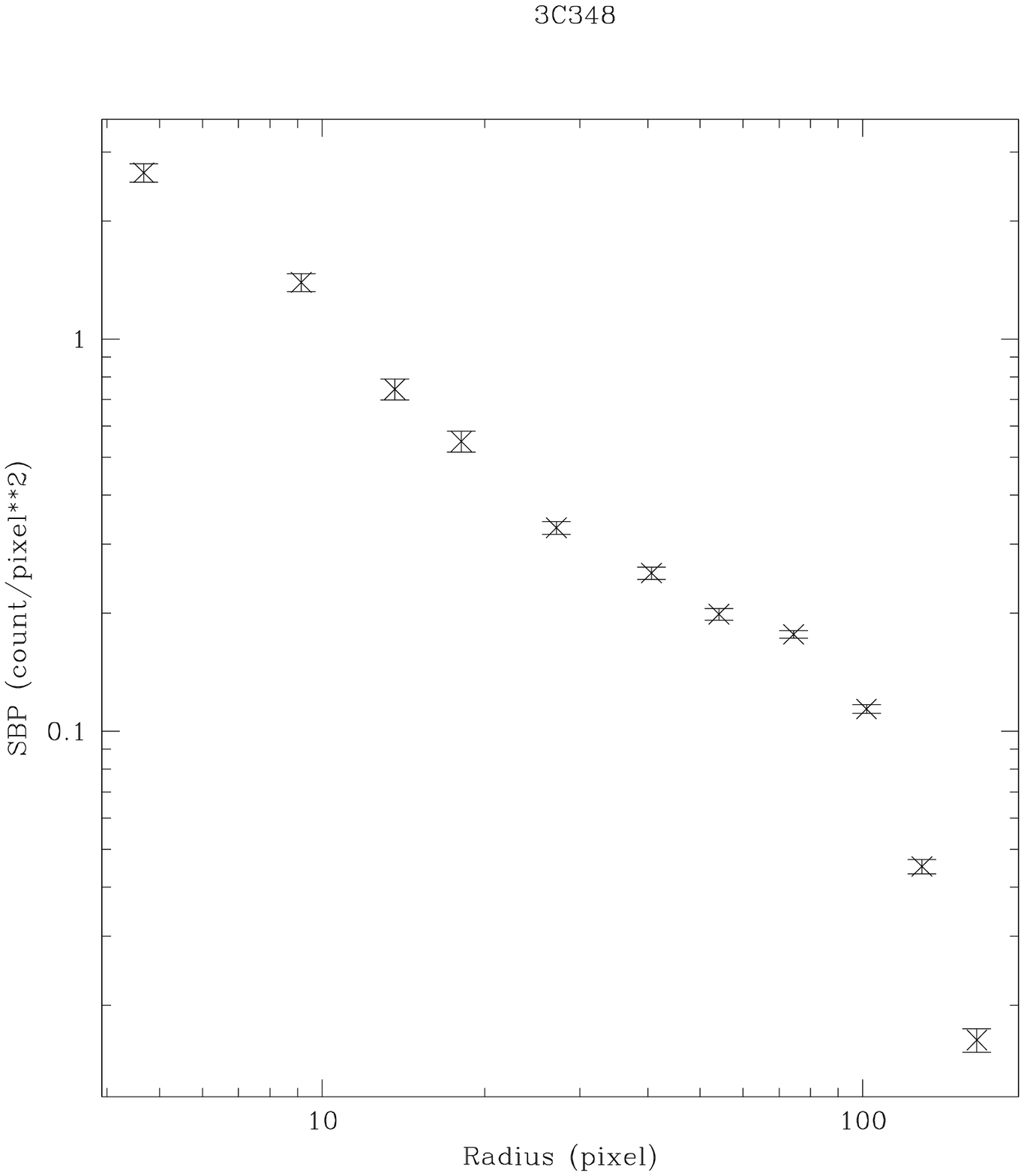,width=0.25\linewidth}
\psfig{figure=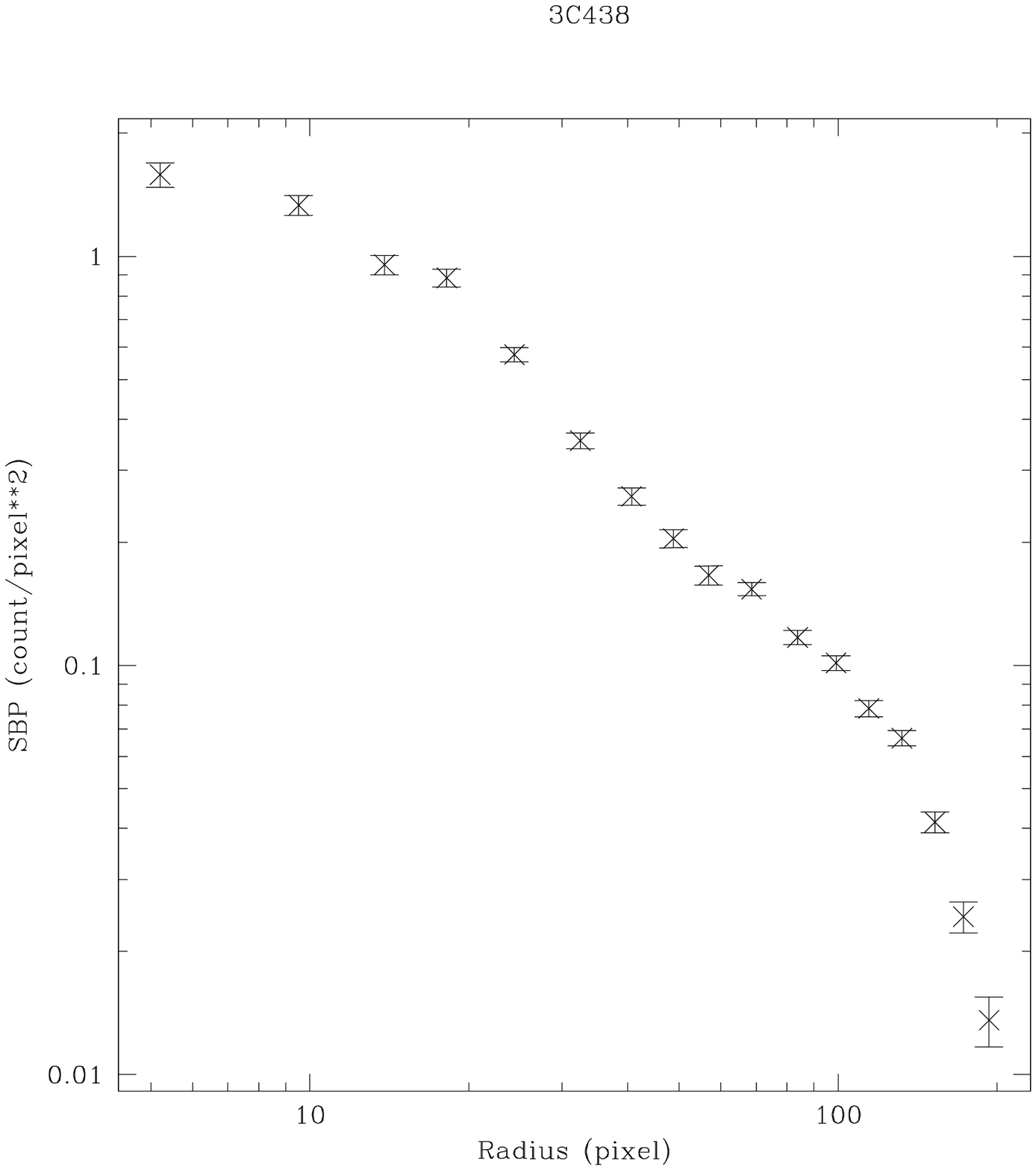,width=0.25\linewidth}
\psfig{figure=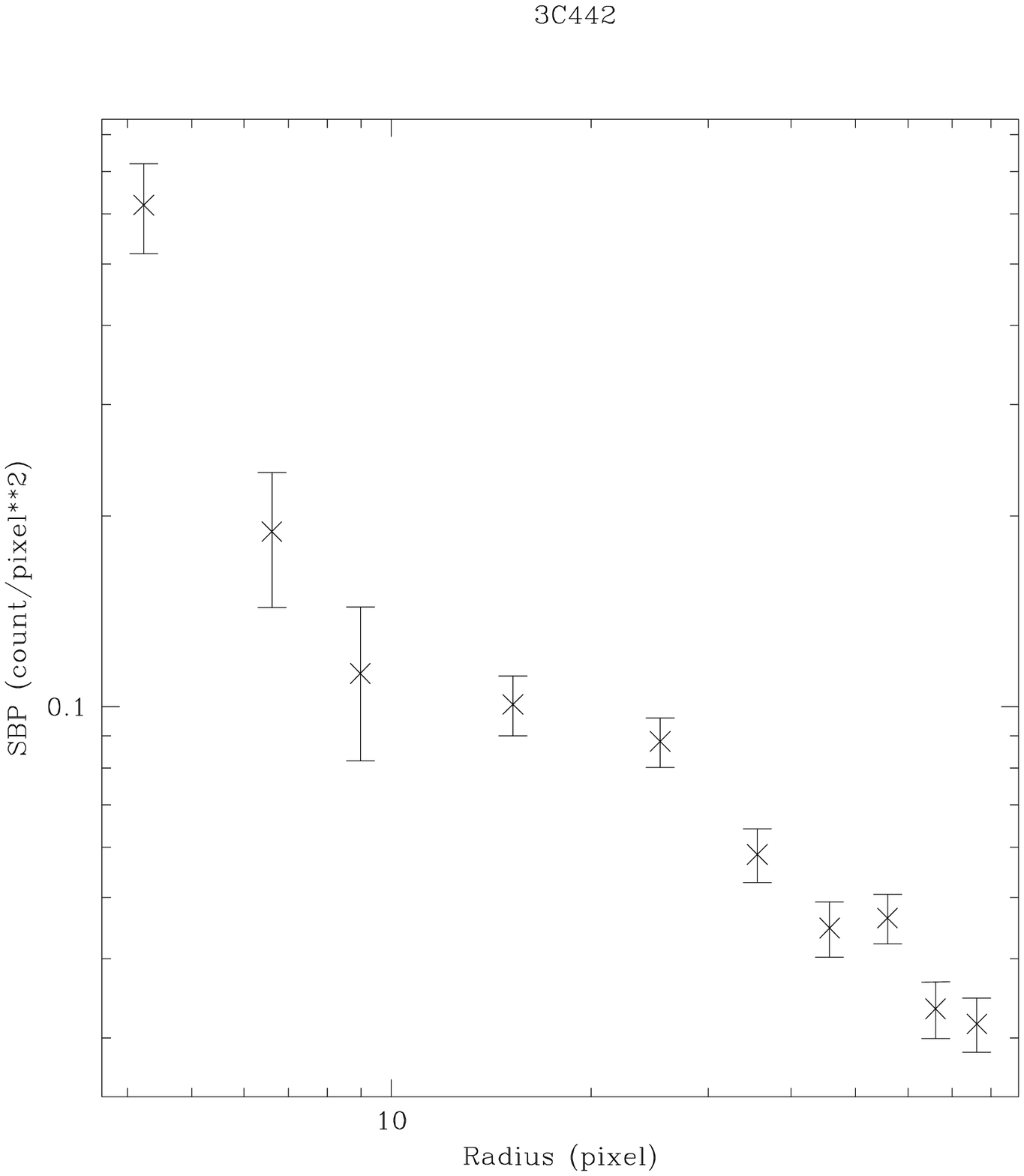,width=0.25\linewidth}}
\end{figure*}
\begin{figure*}
\centerline{
\psfig{figure=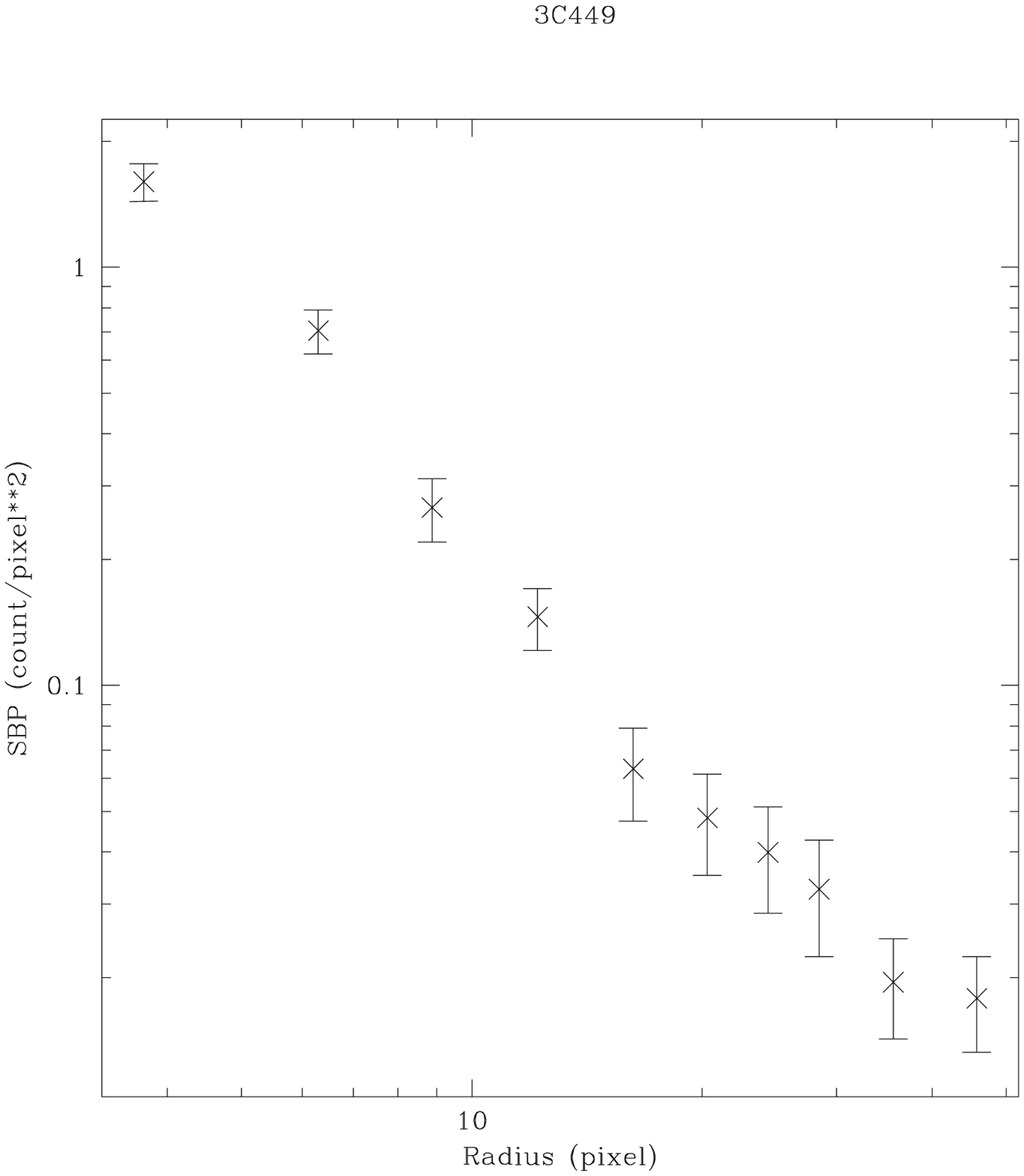,width=0.25\linewidth}
\psfig{figure=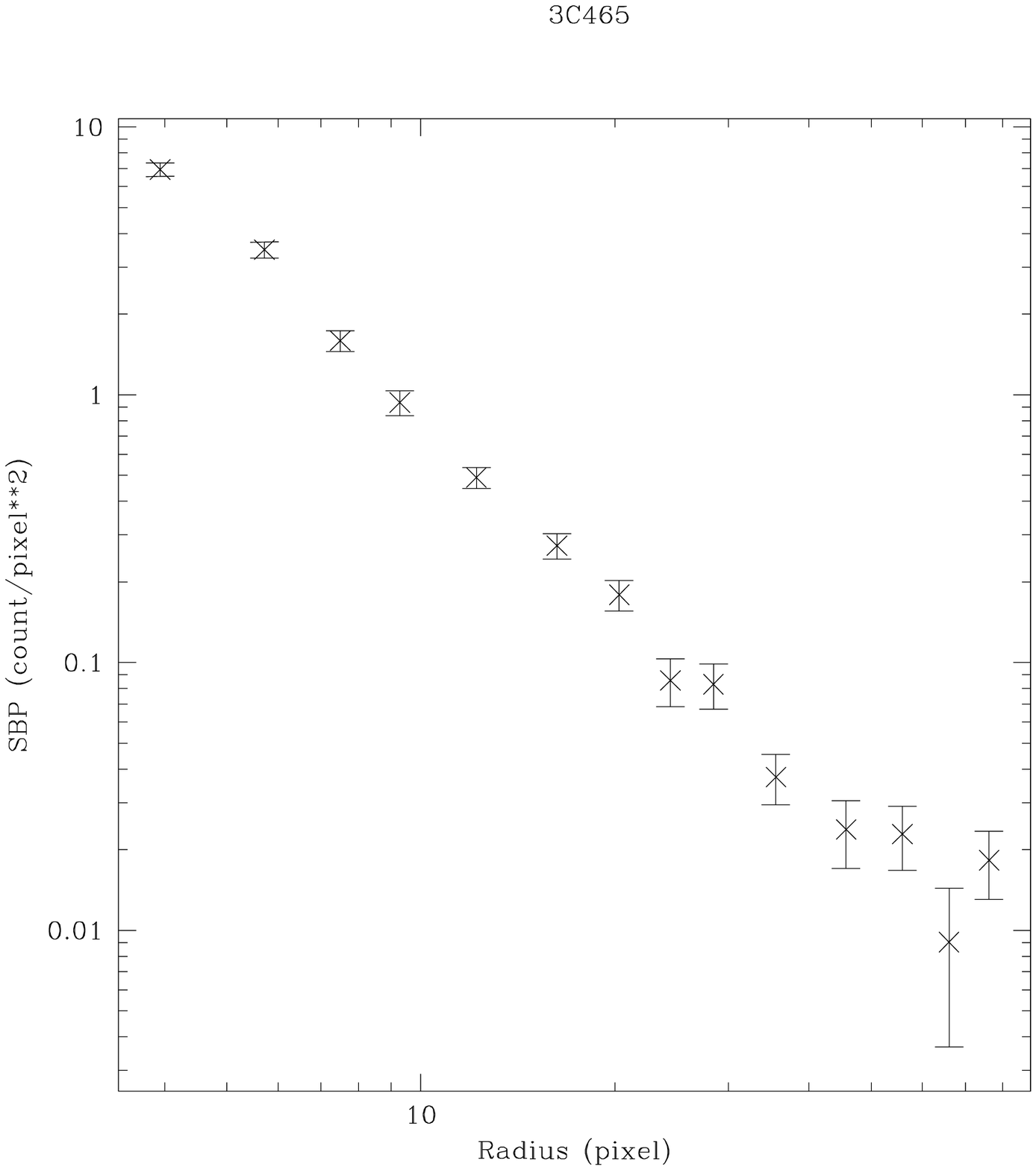,width=0.25\linewidth}
\psfig{figure=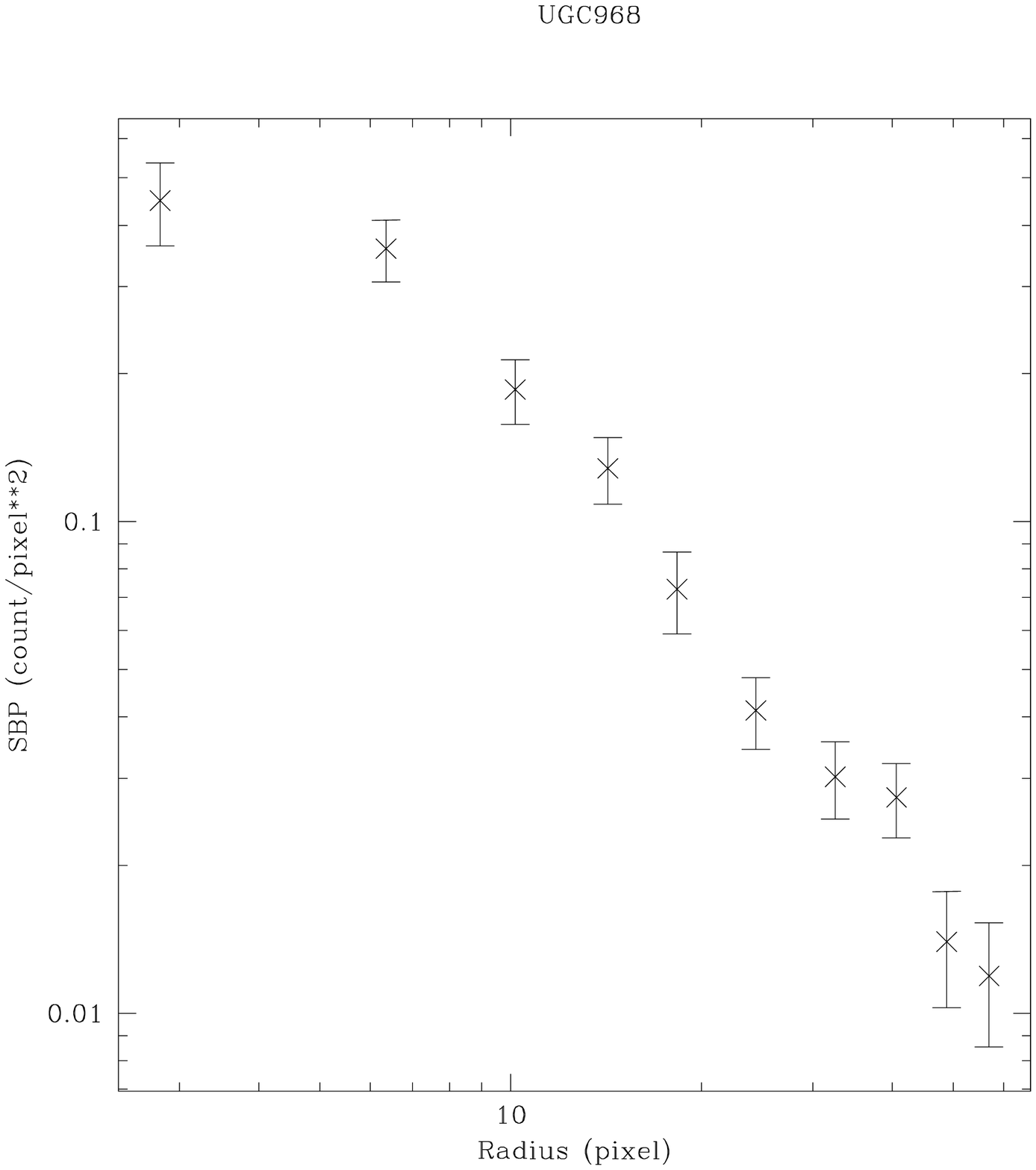,width=0.25\linewidth}
\psfig{figure=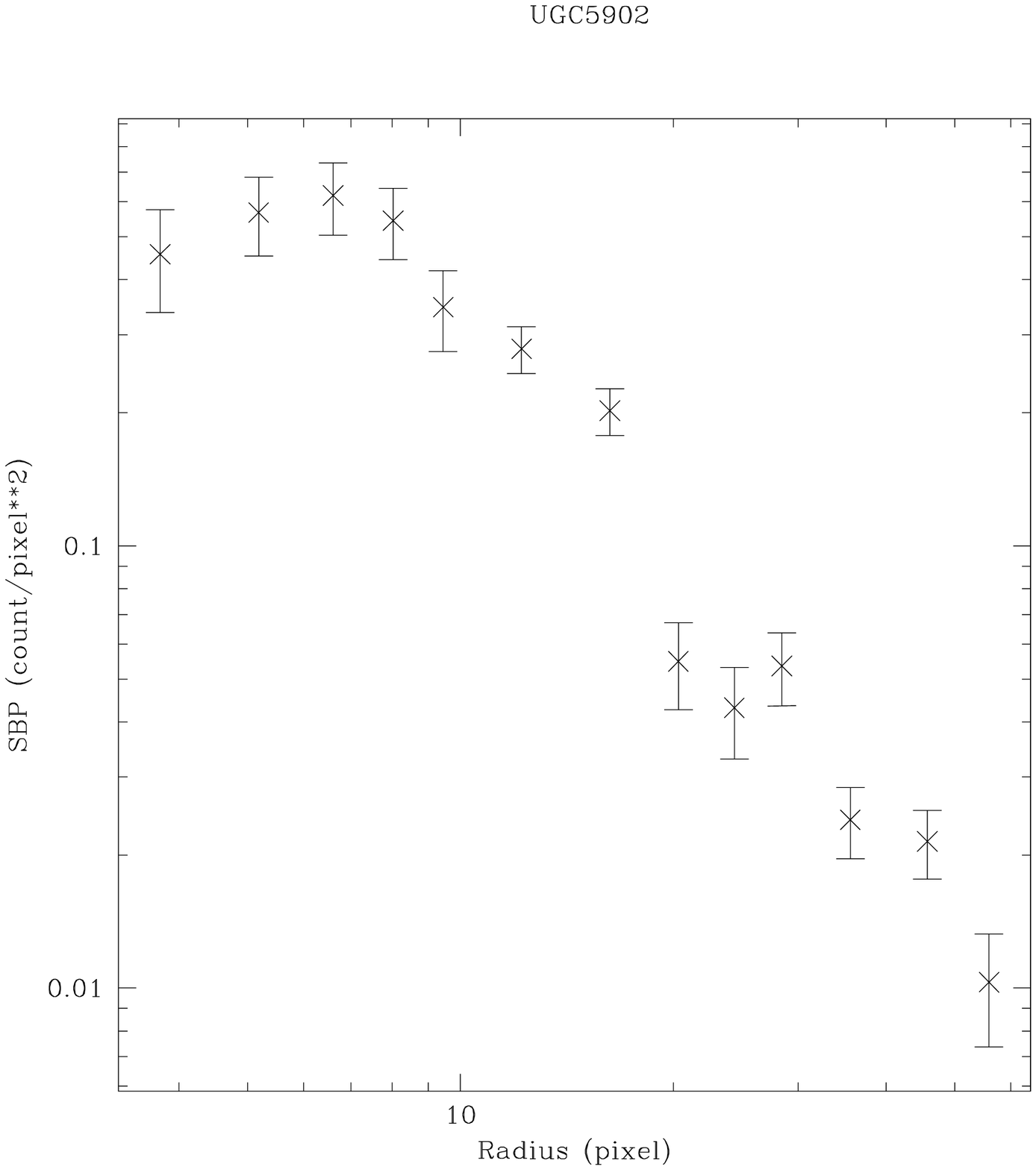,width=0.25\linewidth}}
\centerline{  			
\psfig{figure=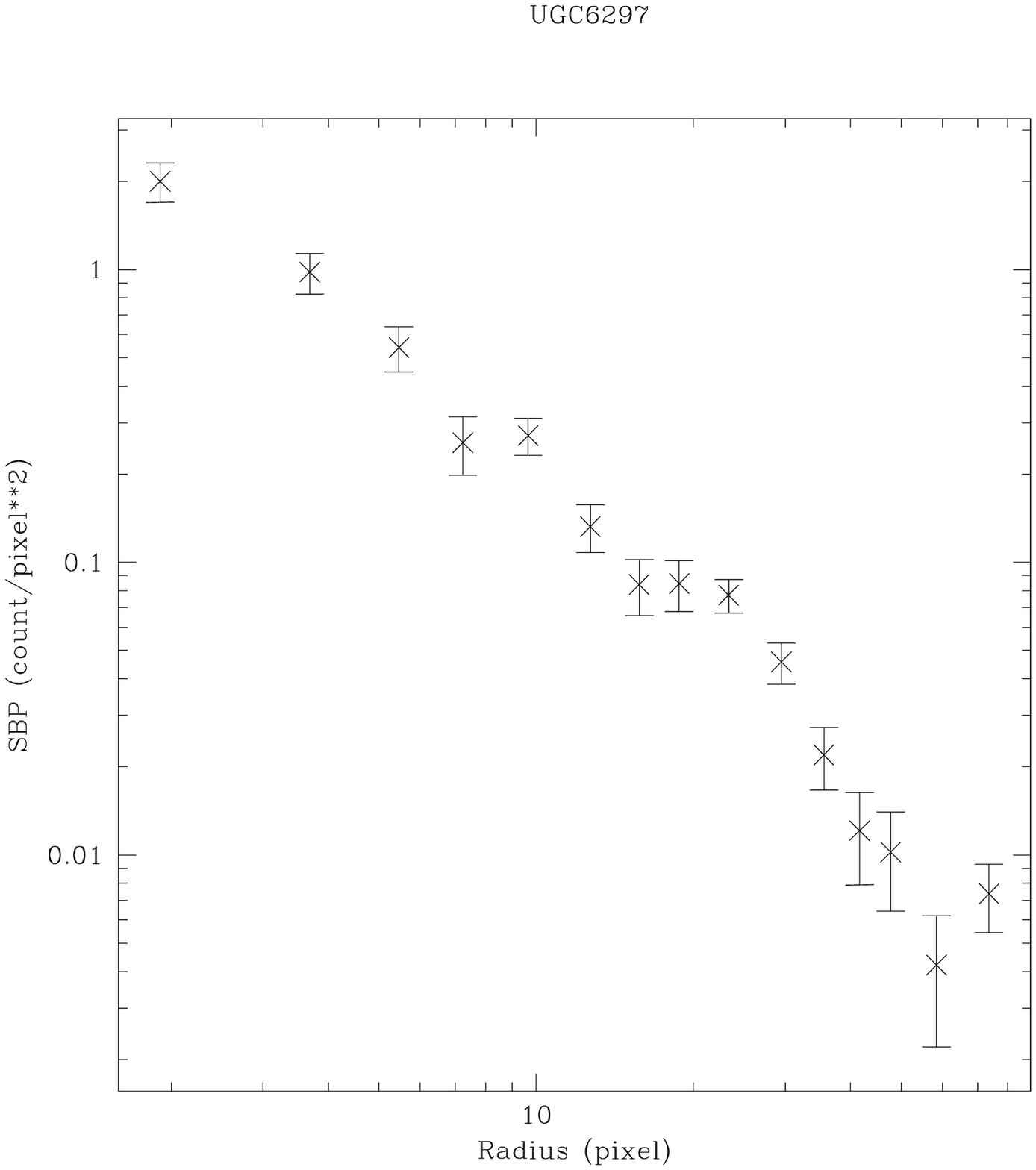,width=0.25\linewidth}
\psfig{figure=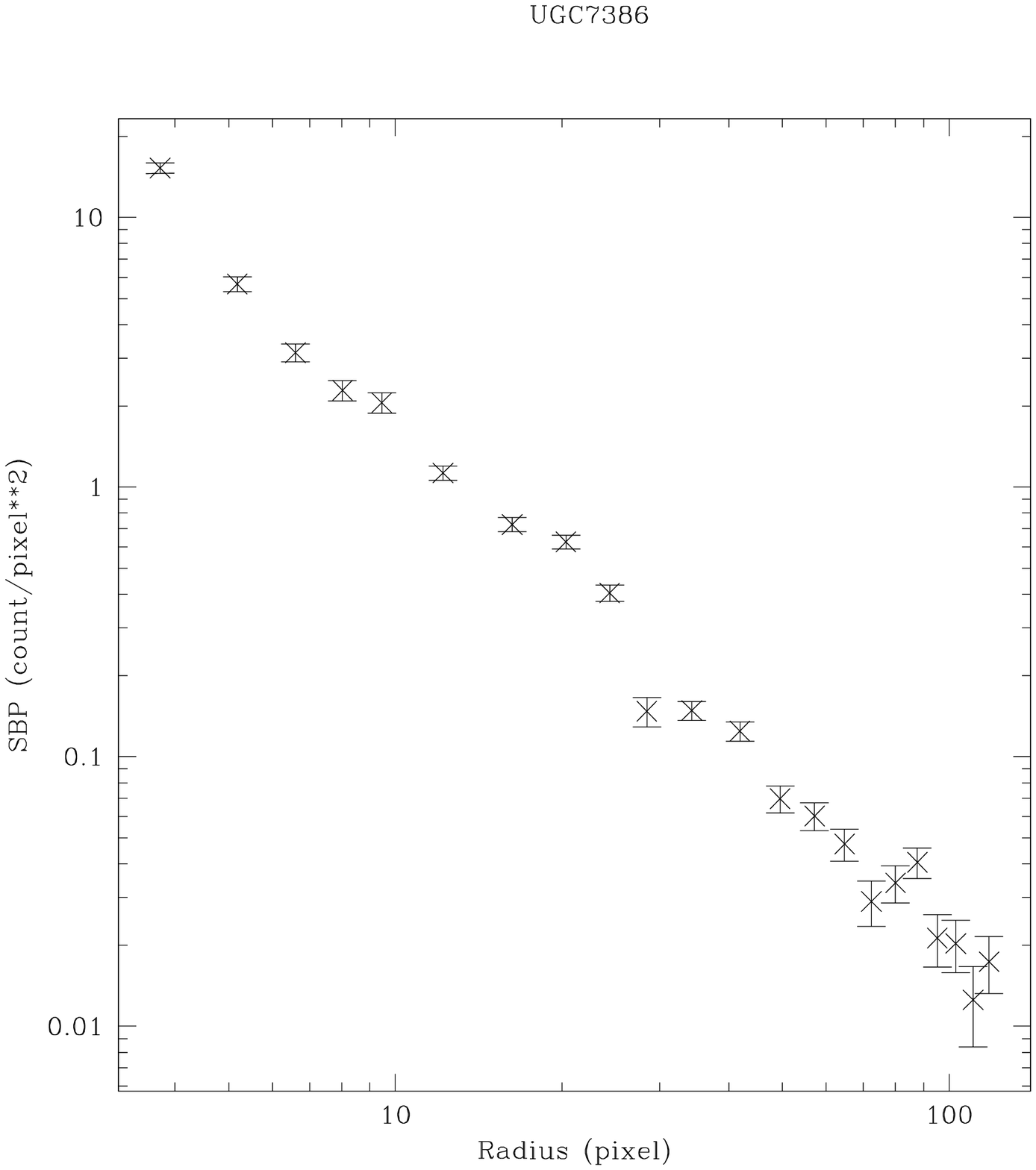,width=0.25\linewidth}
\psfig{figure=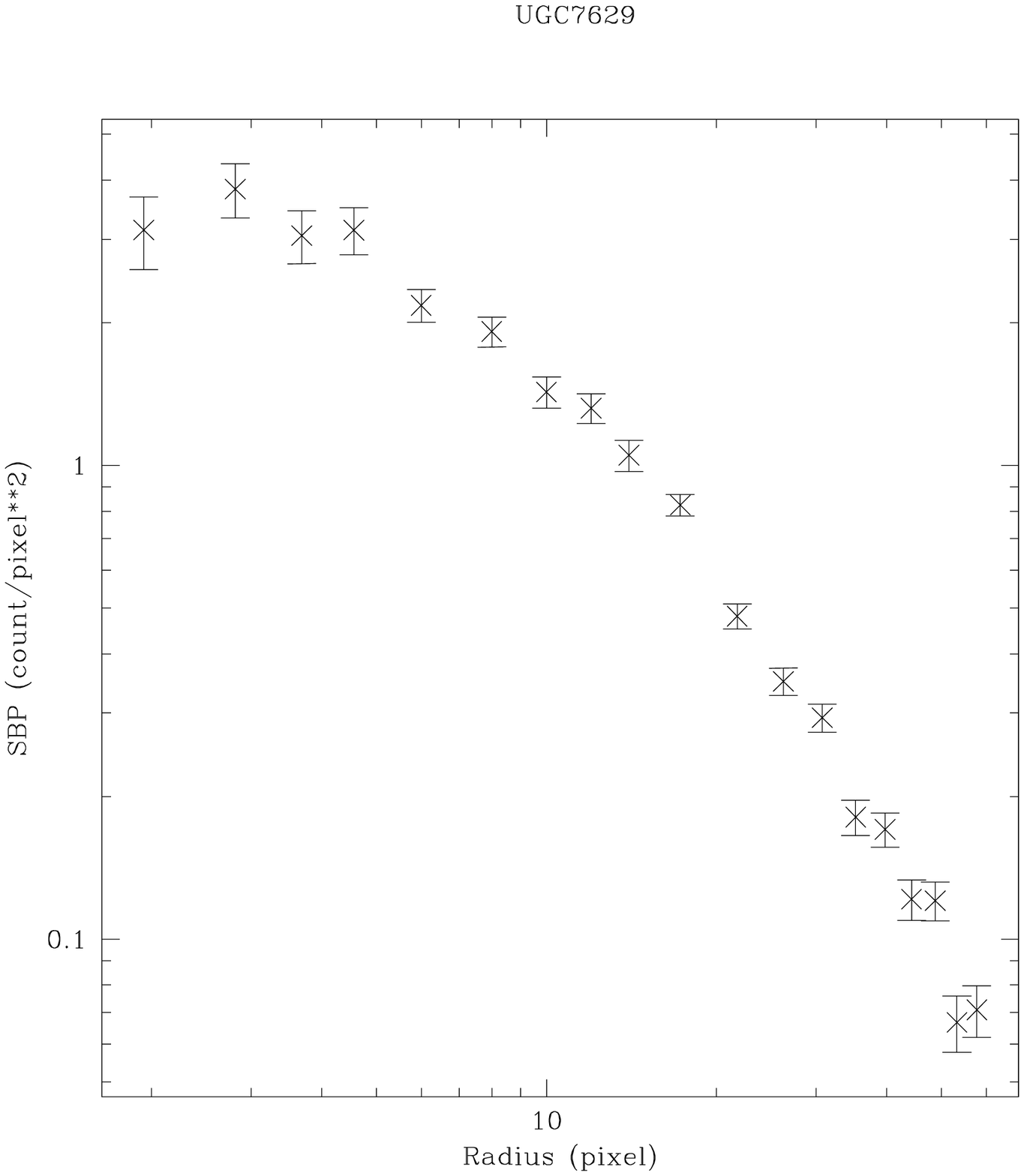,width=0.25\linewidth}
\psfig{figure=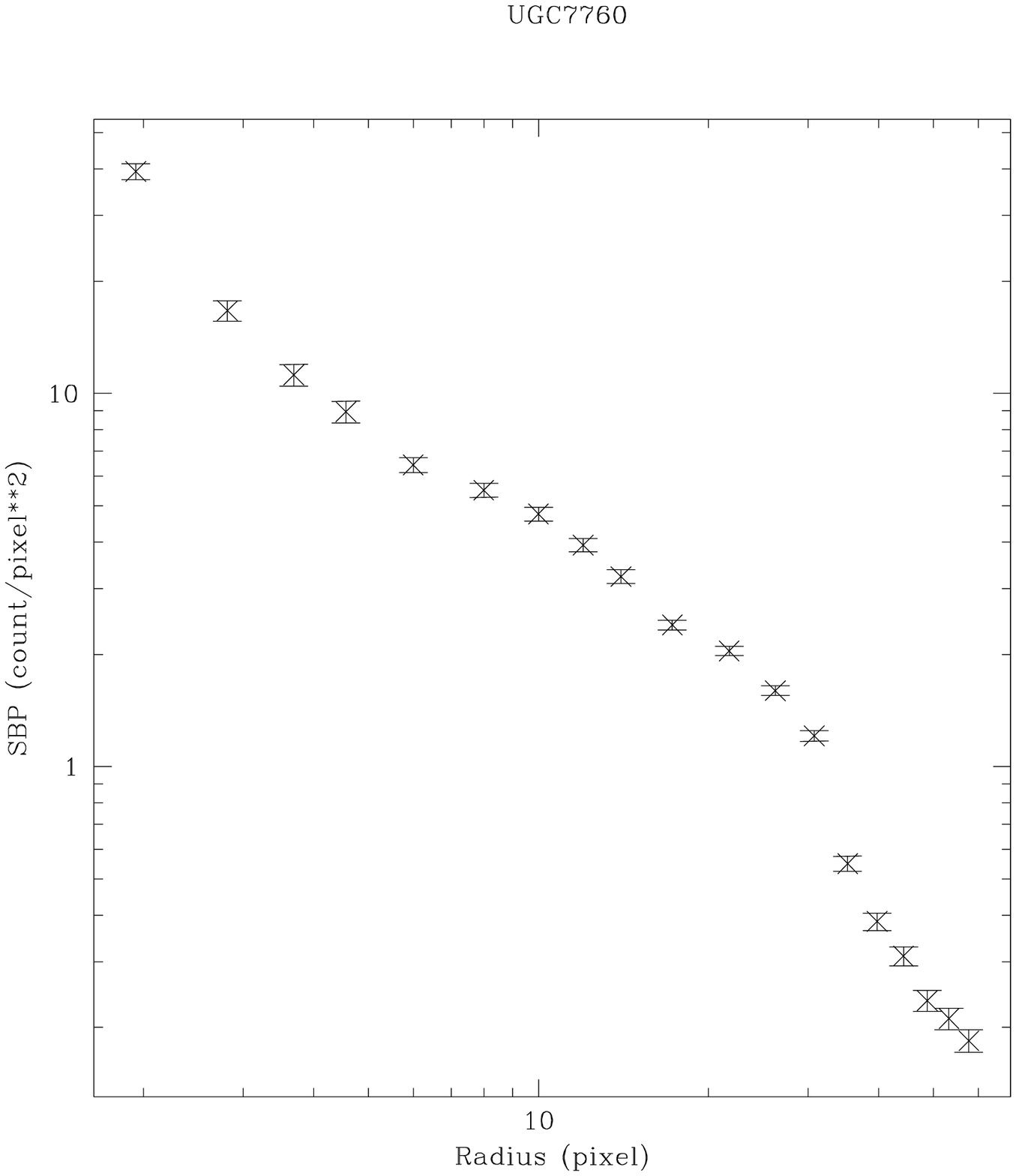,width=0.25\linewidth}}
\centerline{  			
\psfig{figure=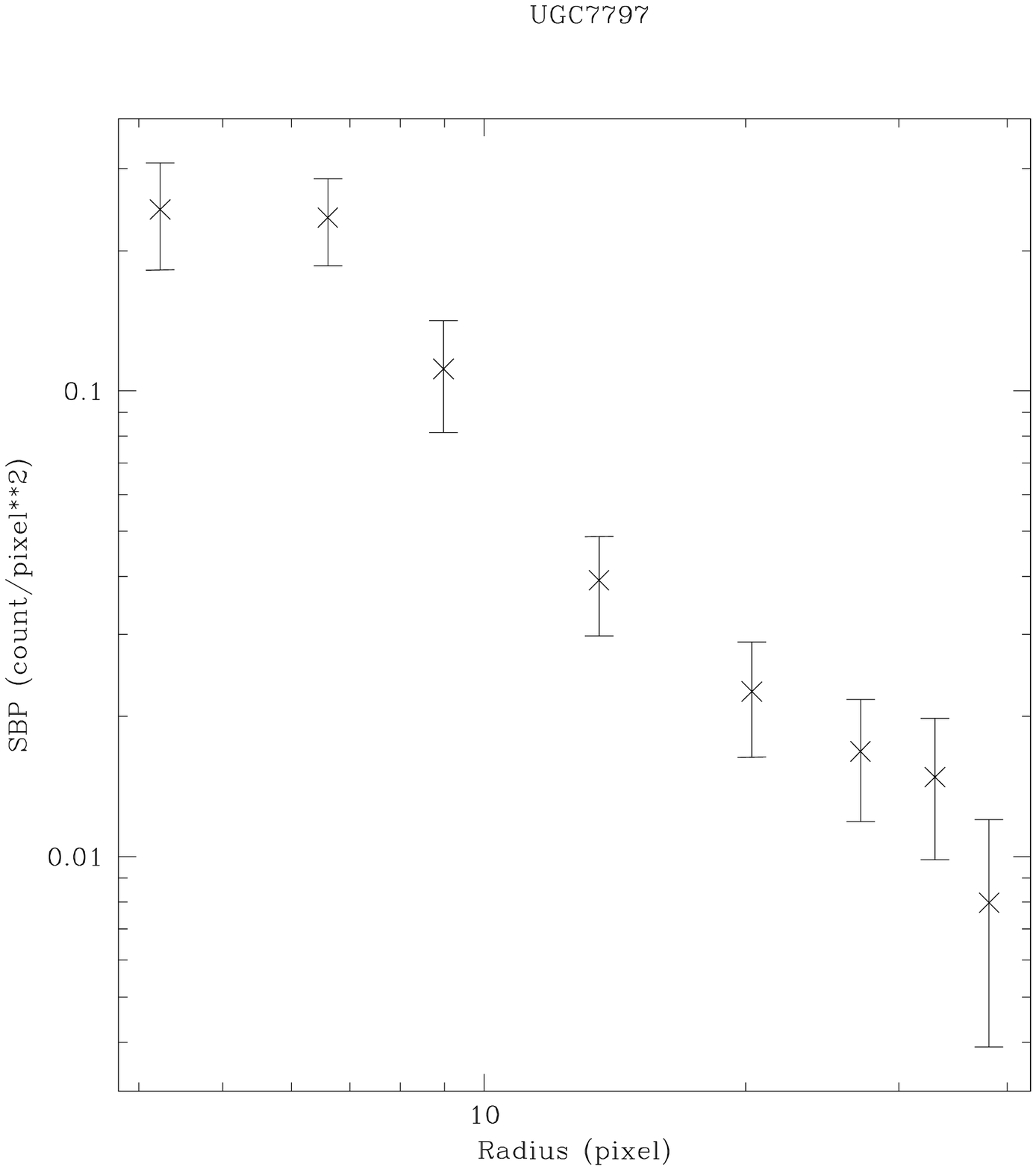,width=0.25\linewidth}
\psfig{figure=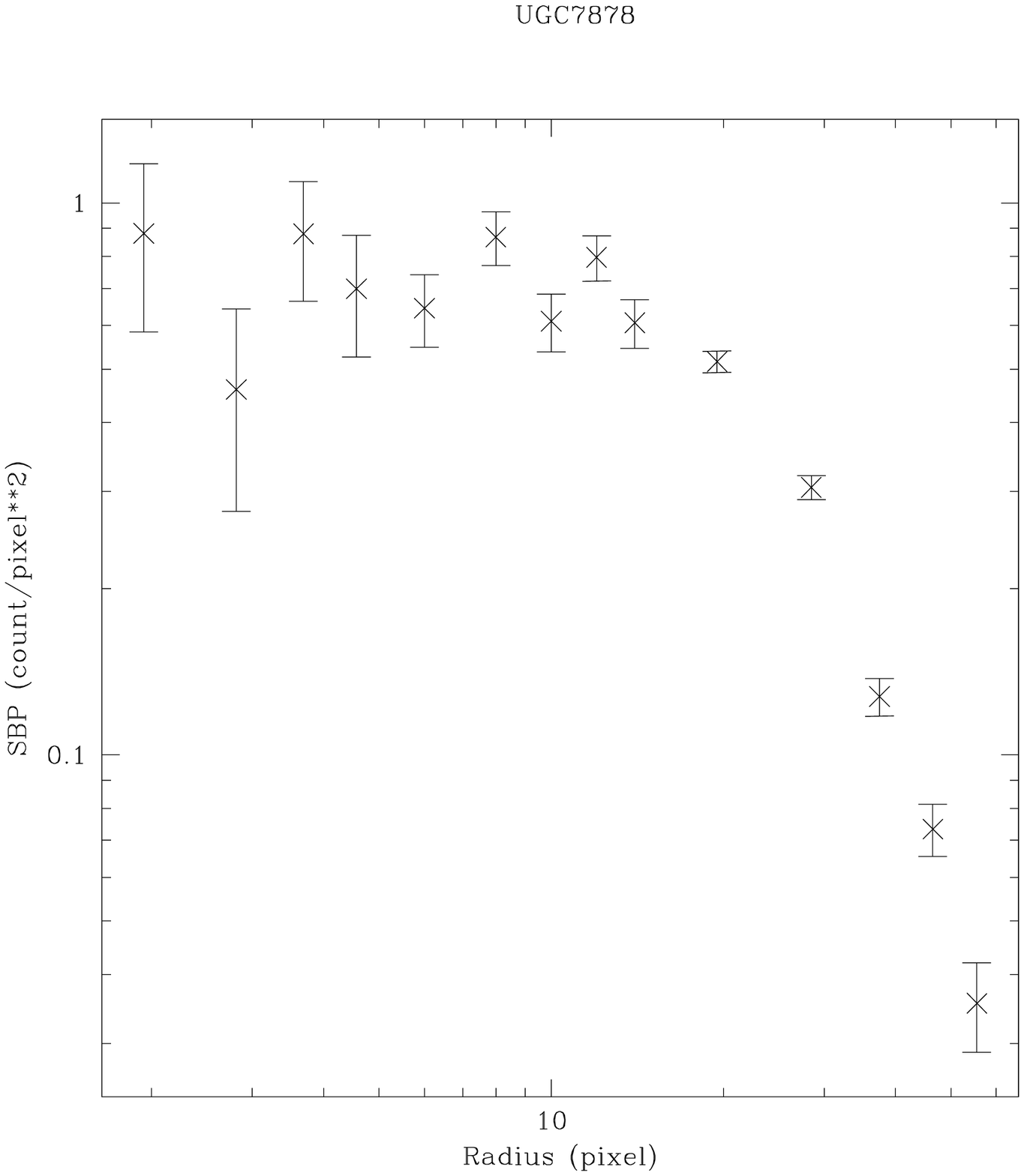,width=0.25\linewidth}
\psfig{figure=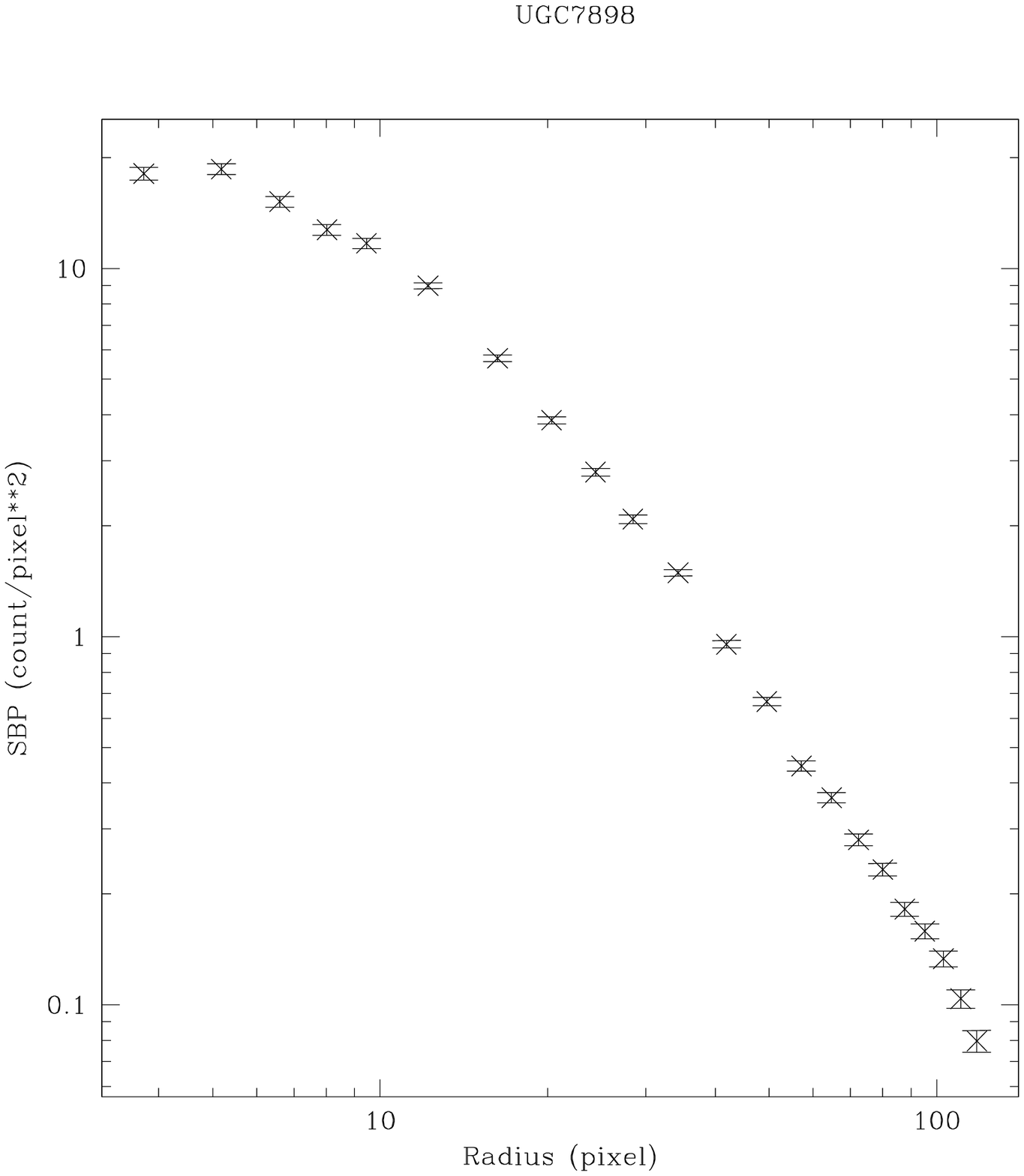,width=0.25\linewidth}
\psfig{figure=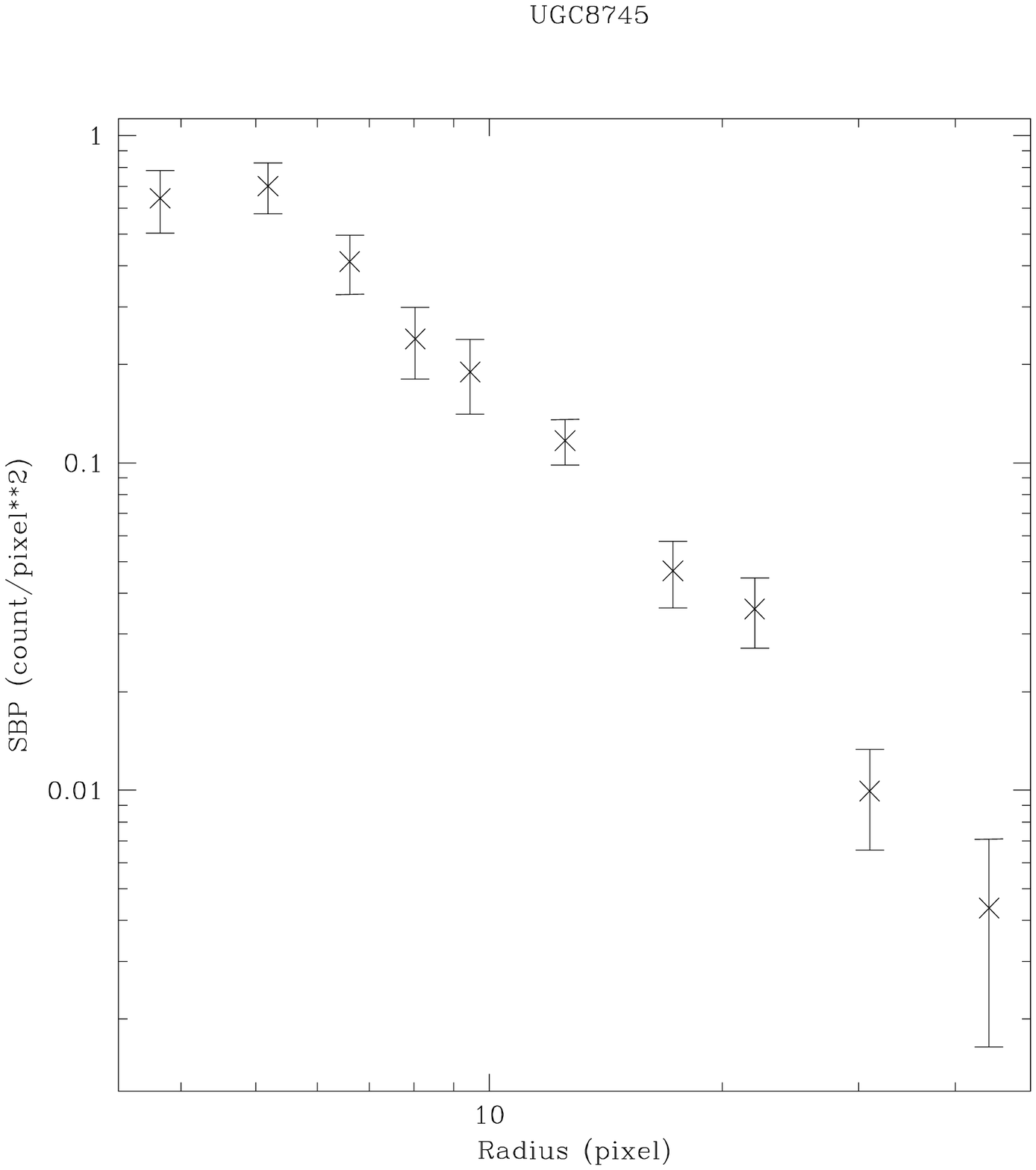,width=0.25\linewidth}}
\centerline{  			
\psfig{figure=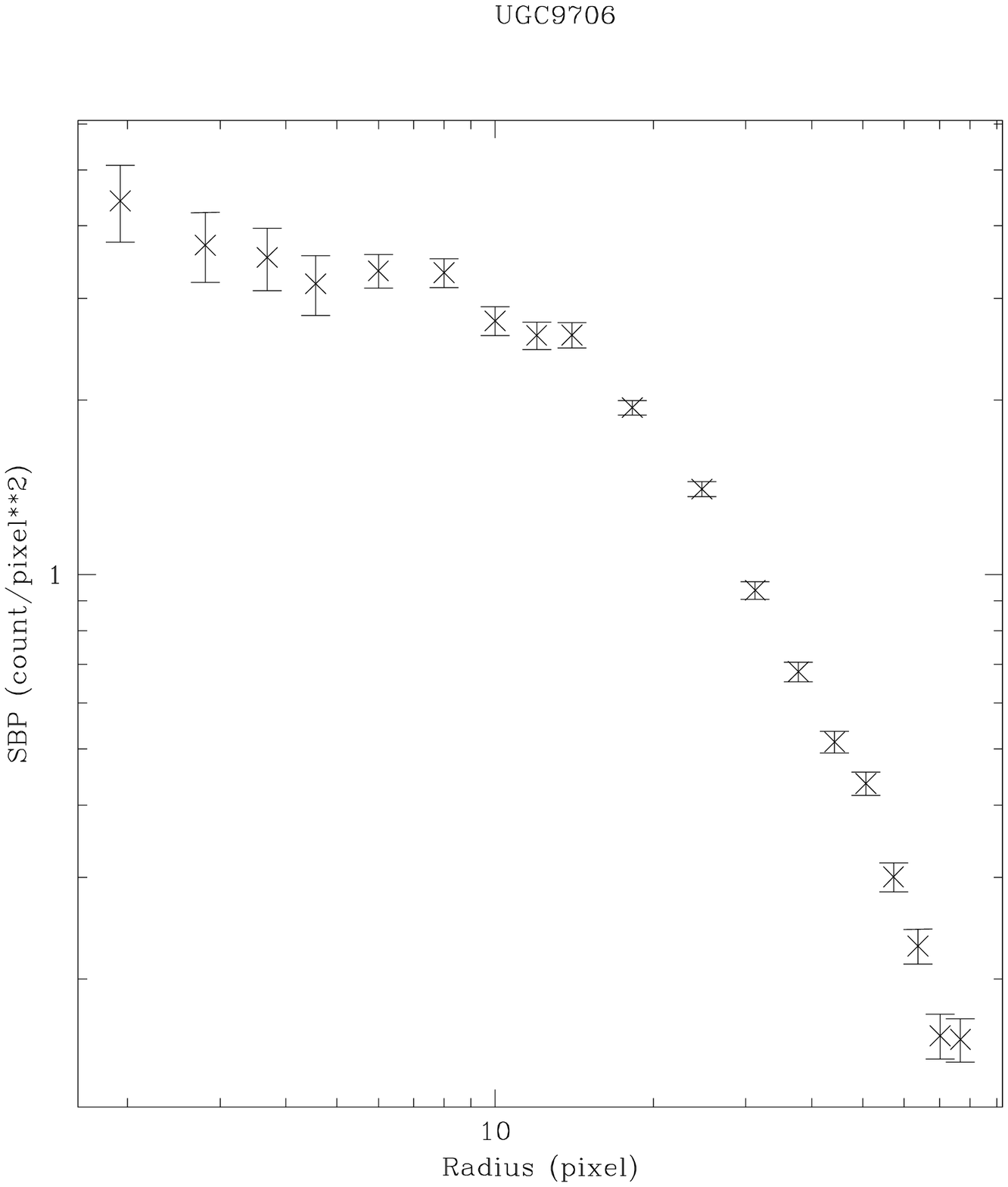,width=0.25\linewidth}
\psfig{figure=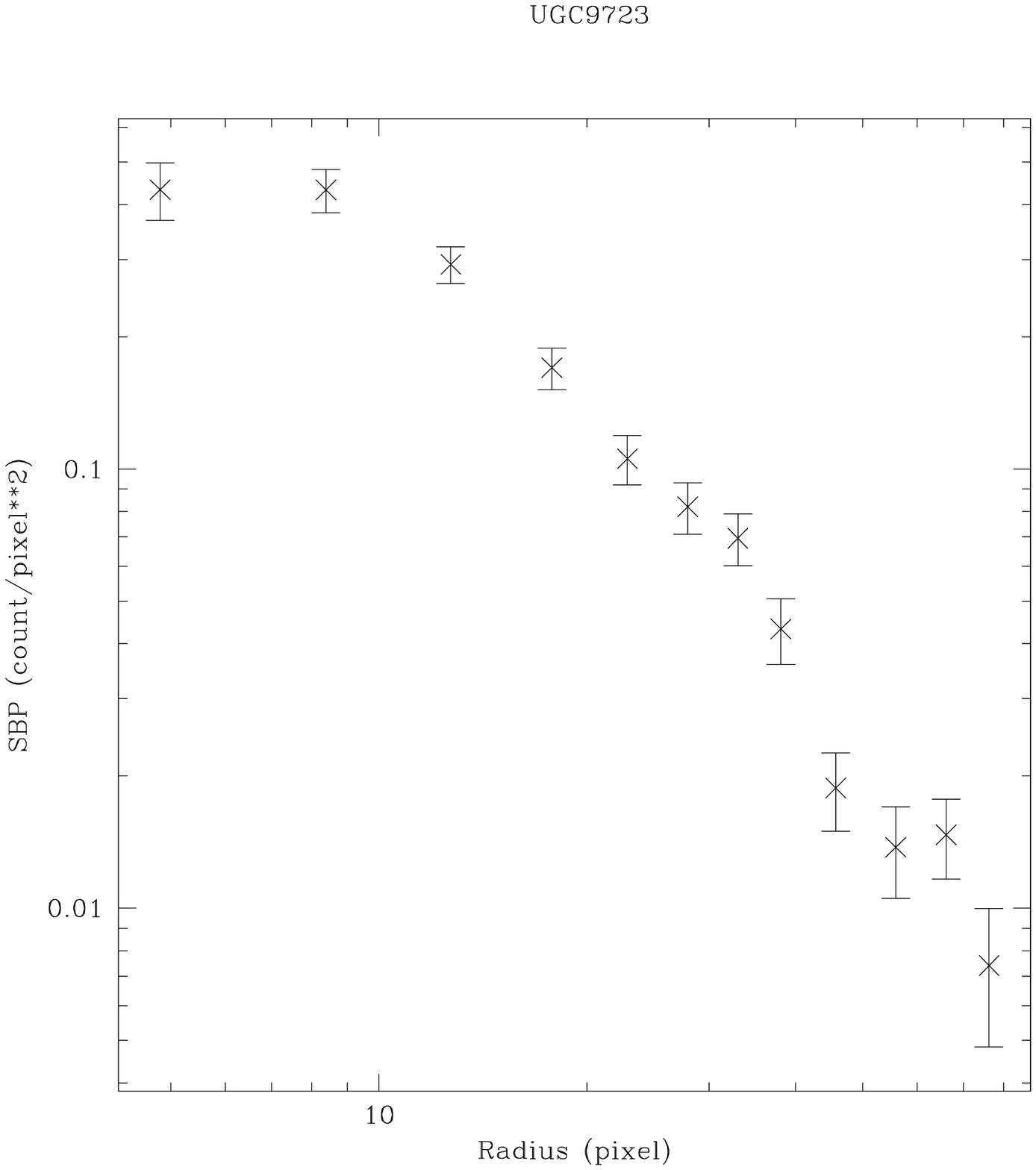,width=0.25\linewidth}
\psfig{figure=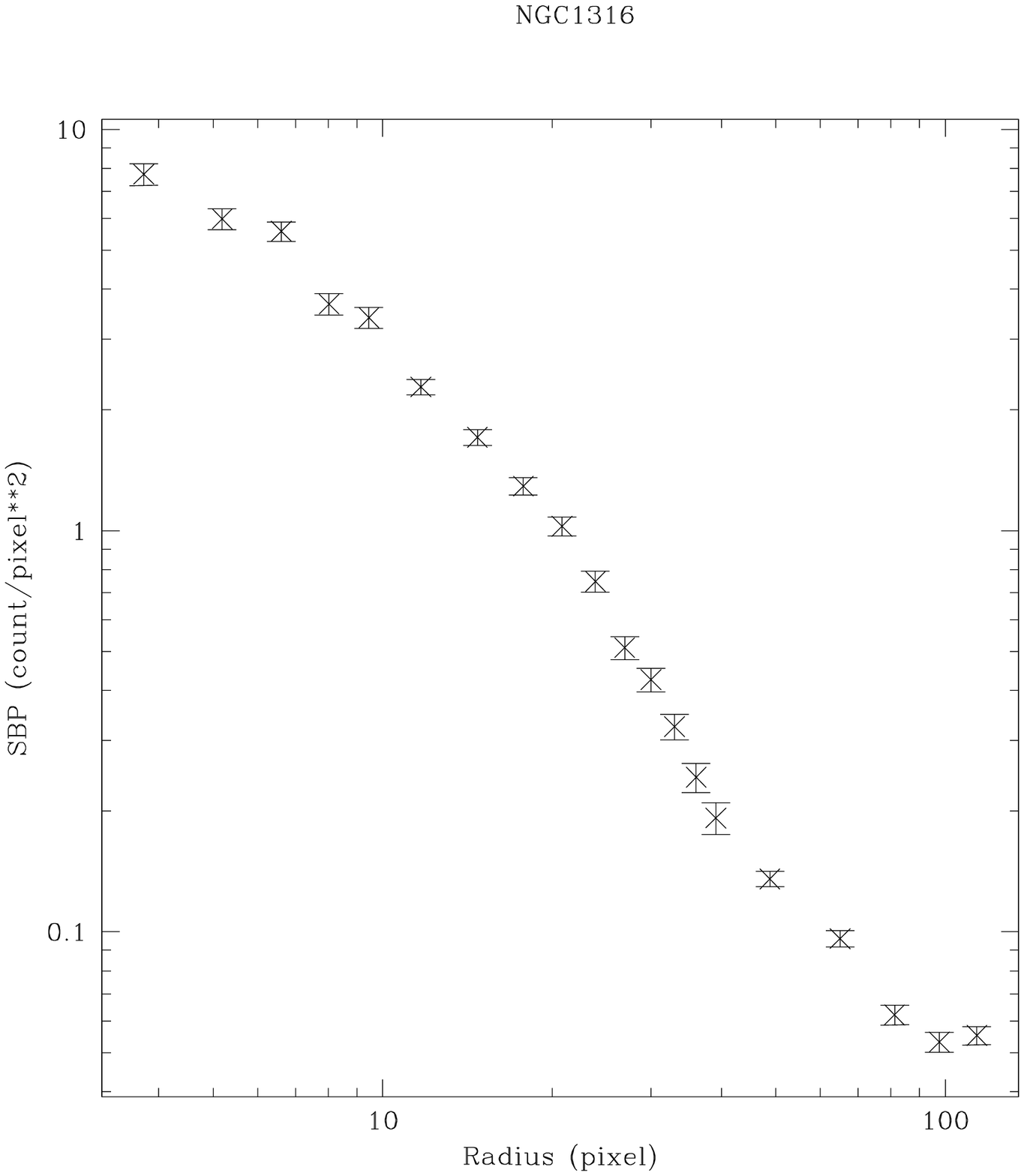,width=0.25\linewidth}
\psfig{figure=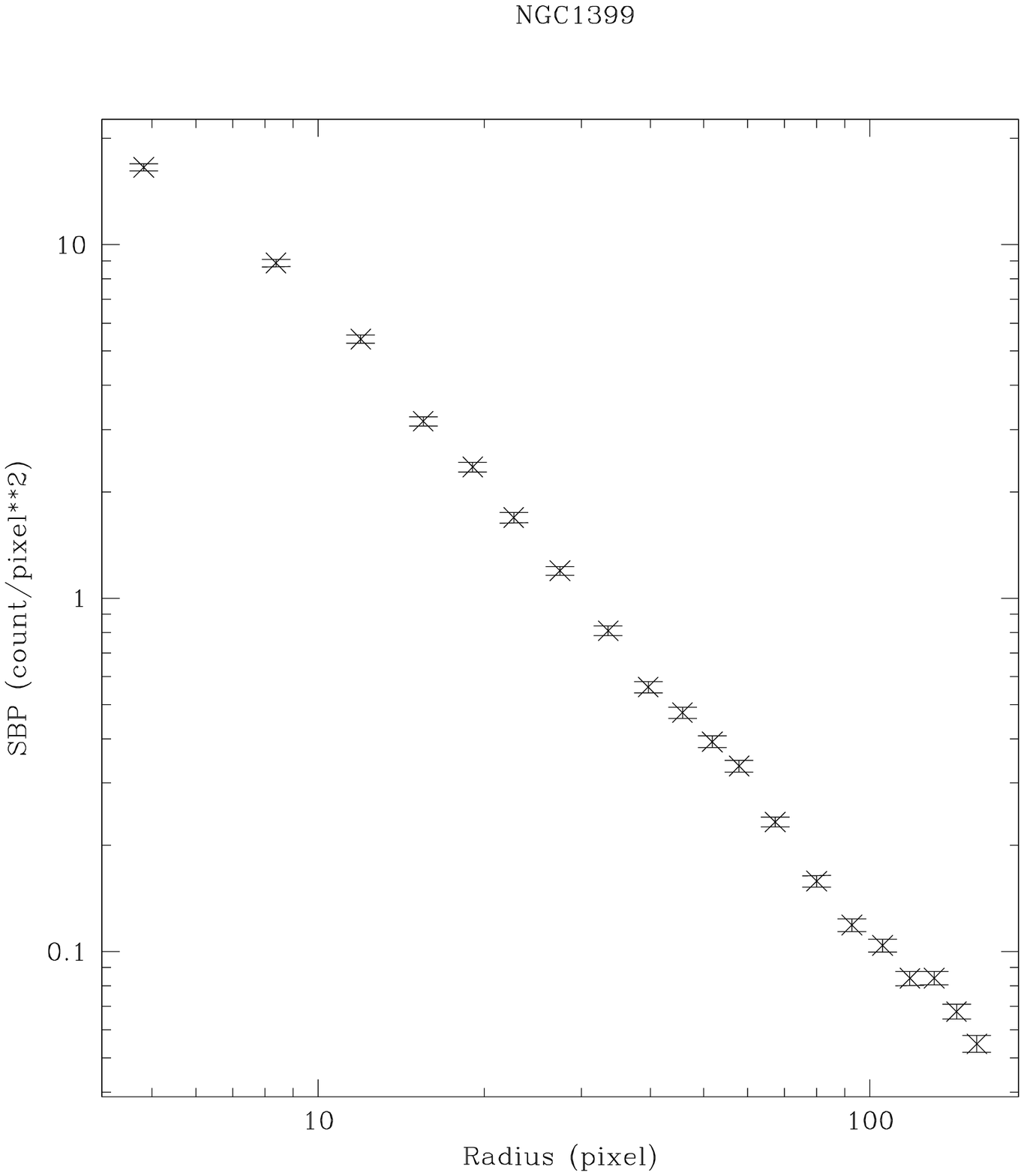,width=0.25\linewidth}}
\centerline{  			
\psfig{figure=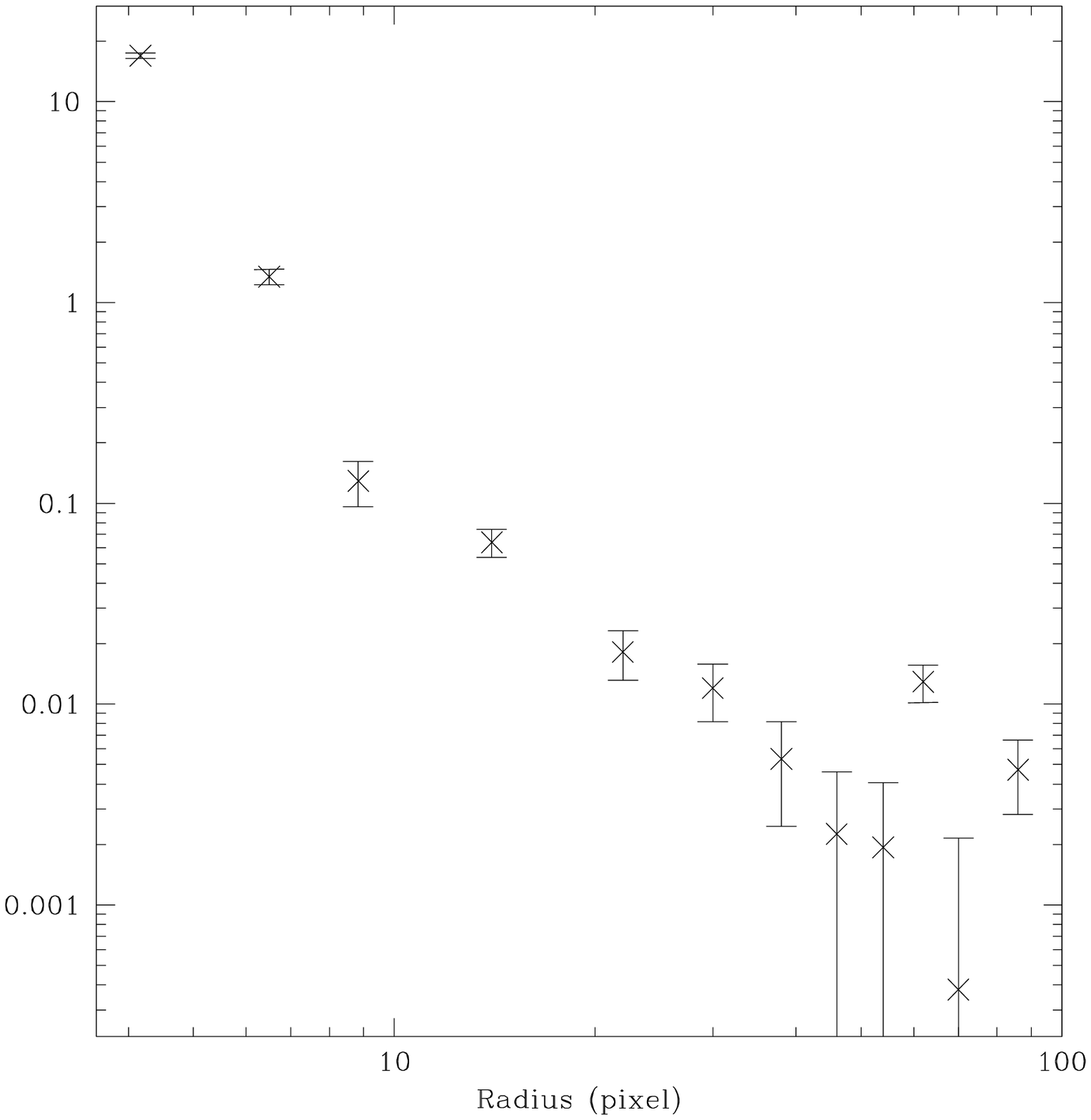,width=0.25\linewidth}
\psfig{figure=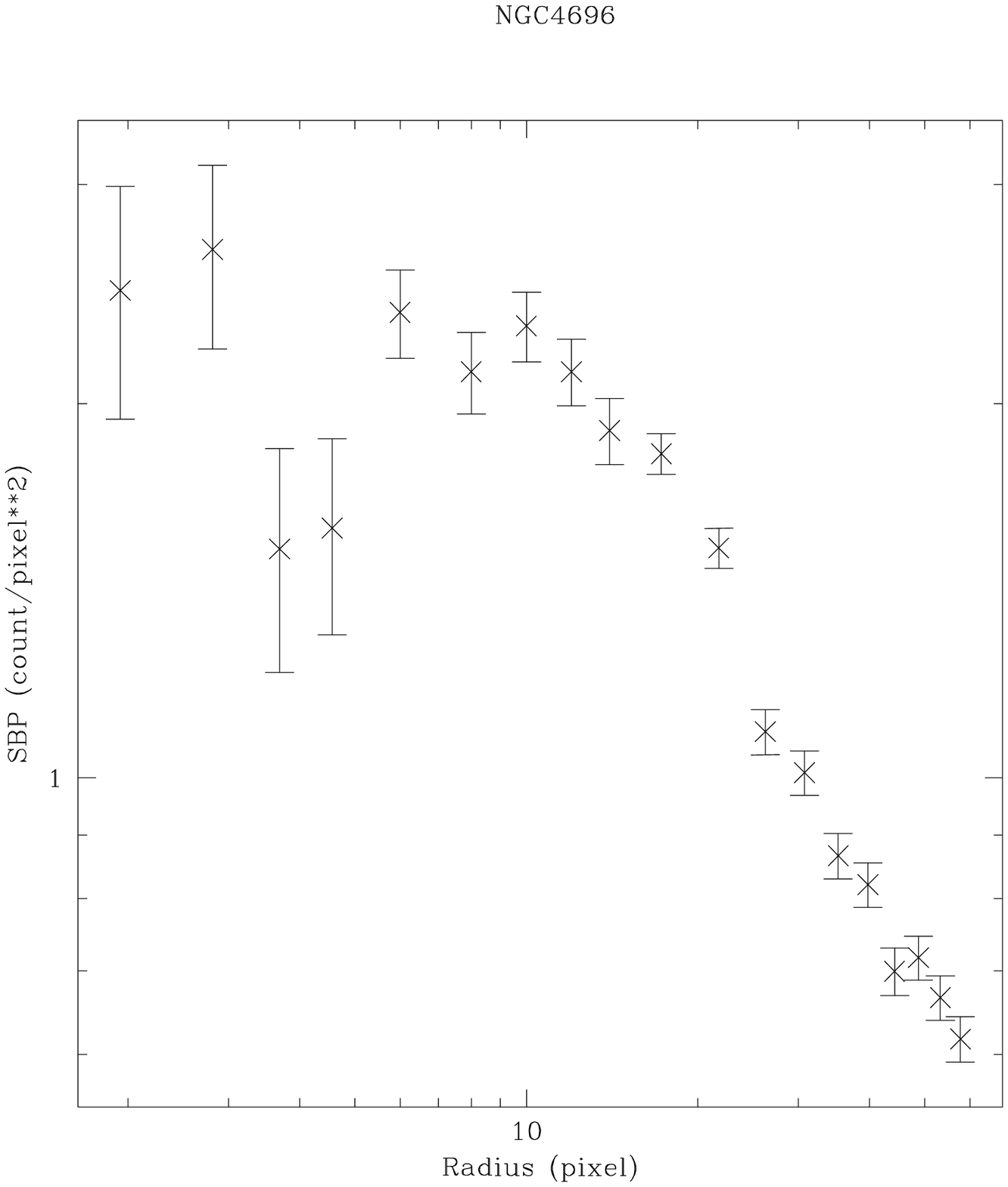,width=0.25\linewidth}
\psfig{figure=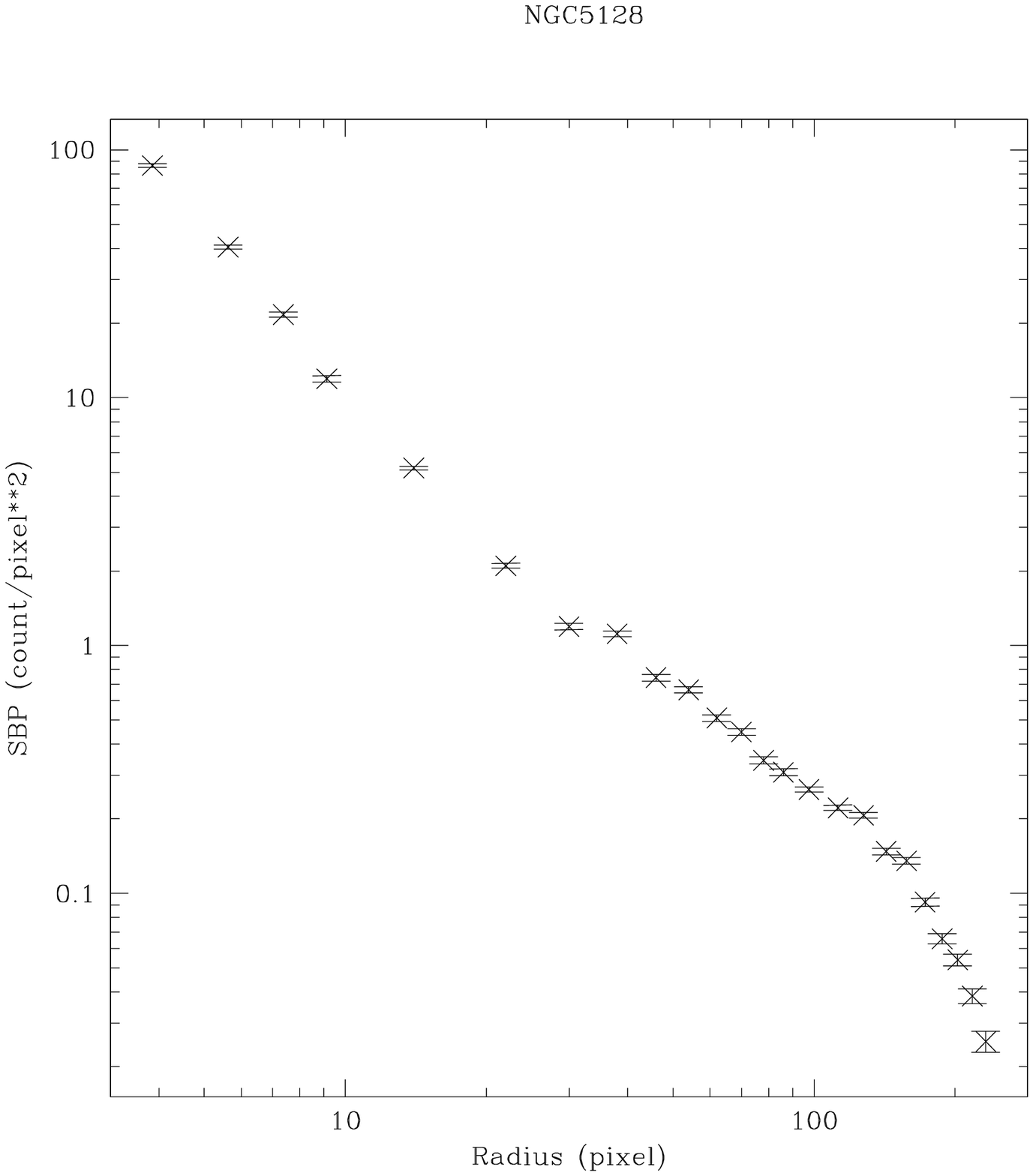,width=0.25\linewidth}
\psfig{figure=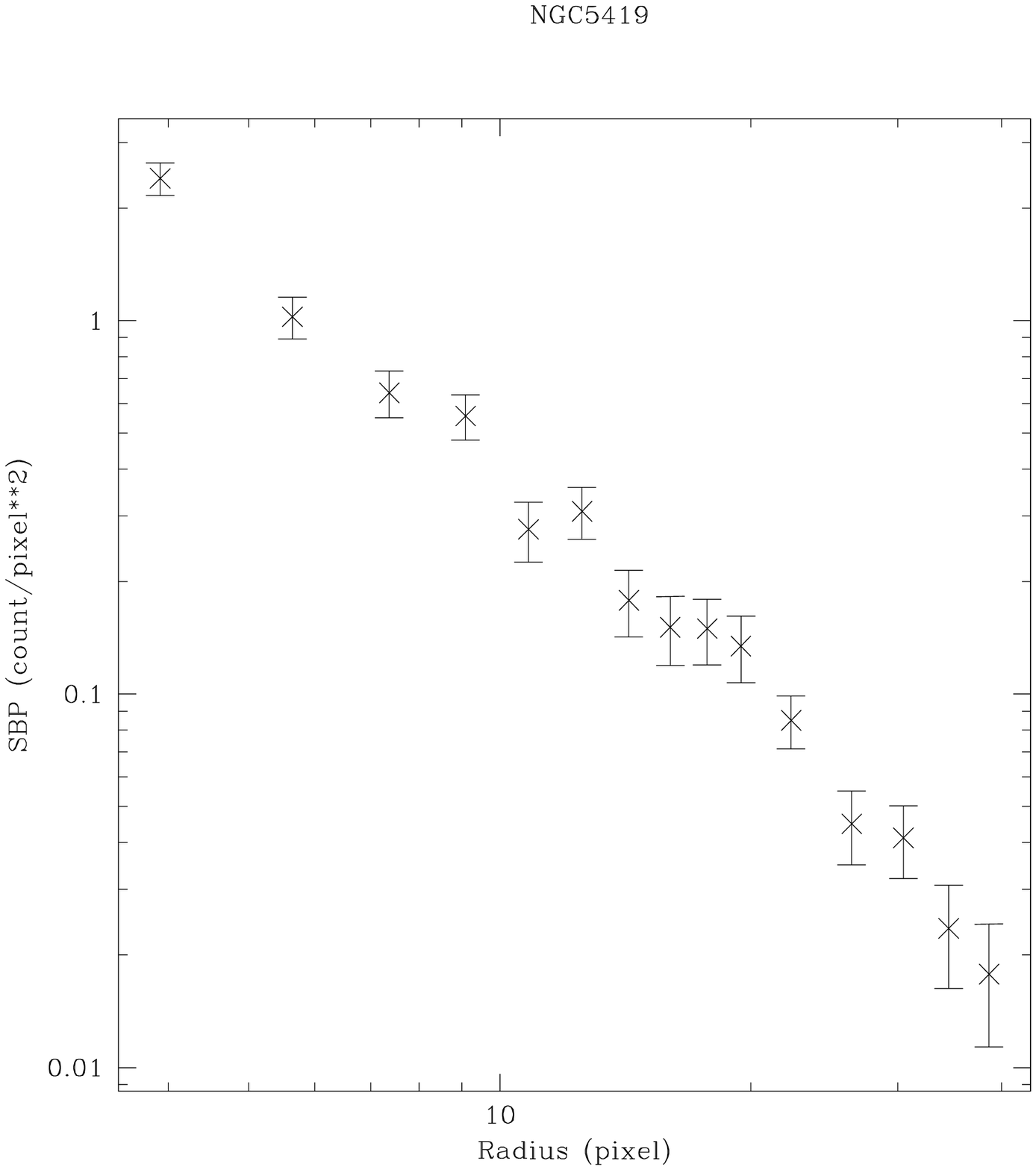,width=0.25\linewidth}}
\end{figure*}
\begin{figure*}
\centerline{
\psfig{figure=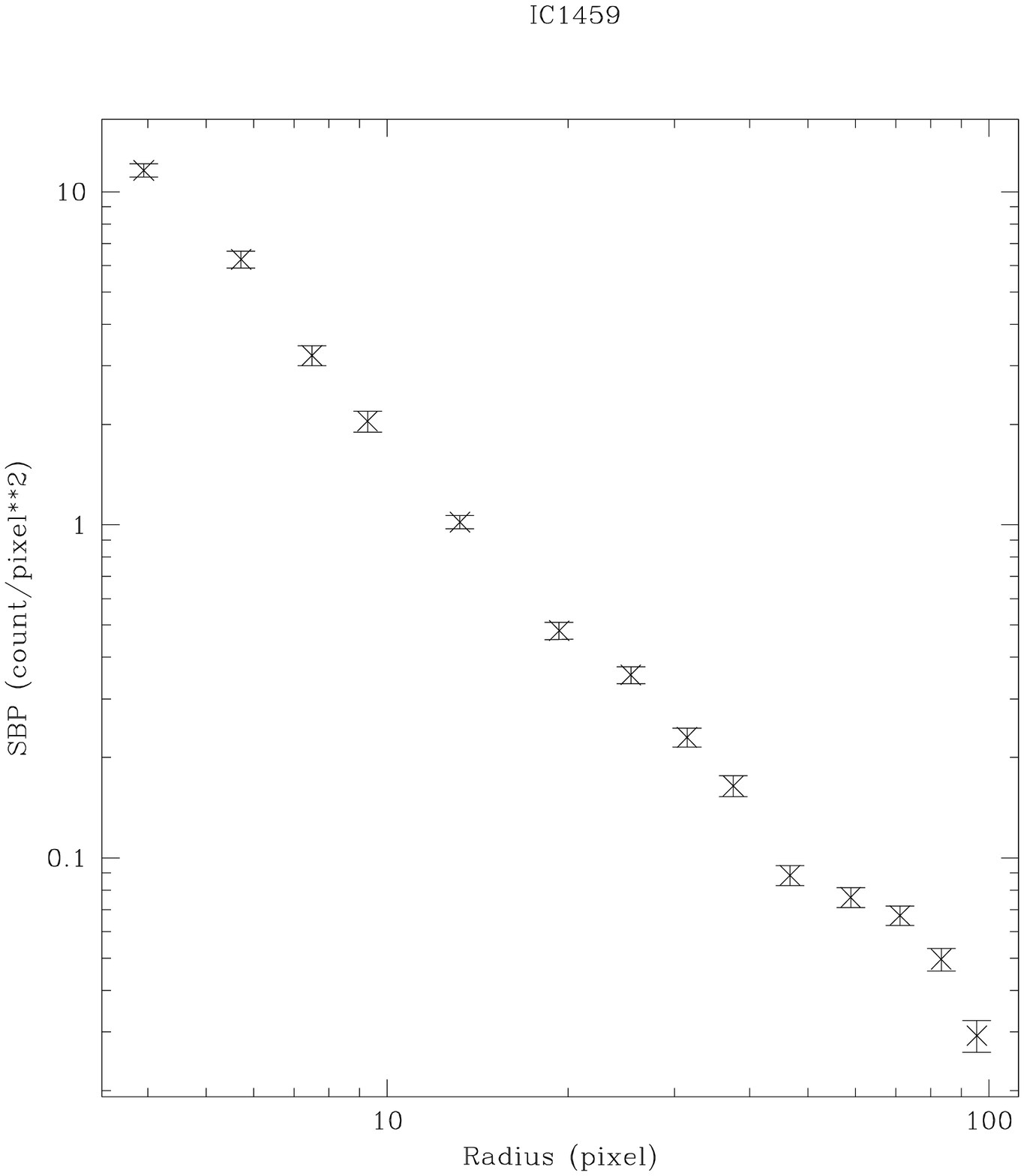,width=0.25\linewidth}
\psfig{figure=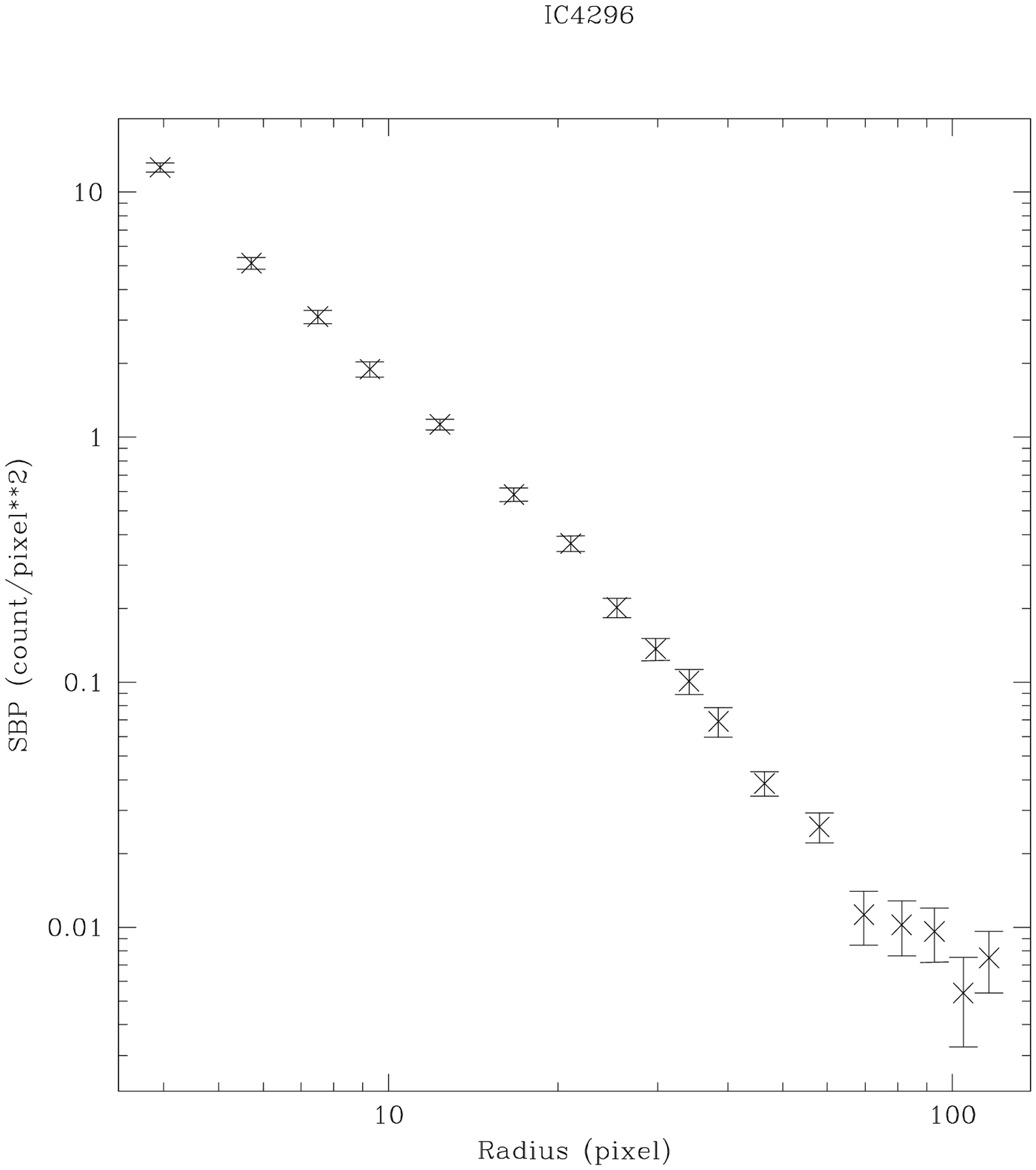,width=0.25\linewidth}
\psfig{figure=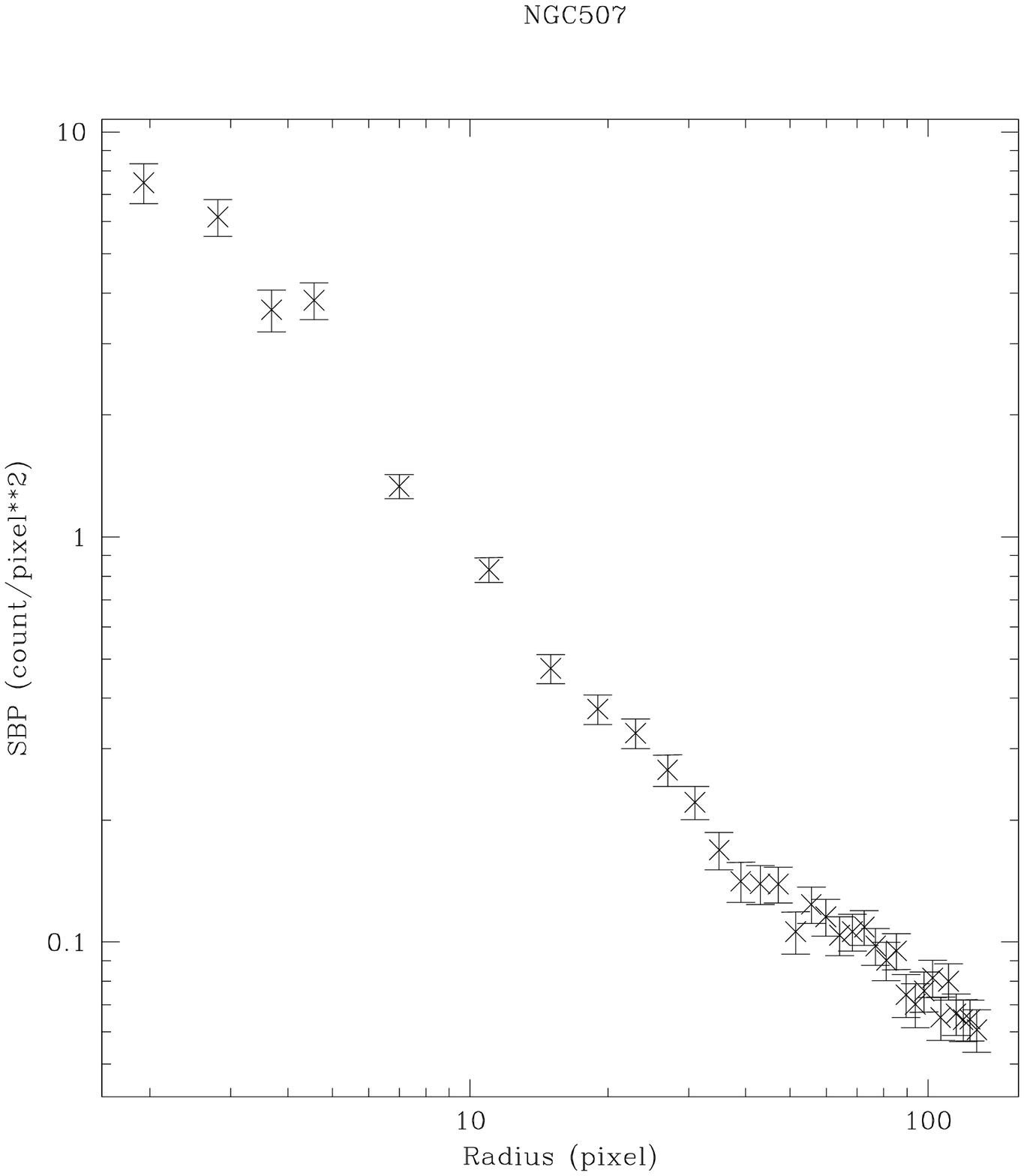,width=0.25\linewidth}}
\caption{The X-ray surface brightness profiles of the 43 galaxies 
of the sample, having excluded UGC~7203.}
\label{sbpprof}
\end{figure*}

\clearpage
\begin{figure*}
\centerline{
\psfig{figure=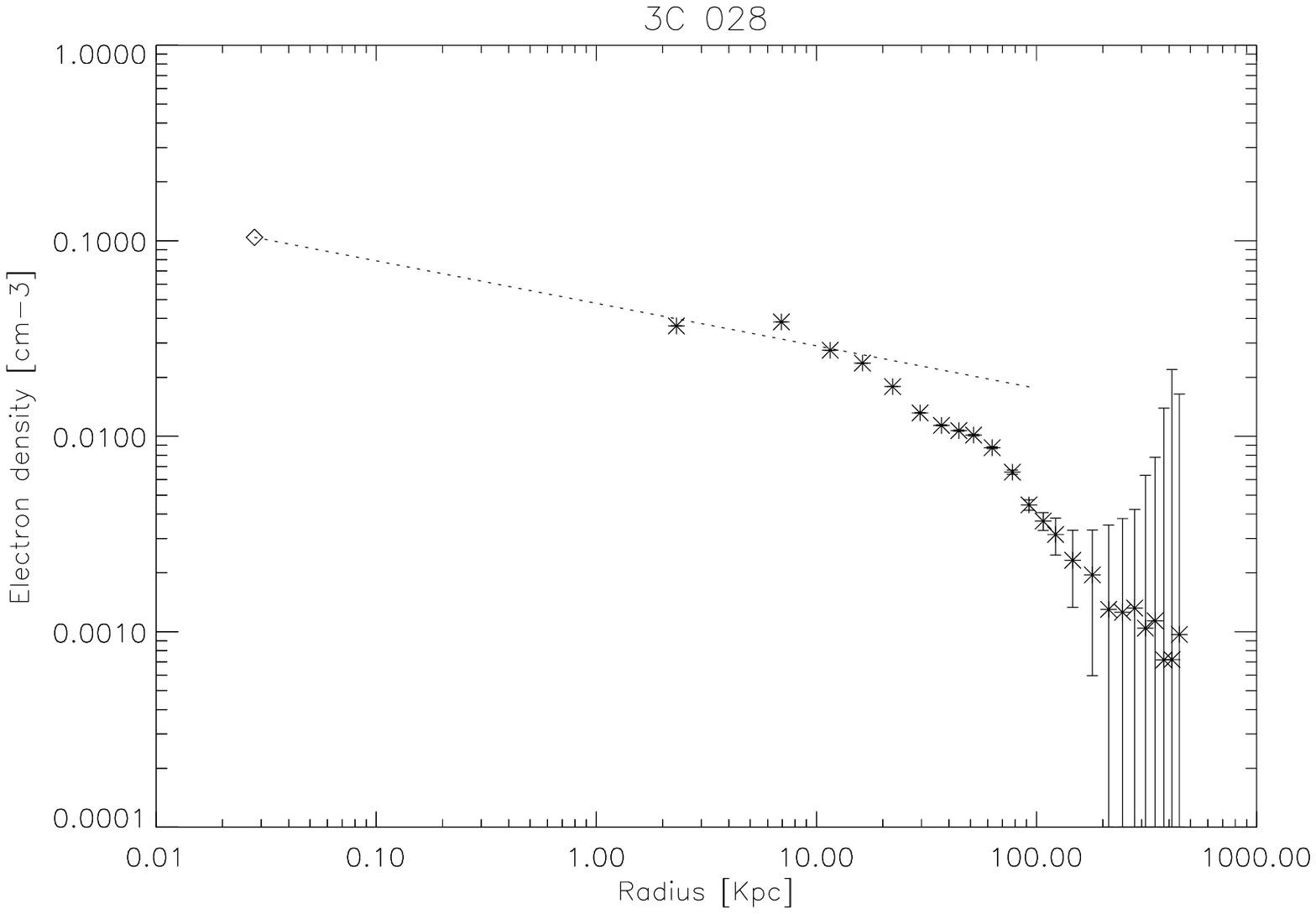,width=0.33\linewidth}
\psfig{figure=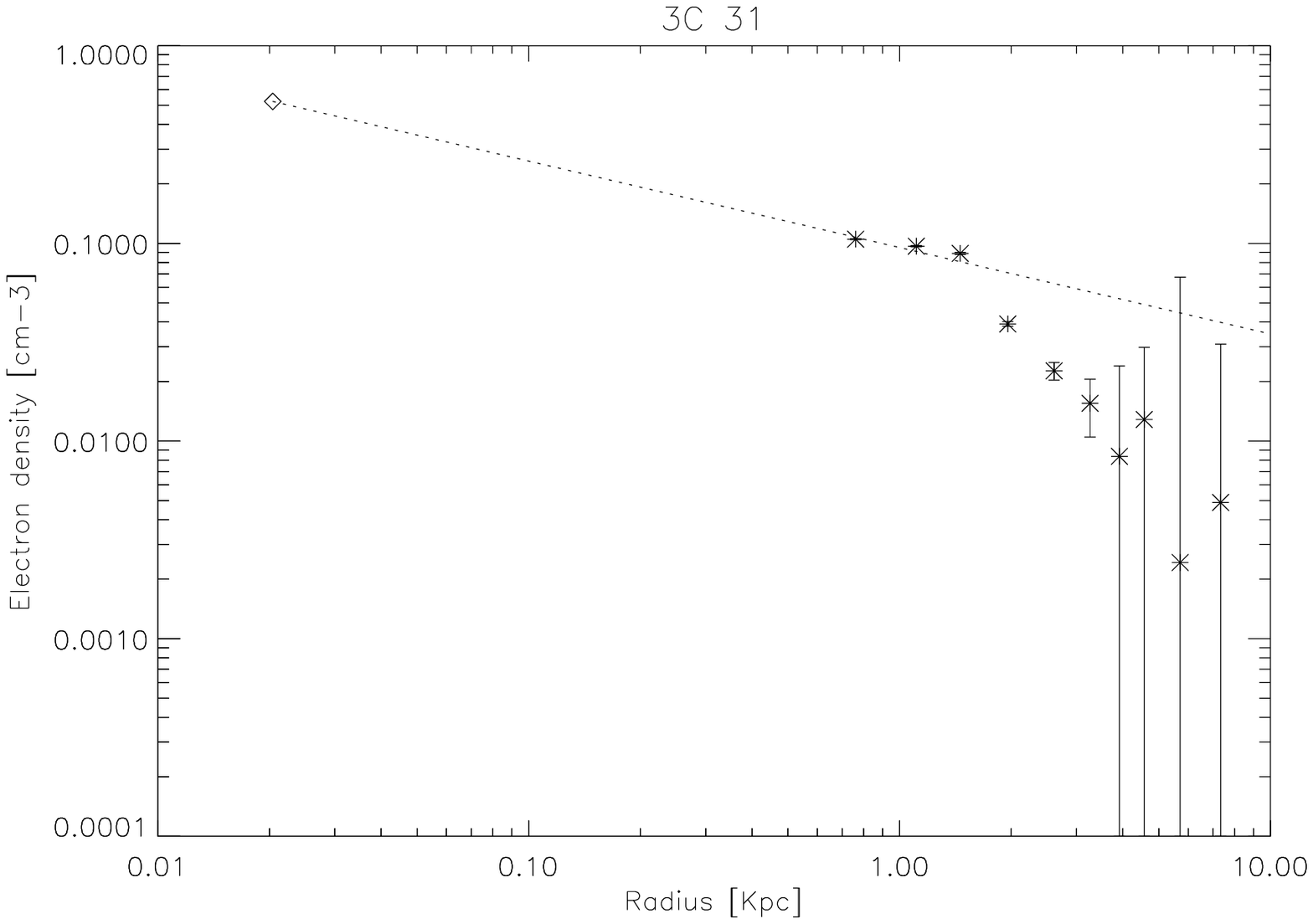,width=0.33\linewidth}
\psfig{figure=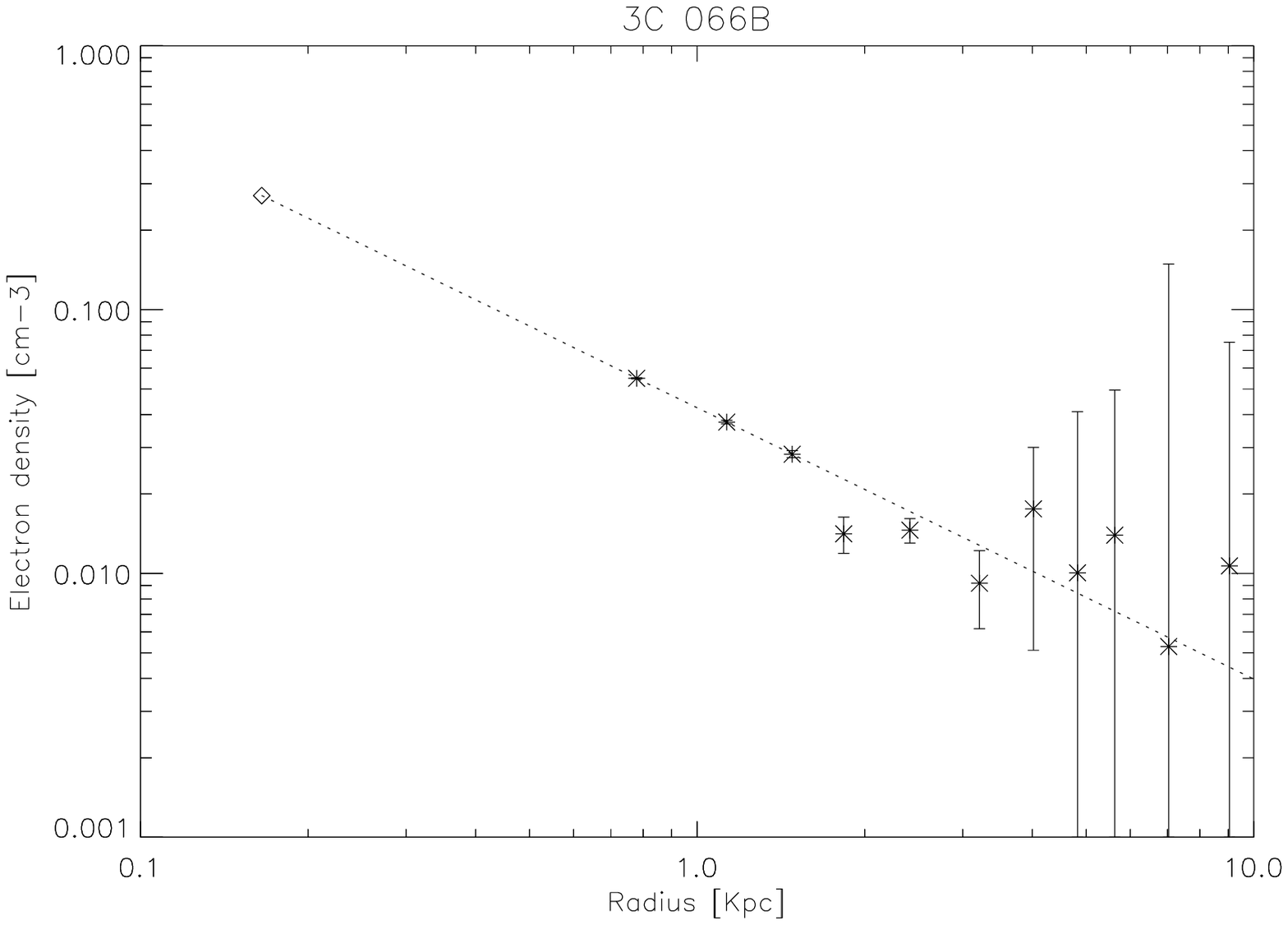,width=0.33\linewidth}}
\centerline{  		     
\psfig{figure=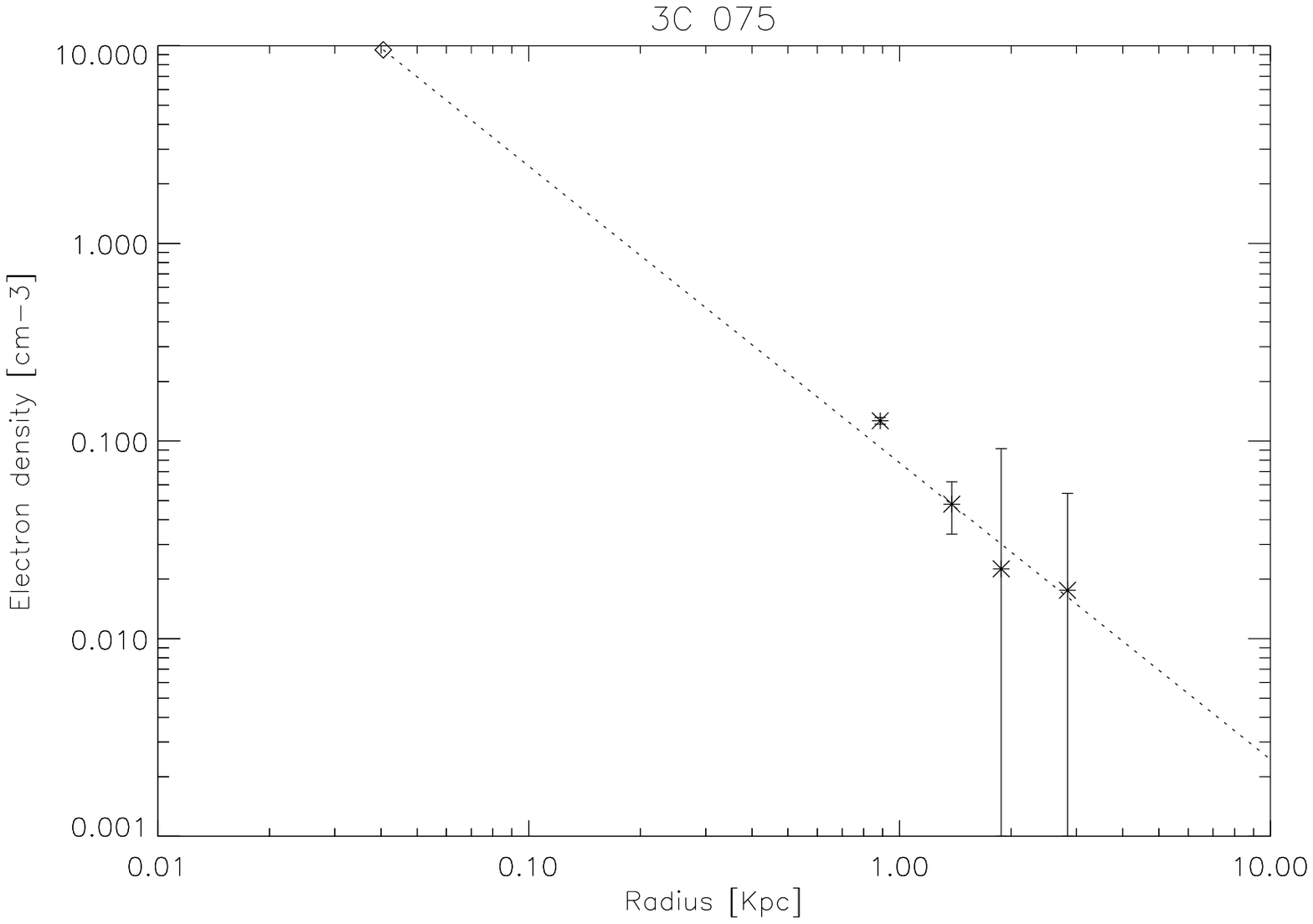,width=0.33\linewidth}
\psfig{figure=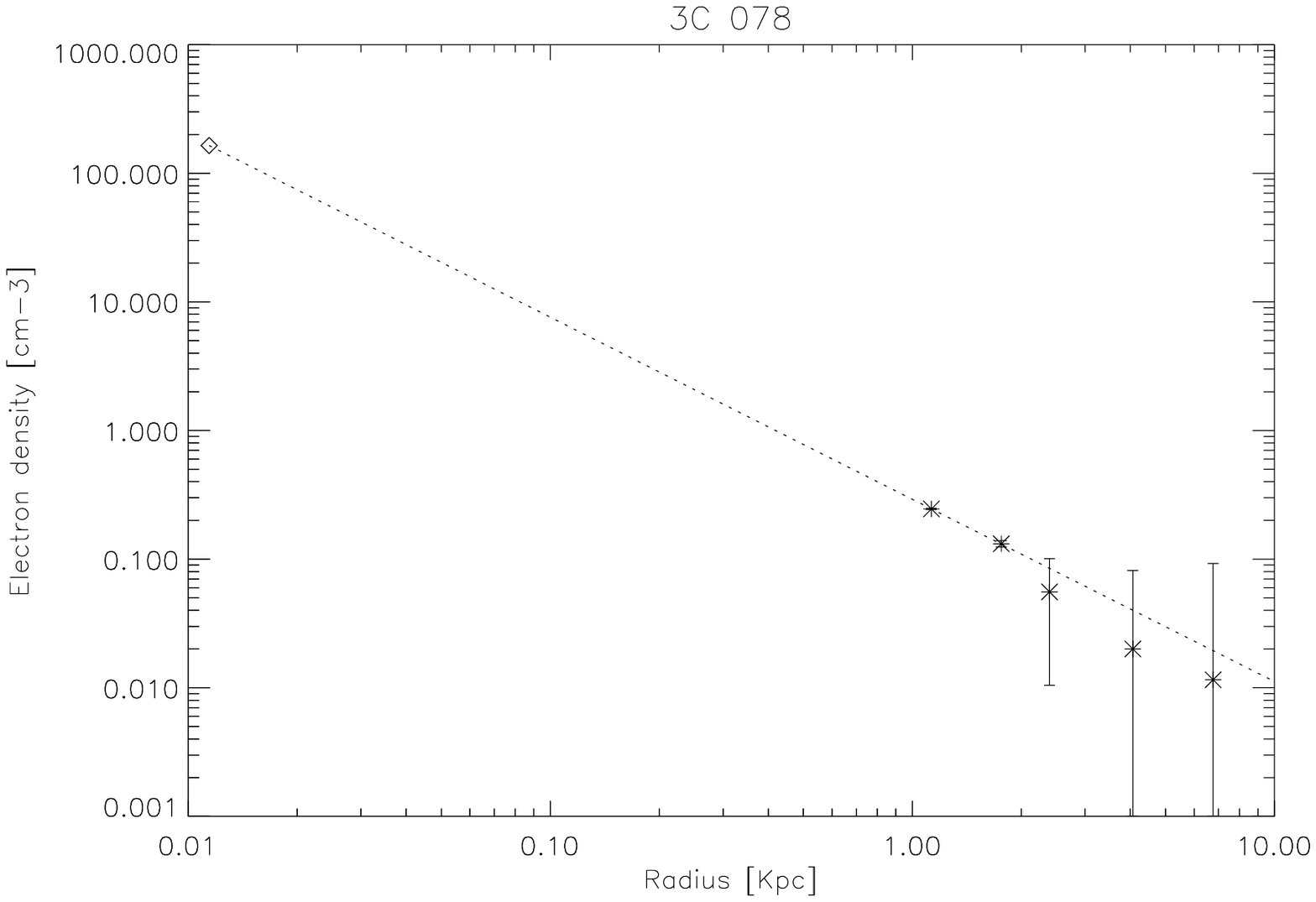,width=0.33\linewidth}
\psfig{figure=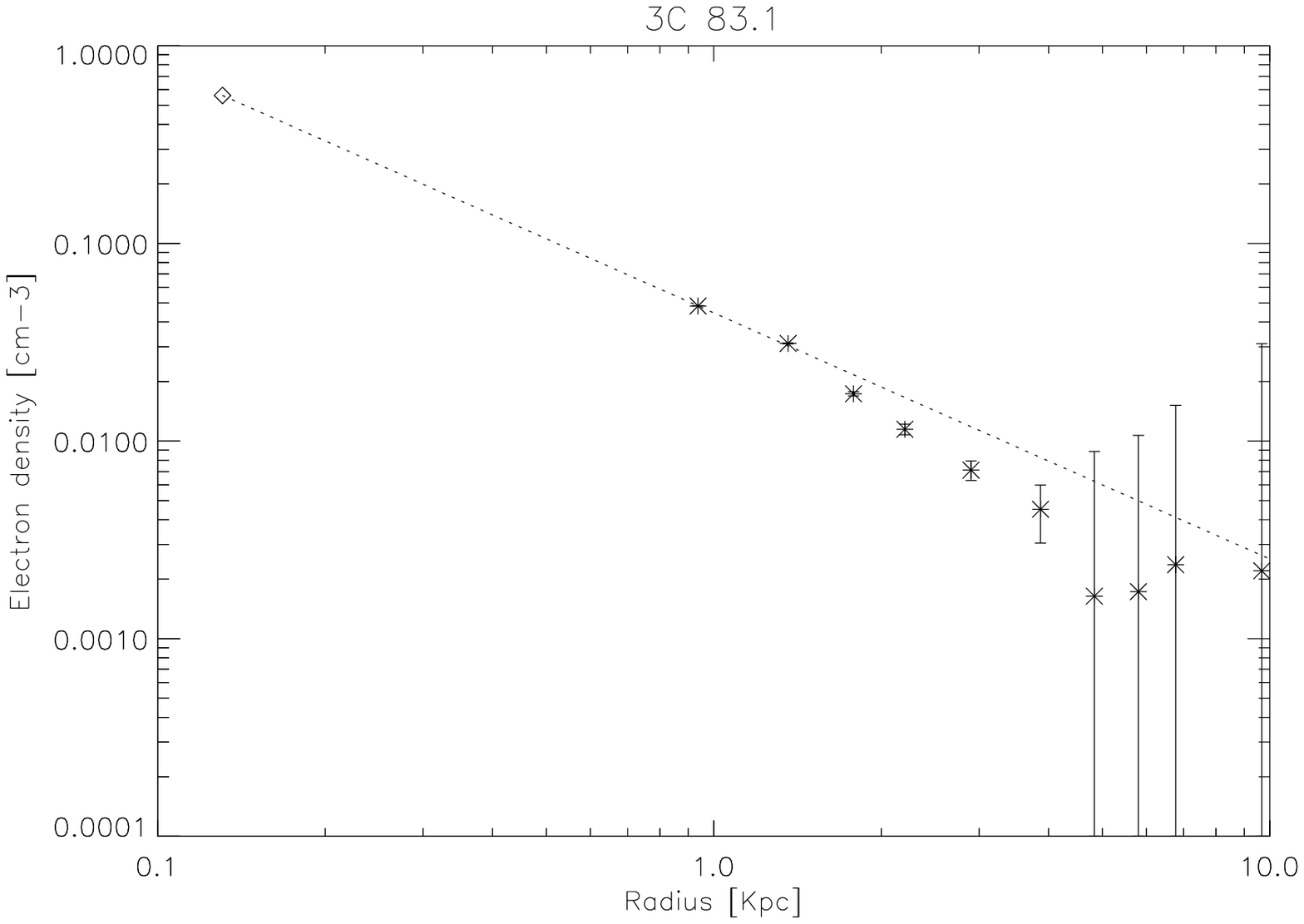,width=0.33\linewidth}}
\centerline{  		     
\psfig{figure=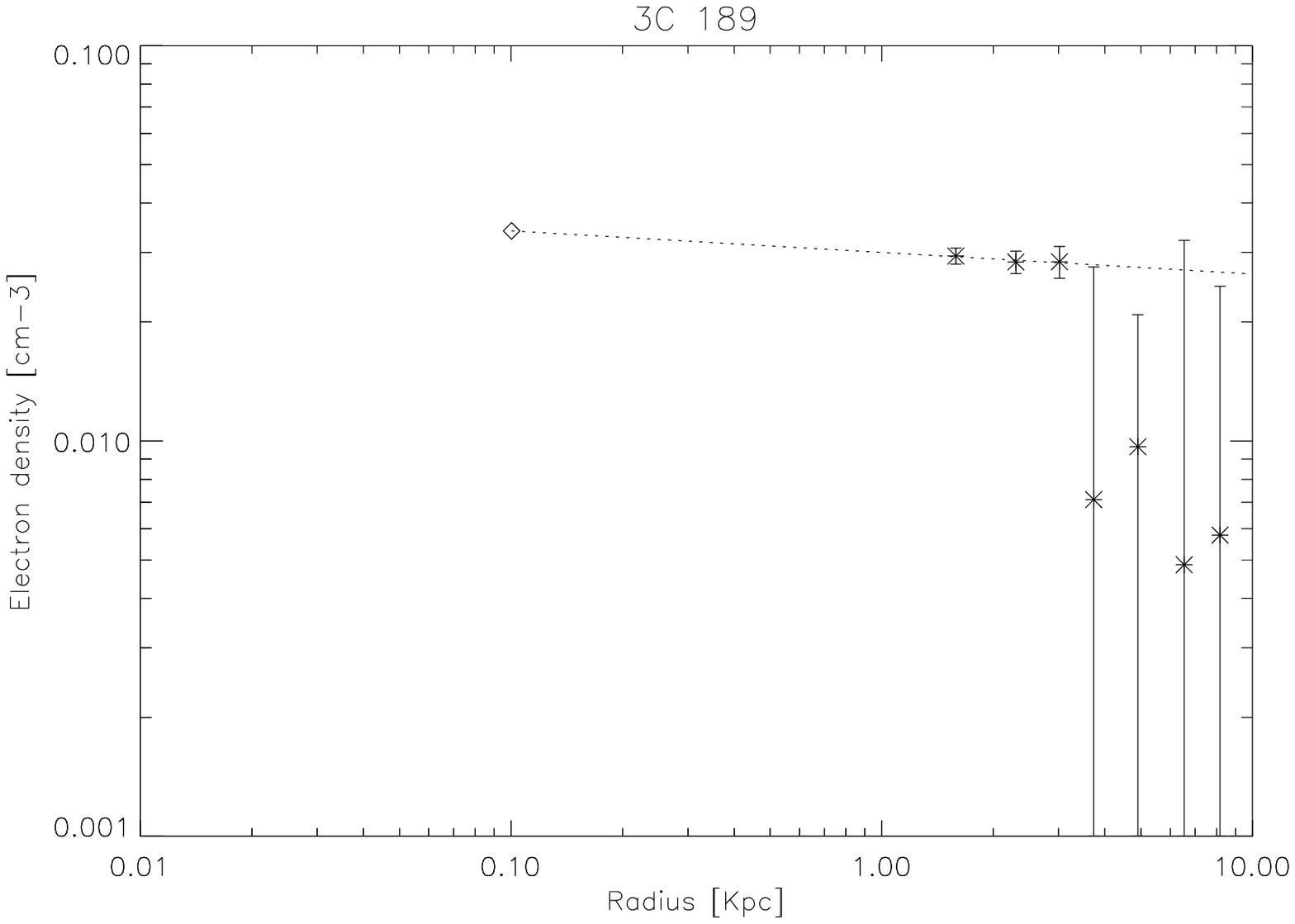,width=0.33\linewidth}
\psfig{figure=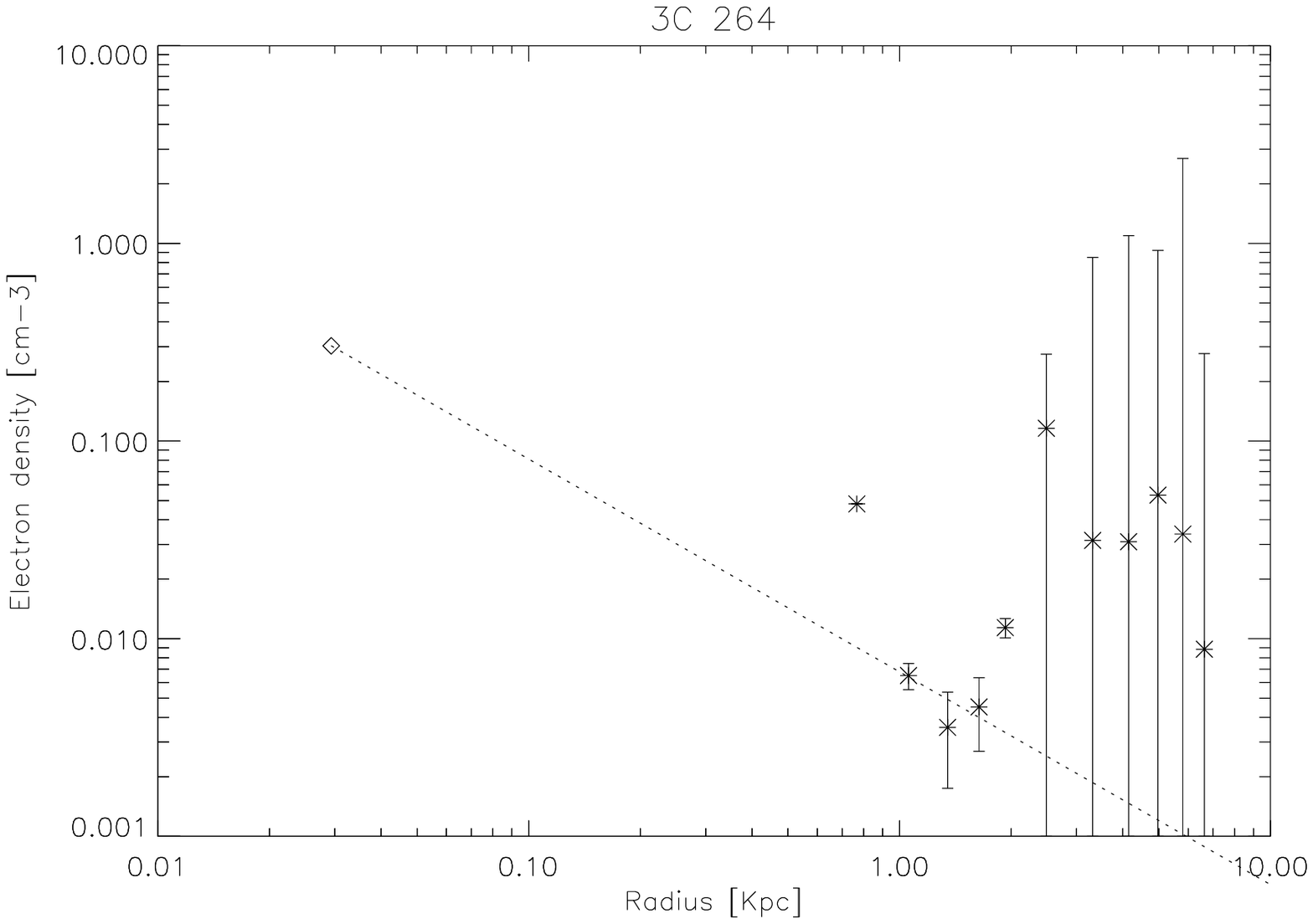,width=0.33\linewidth}
\psfig{figure=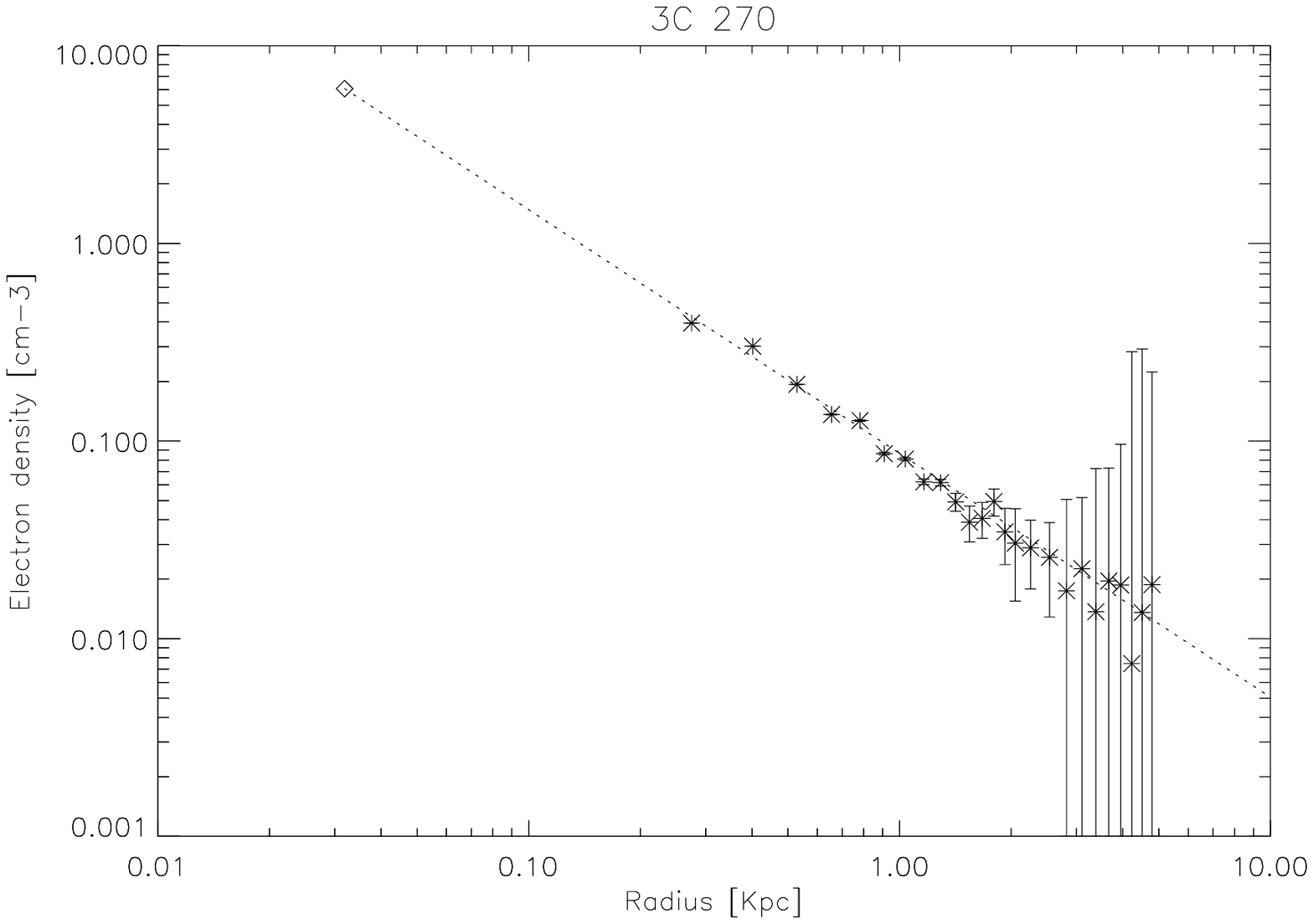,width=0.33\linewidth}}
\centerline{  		     
\psfig{figure=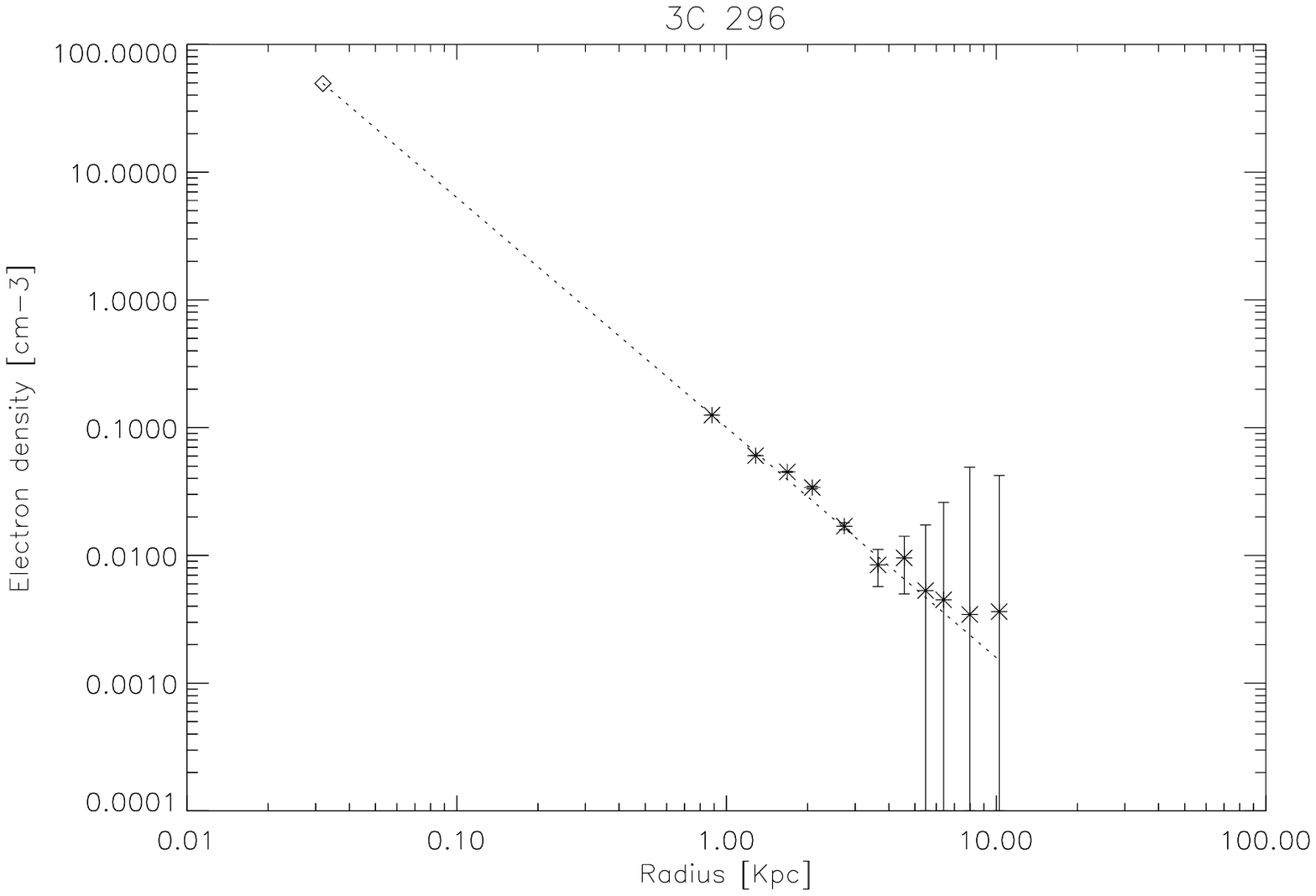,width=0.33\linewidth}
\psfig{figure=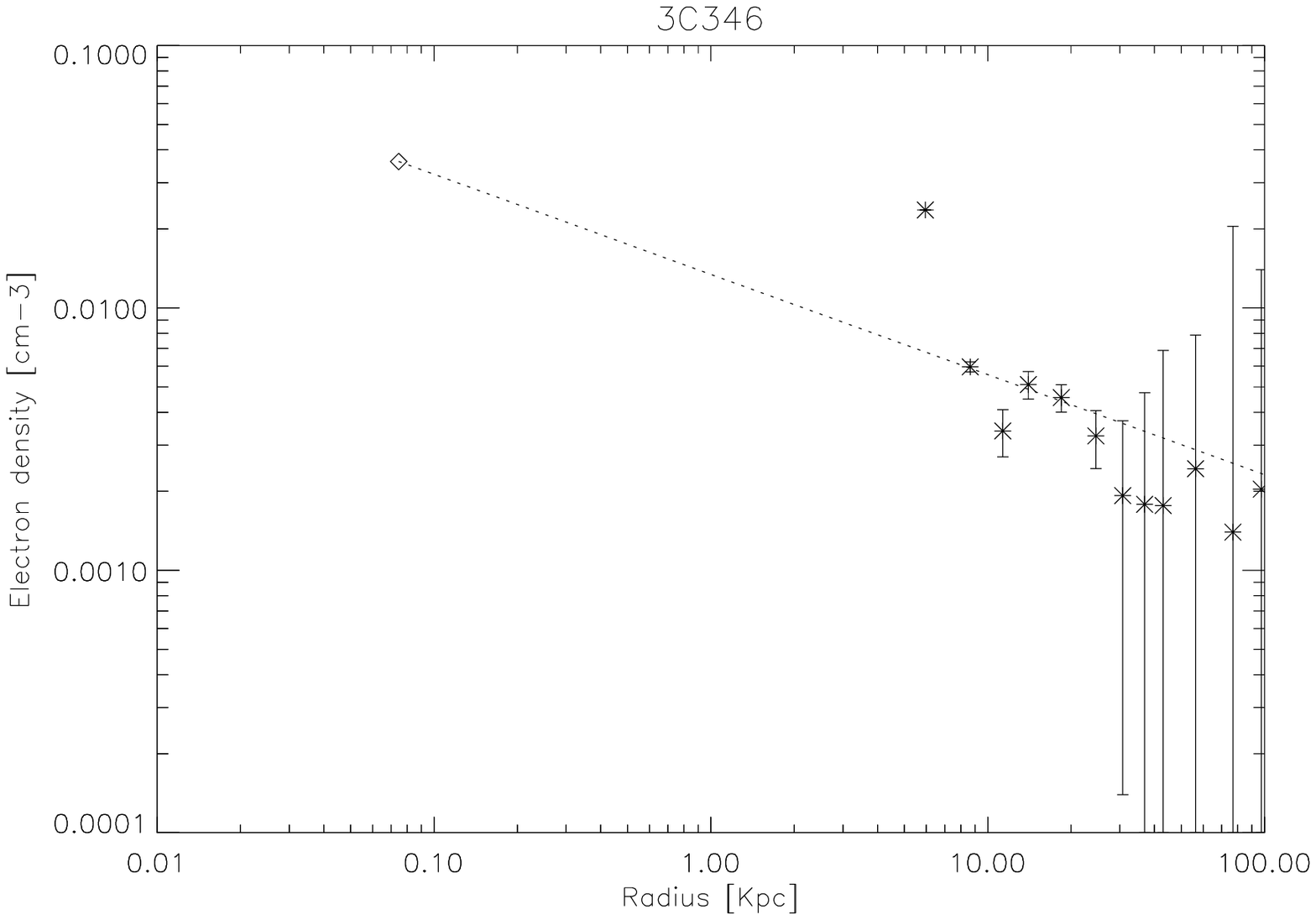,width=0.33\linewidth}
\psfig{figure=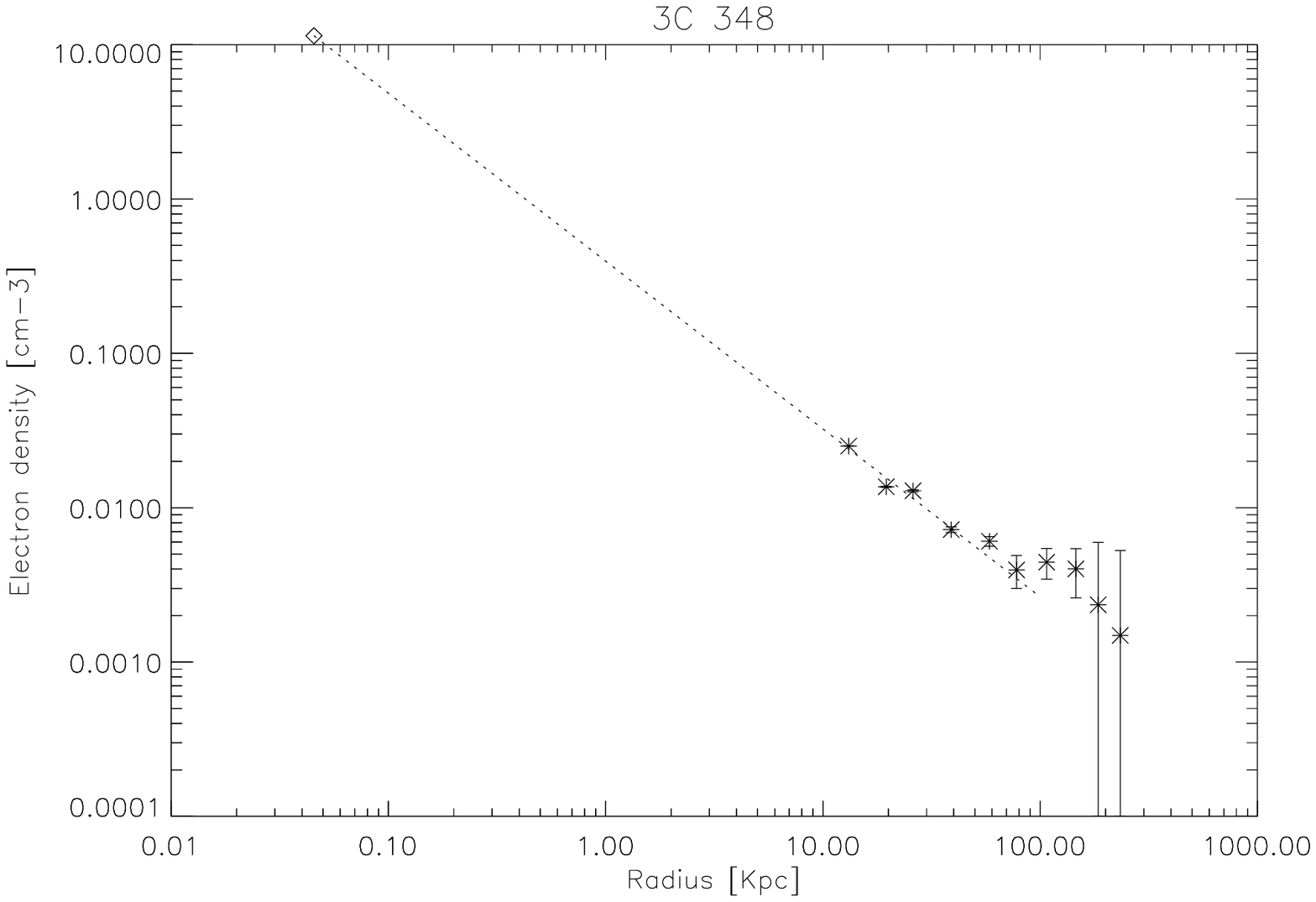,width=0.33\linewidth}}
\centerline{
\psfig{figure=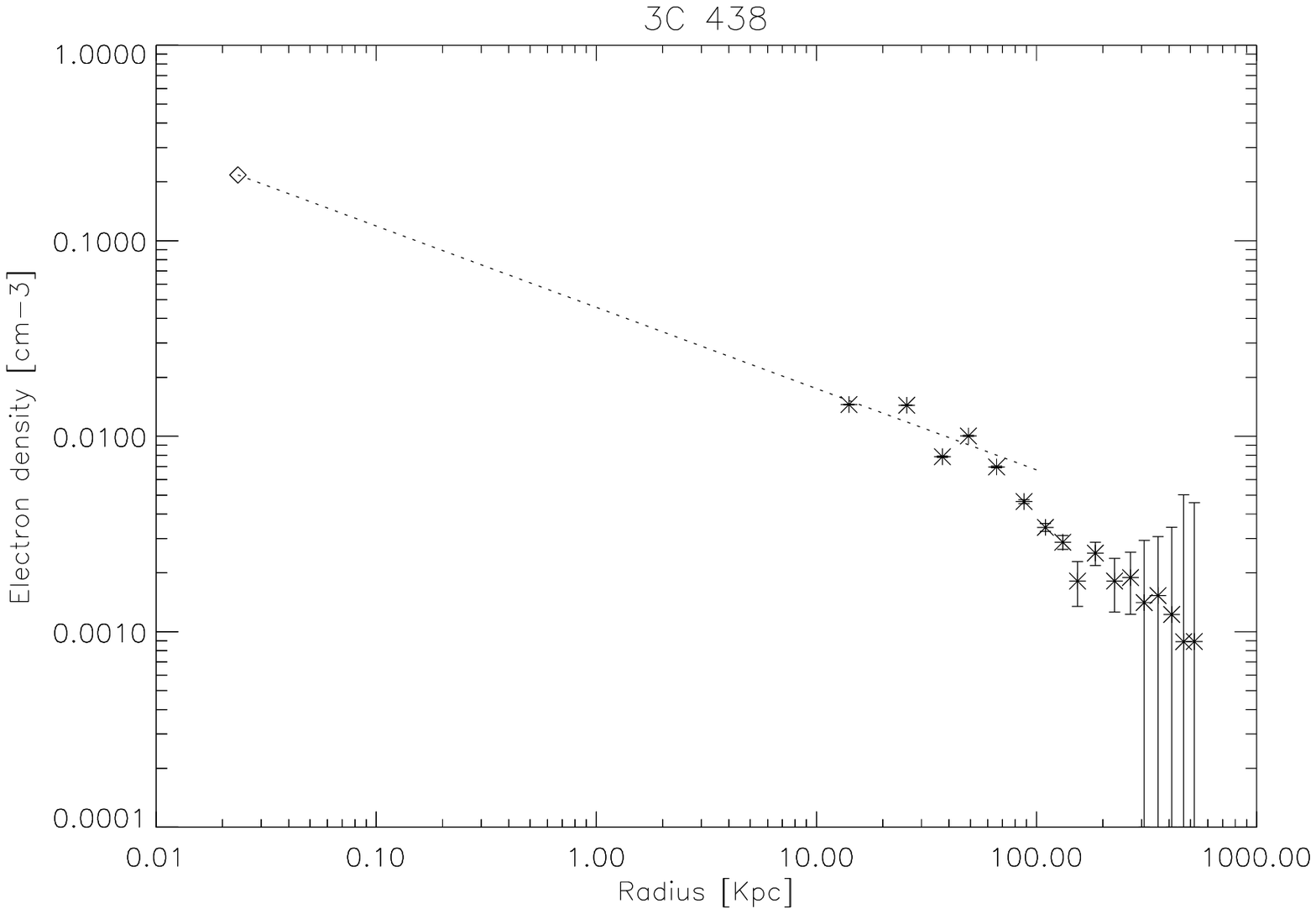,width=0.33\linewidth}
\psfig{figure=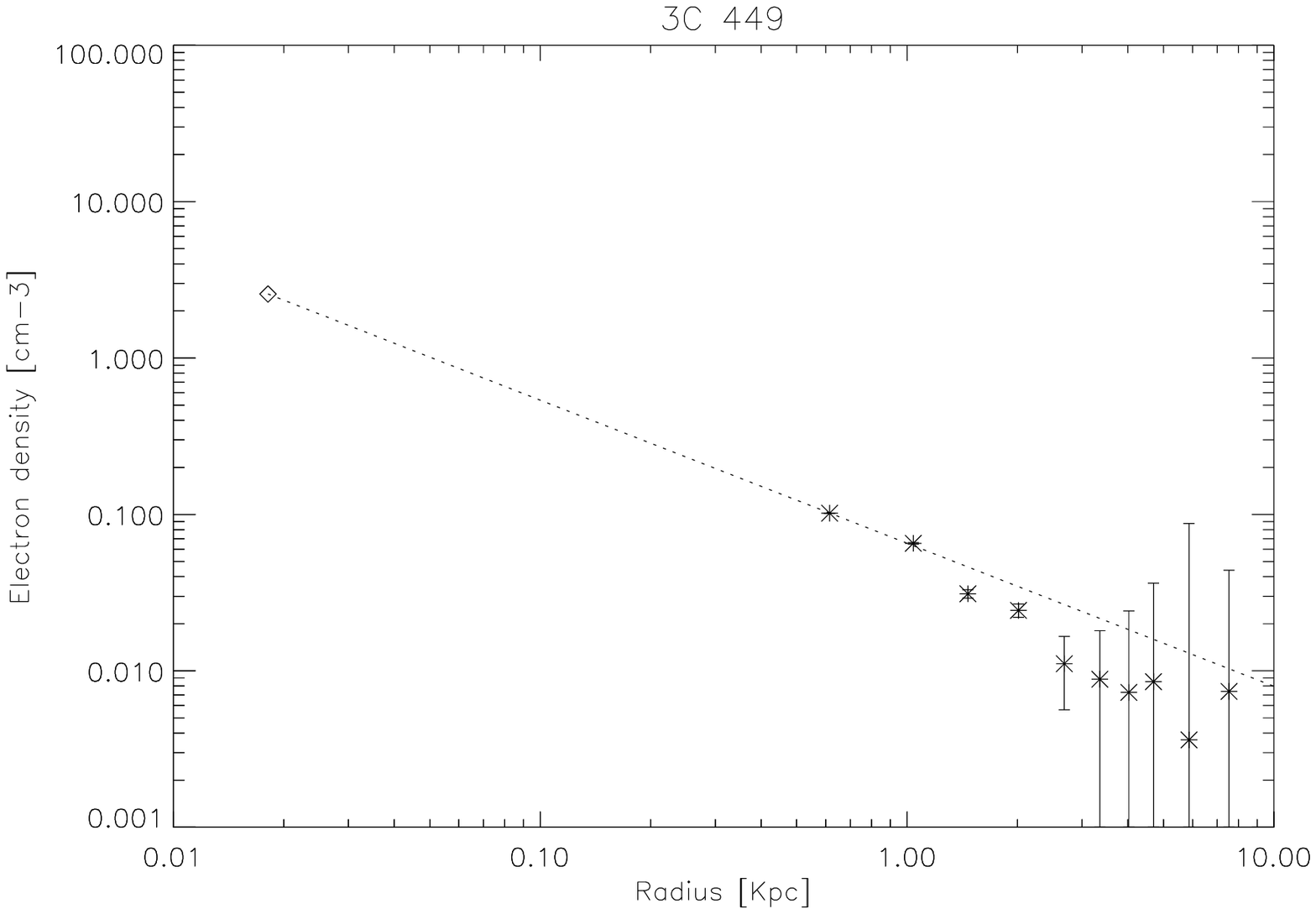,width=0.33\linewidth}
\psfig{figure=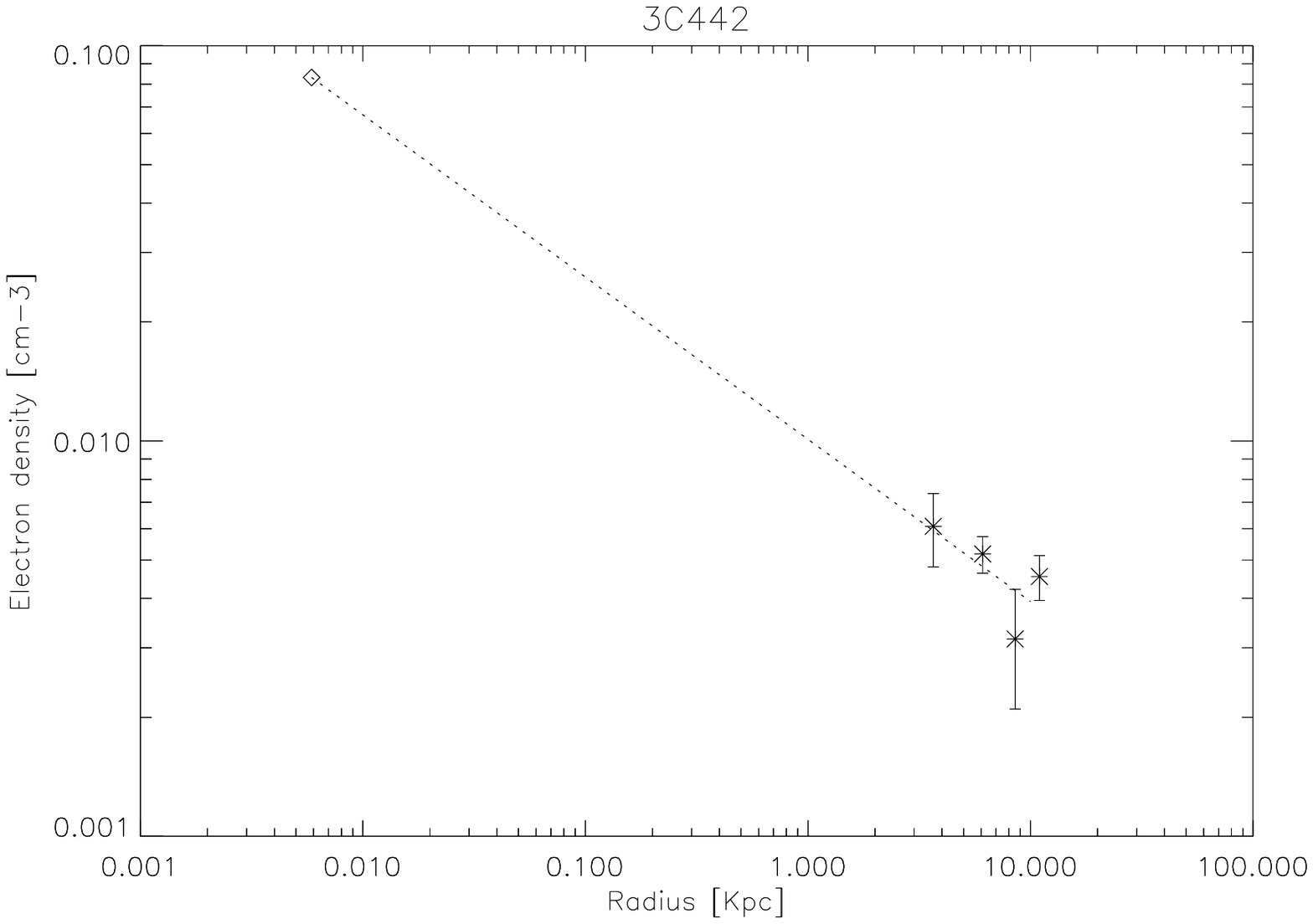,width=0.33\linewidth}}
\end{figure*}
\begin{figure*}
\centerline{  		     
\psfig{figure=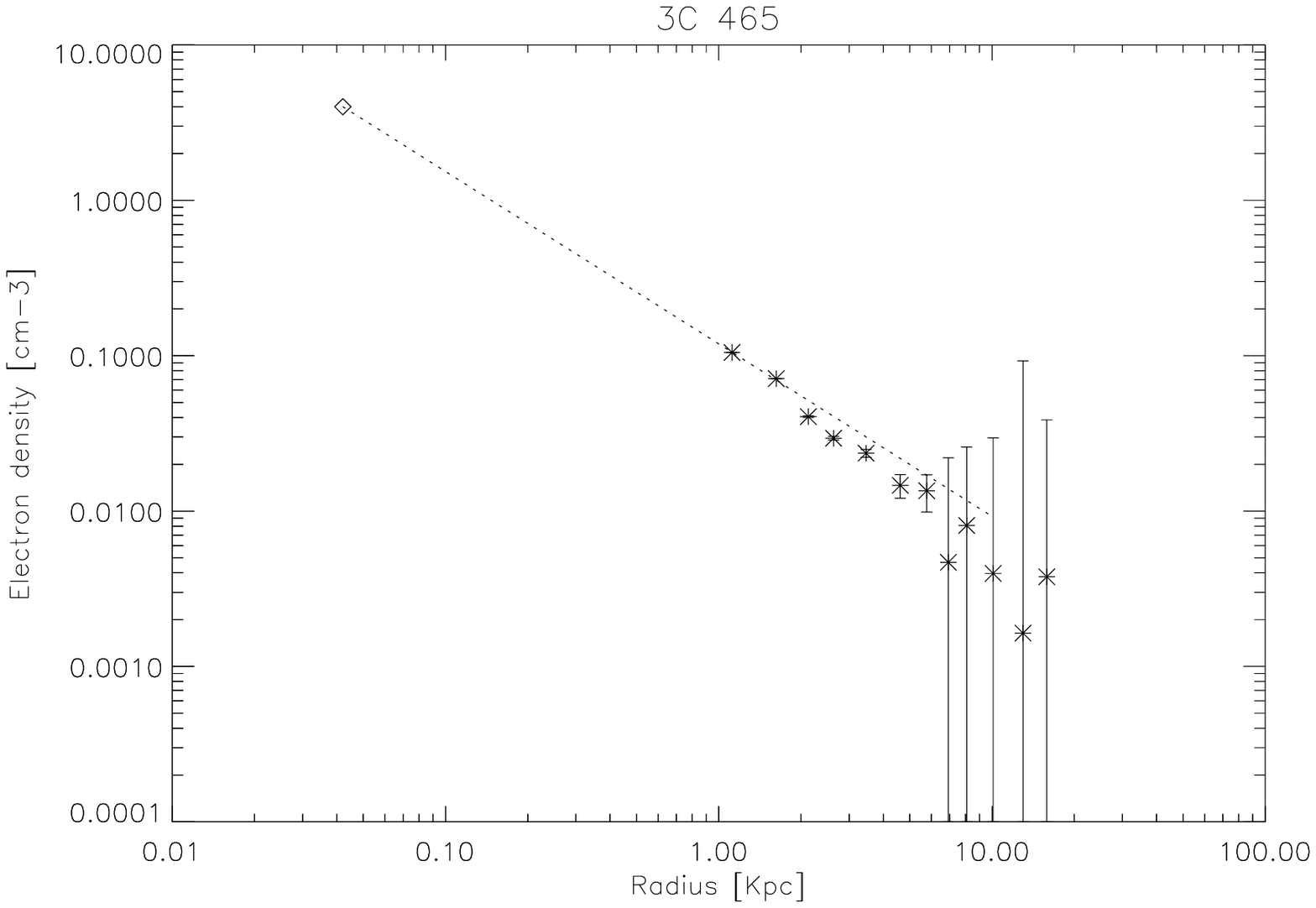,width=0.33\linewidth}
\psfig{figure=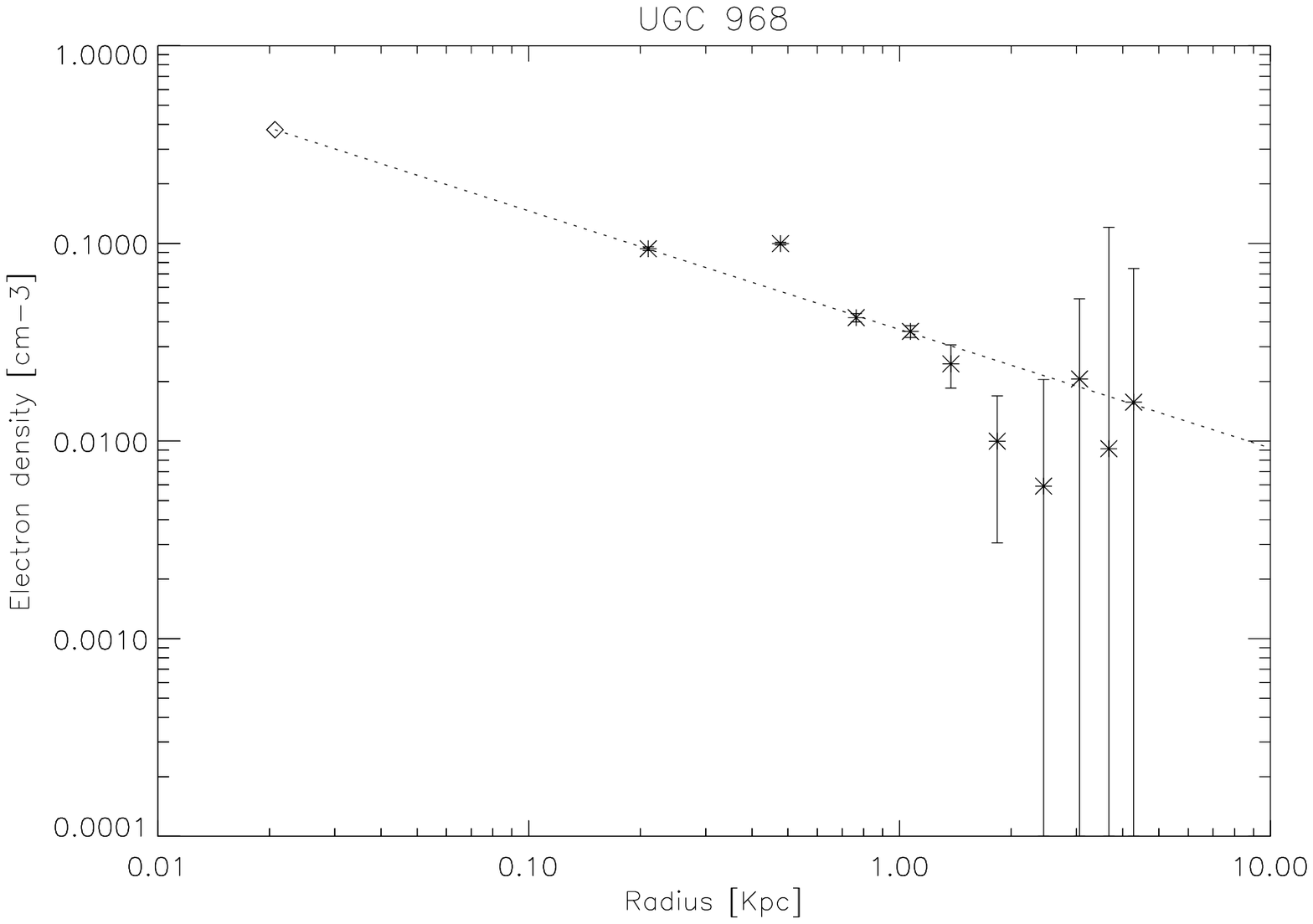,width=0.33\linewidth}
\psfig{figure=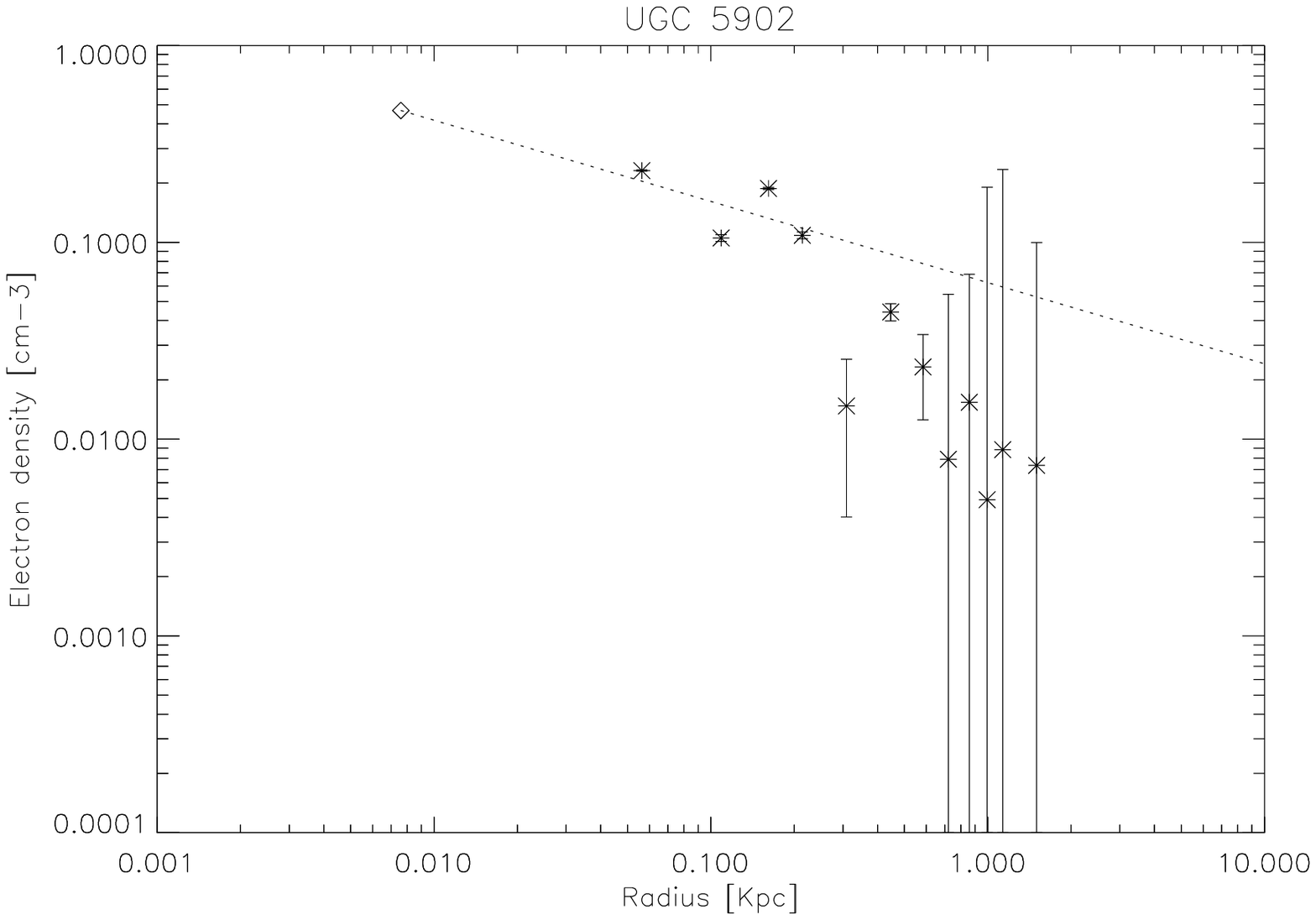,width=0.33\linewidth}}
\centerline{  		     
\psfig{figure=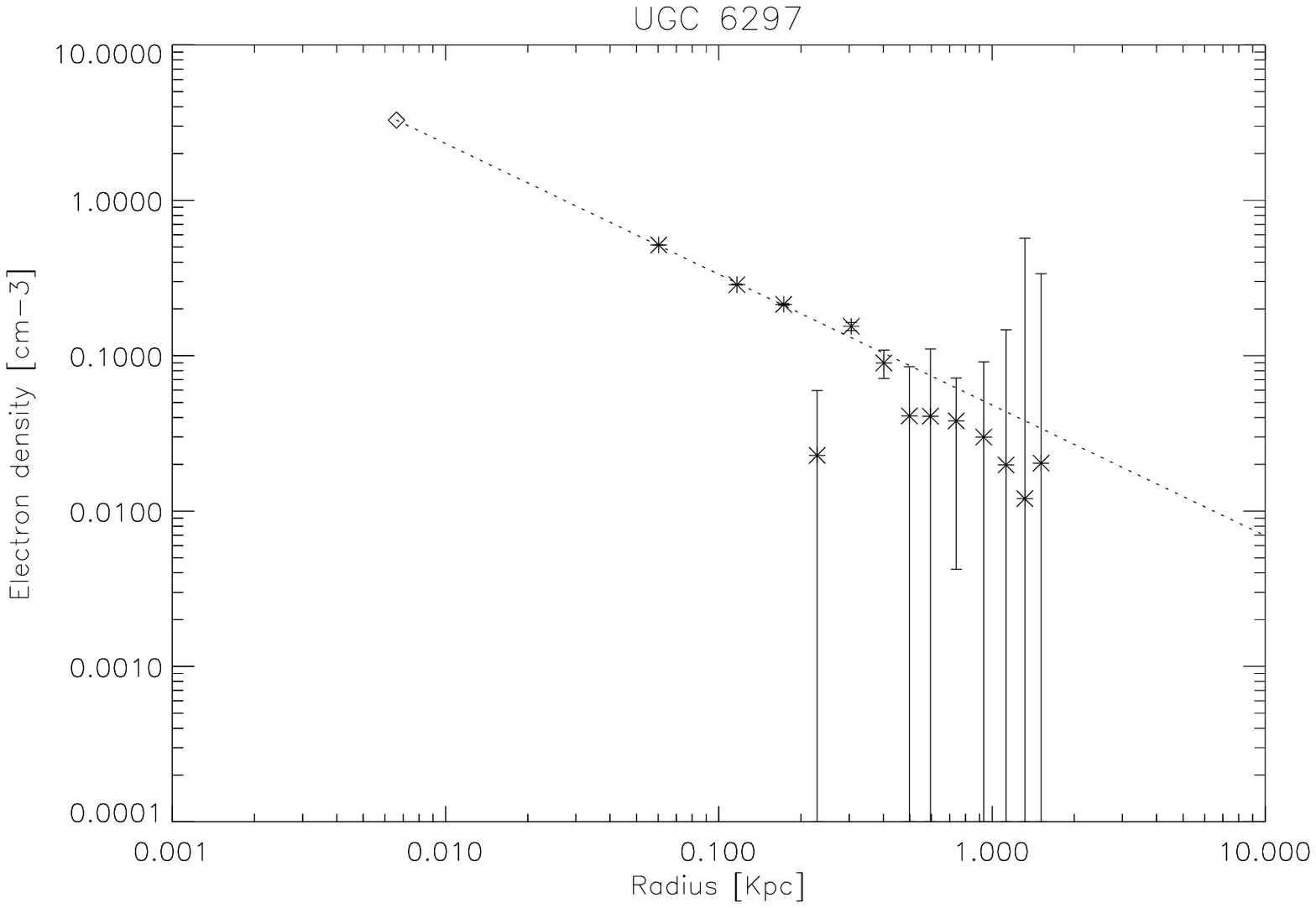,width=0.33\linewidth}
\psfig{figure=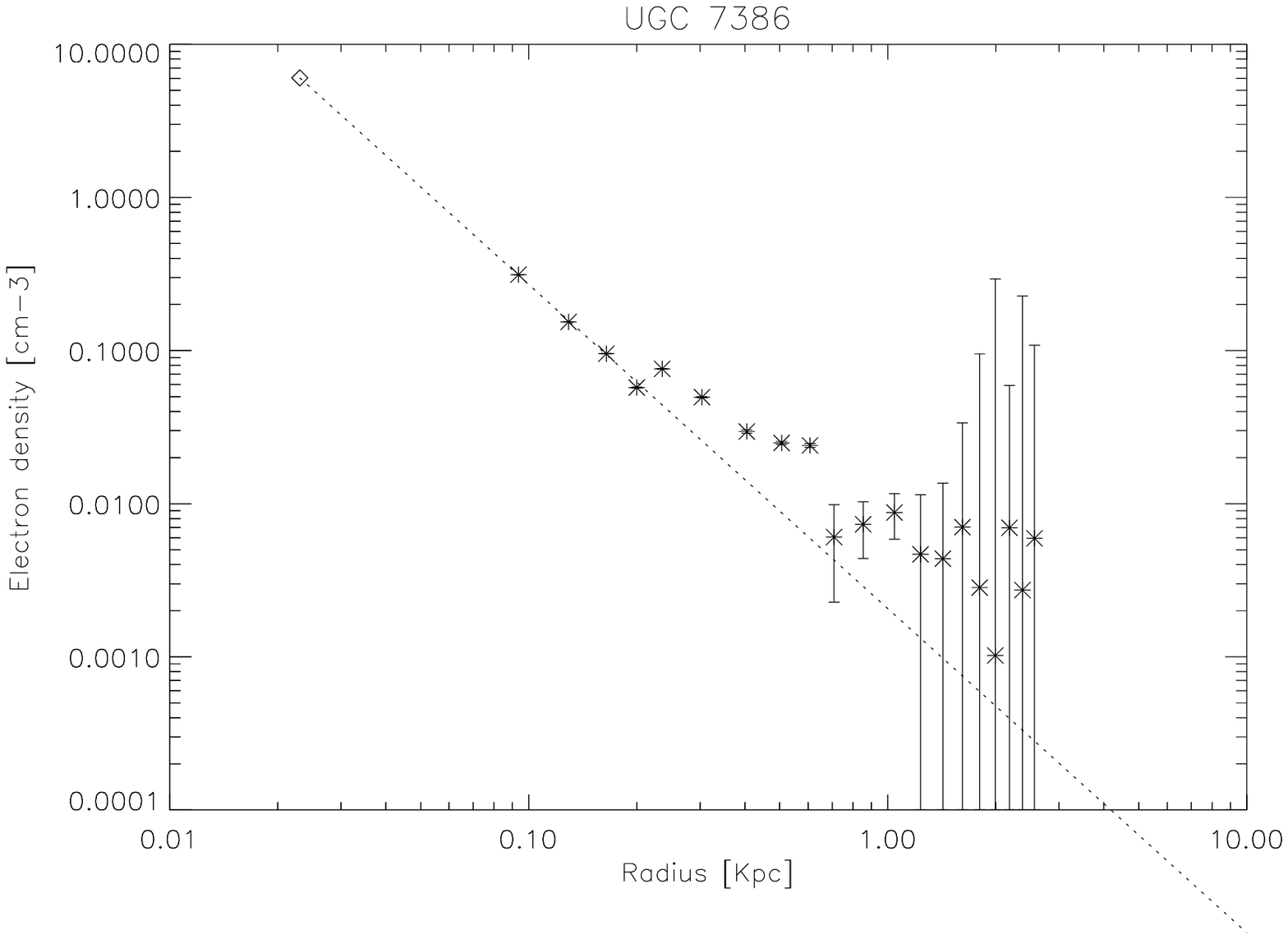,width=0.33\linewidth}
\psfig{figure=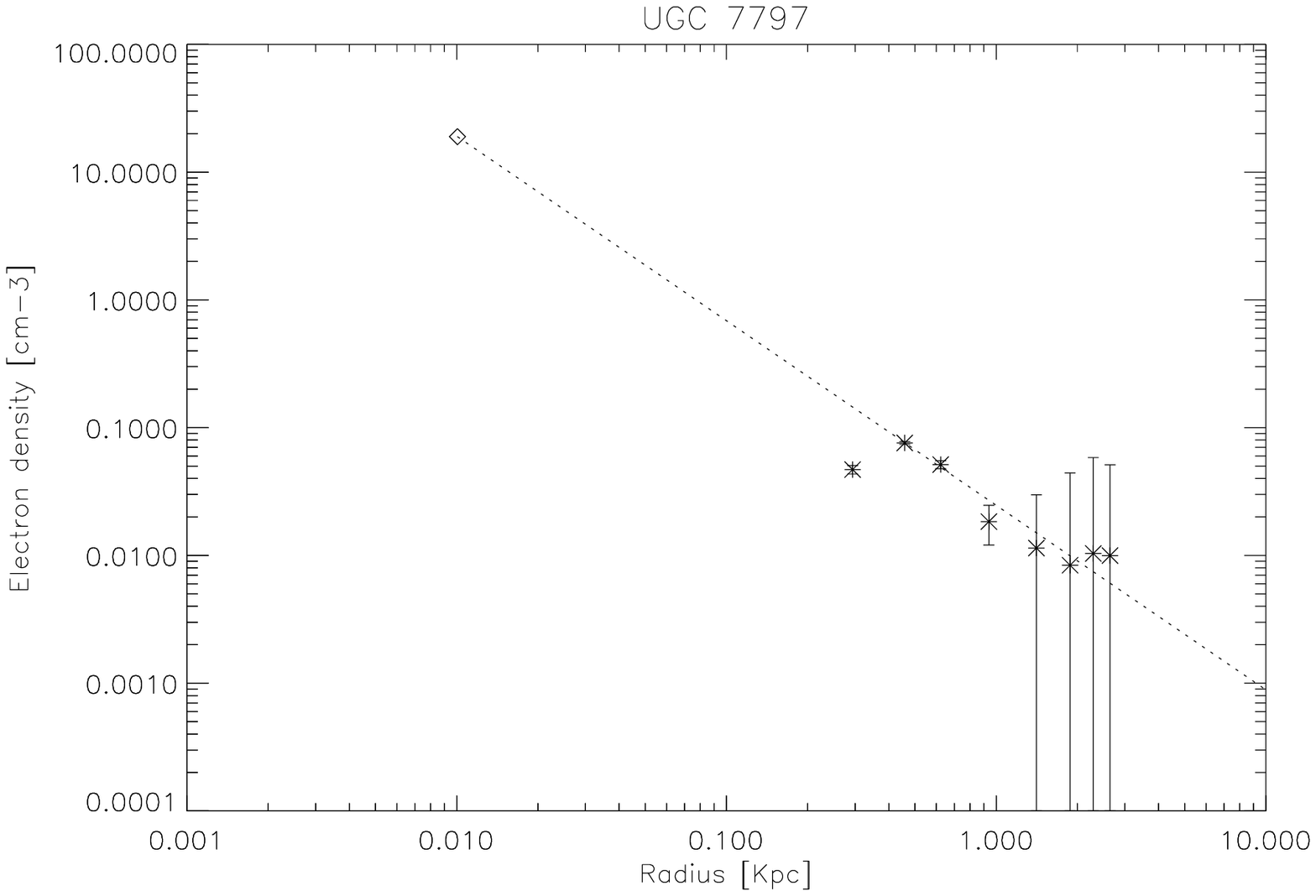,width=0.33\linewidth}}
\centerline{  		     
\psfig{figure=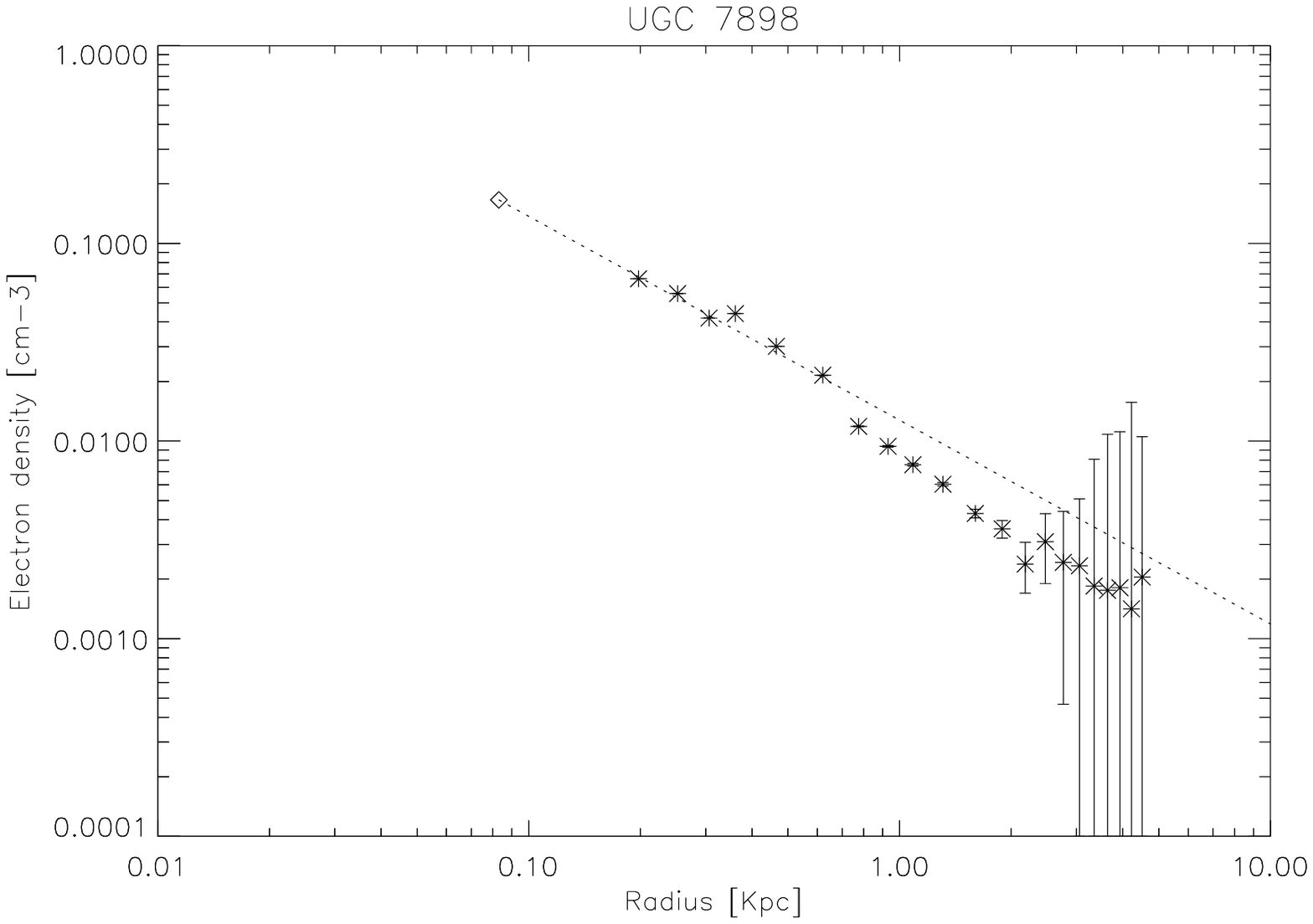,width=0.33\linewidth}
\psfig{figure=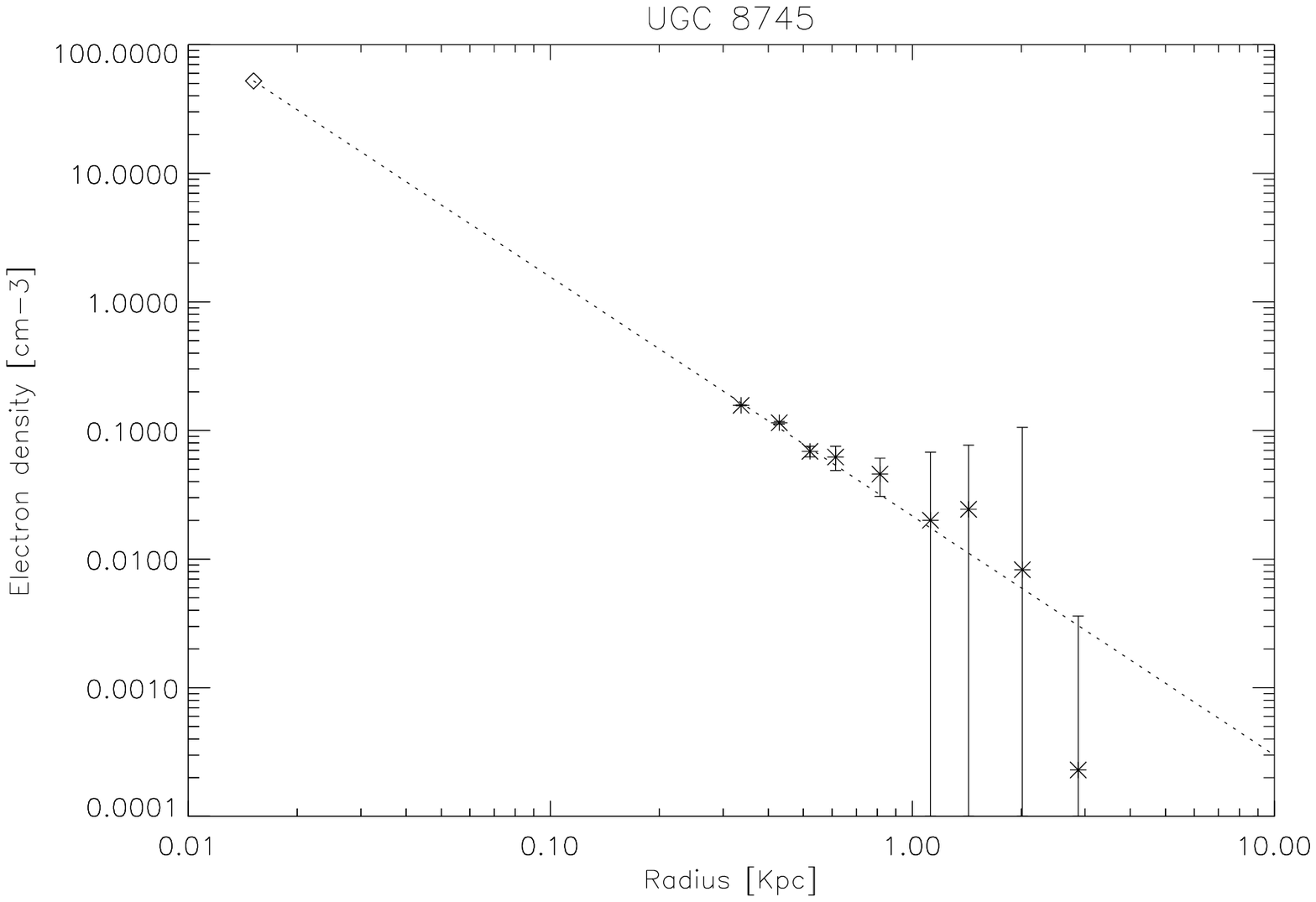,width=0.33\linewidth}
\psfig{figure=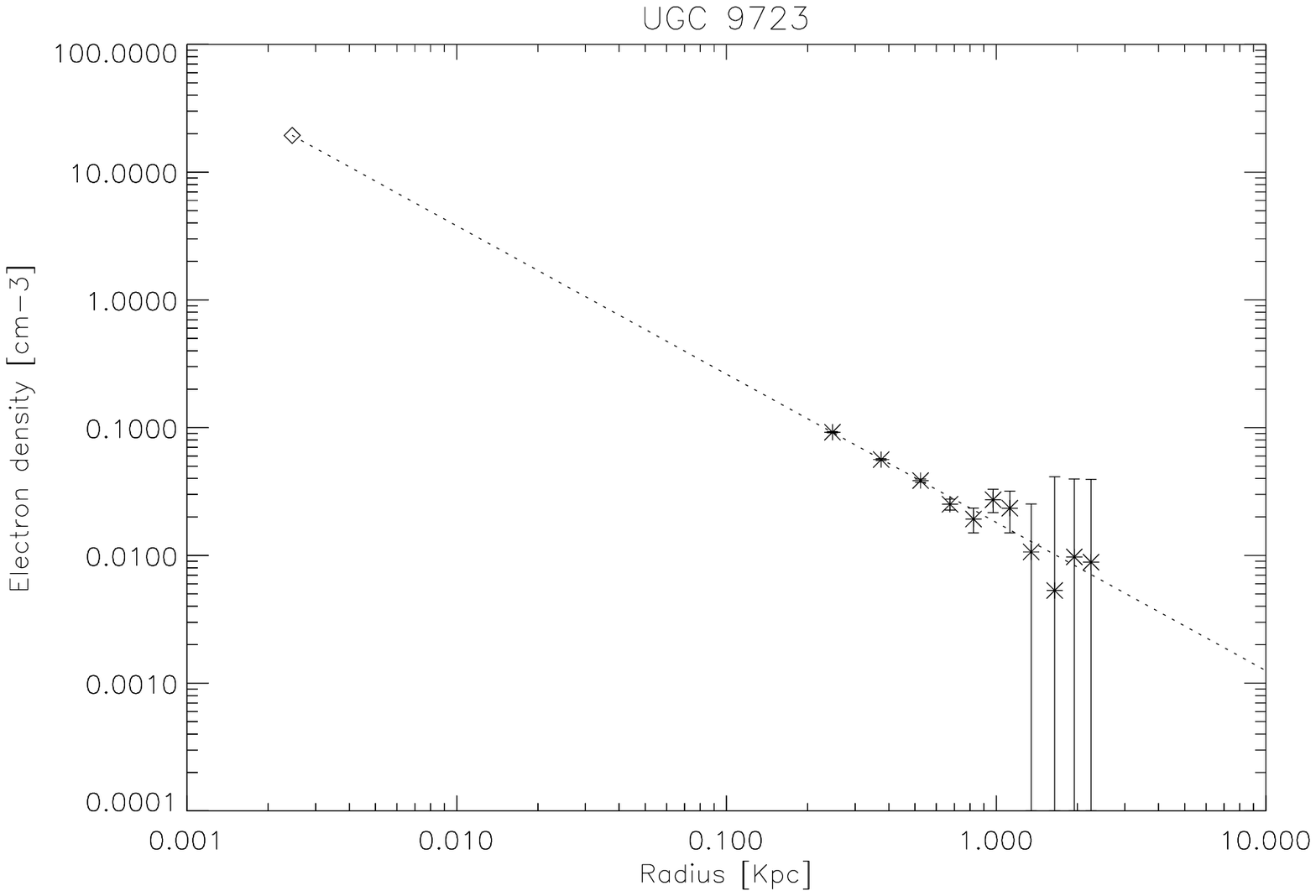,width=0.33\linewidth}}
\centerline{
\psfig{figure=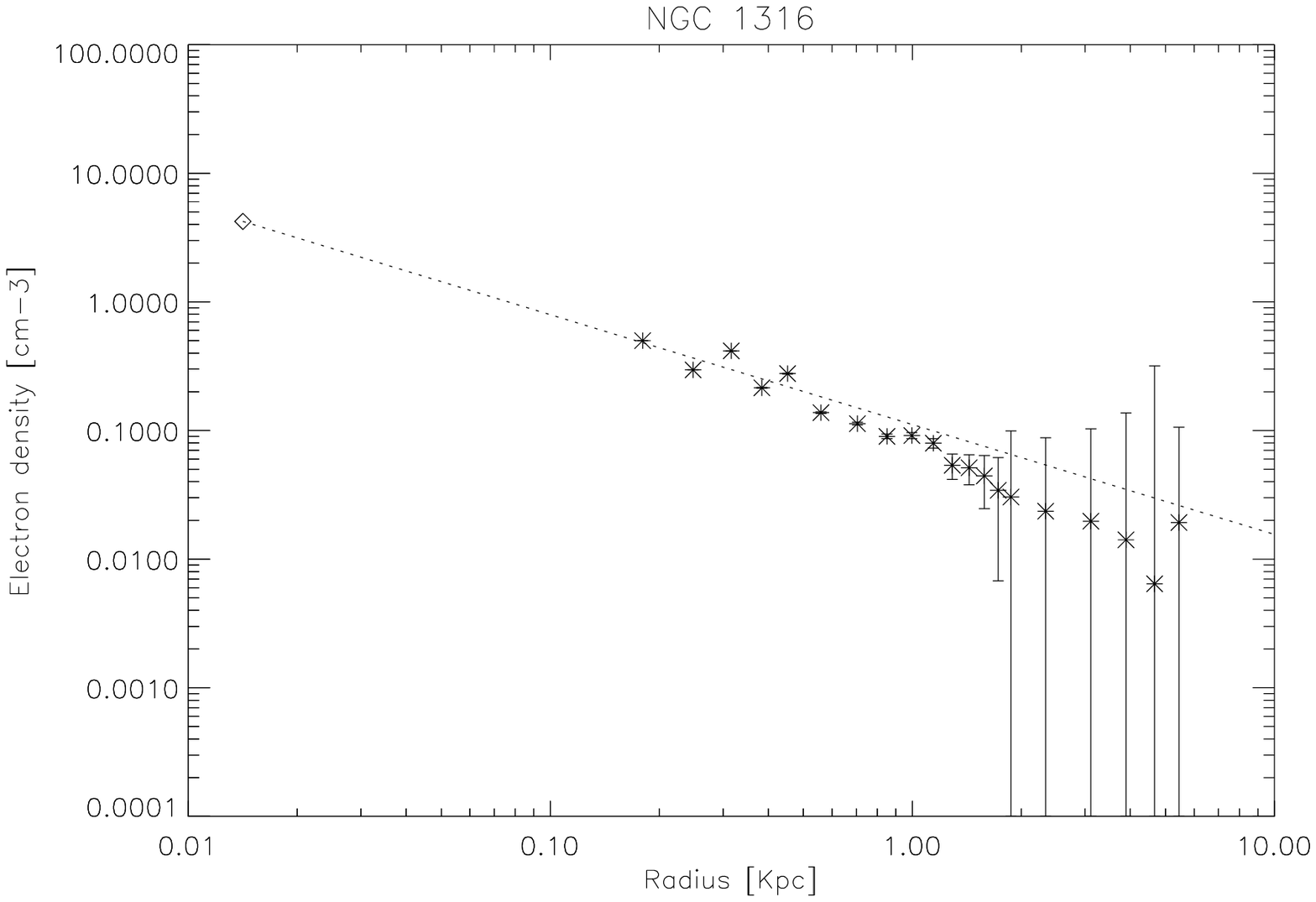,width=0.33\linewidth}
\psfig{figure=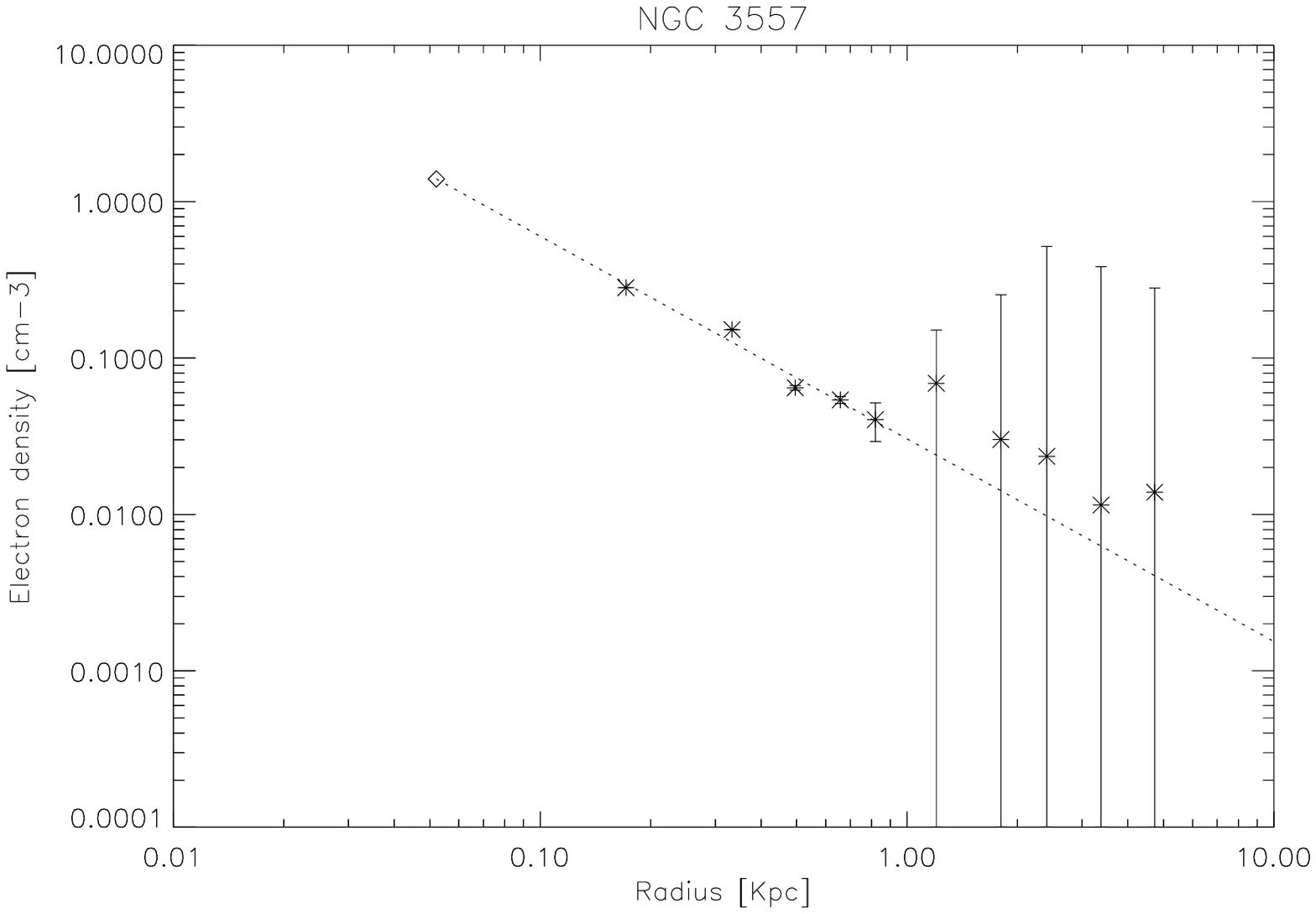,width=0.33\linewidth}
\psfig{figure=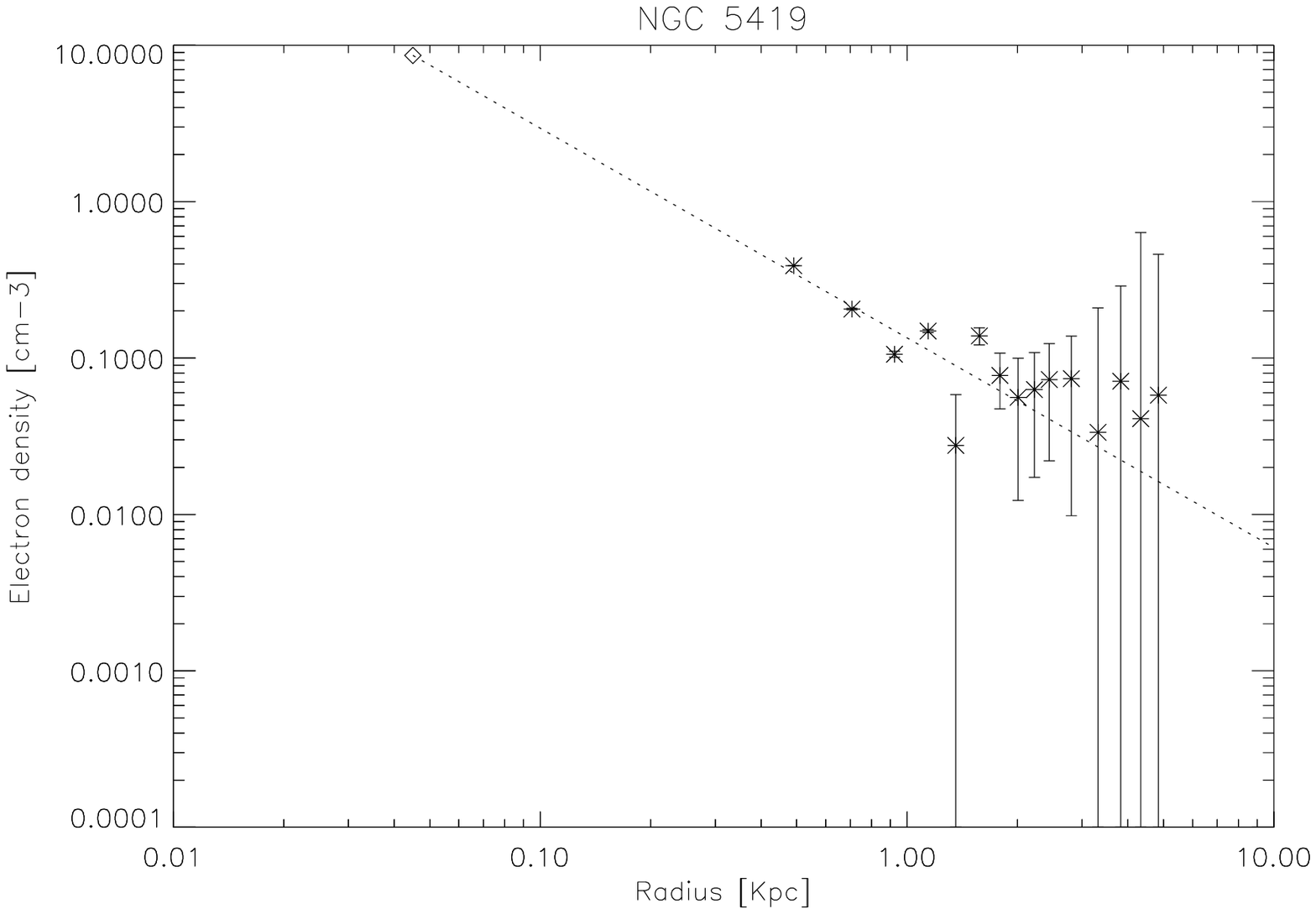,width=0.33\linewidth}}
\end{figure*}
\begin{figure*}
\centerline{
\psfig{figure=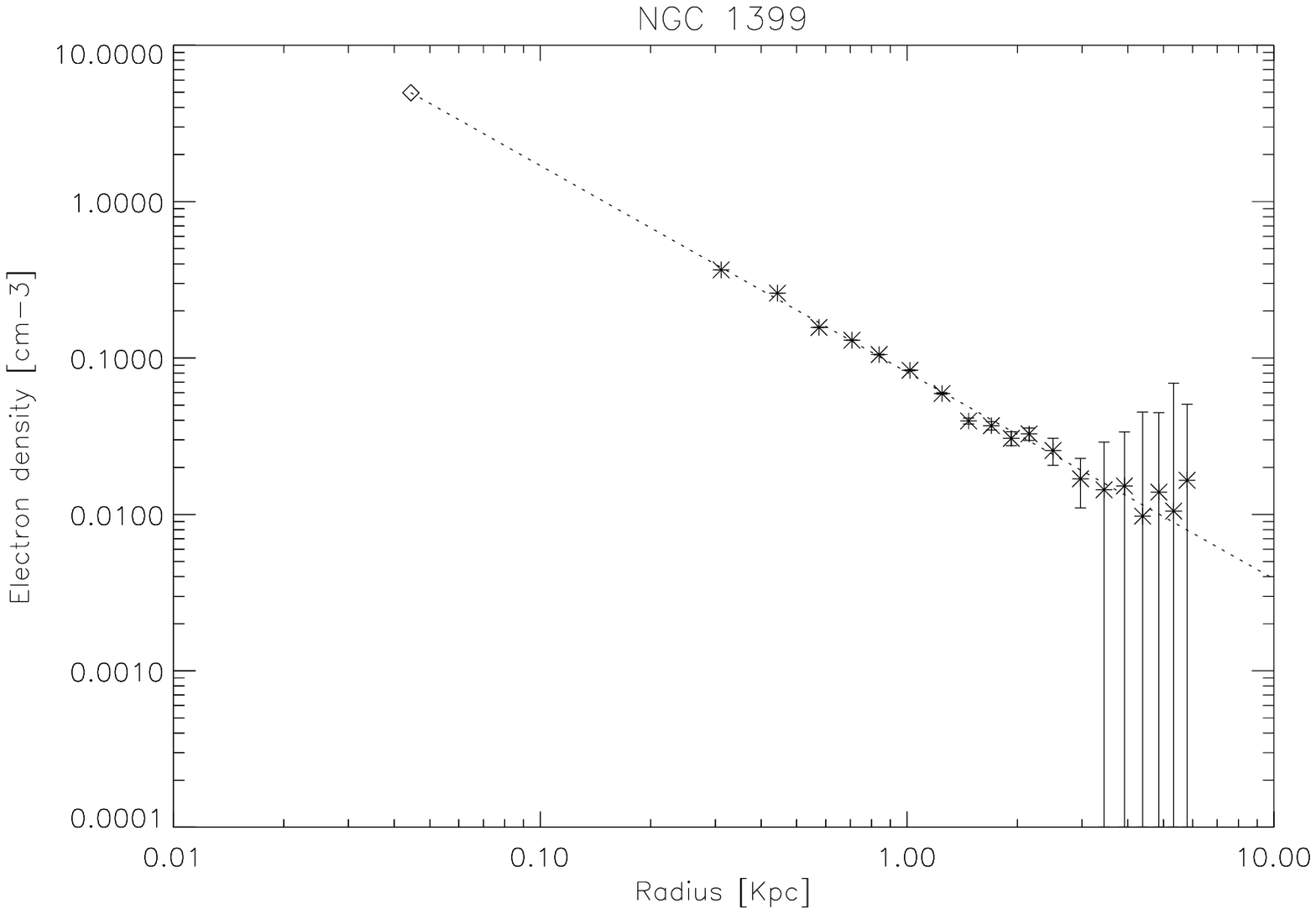,width=0.33\linewidth}
\psfig{figure=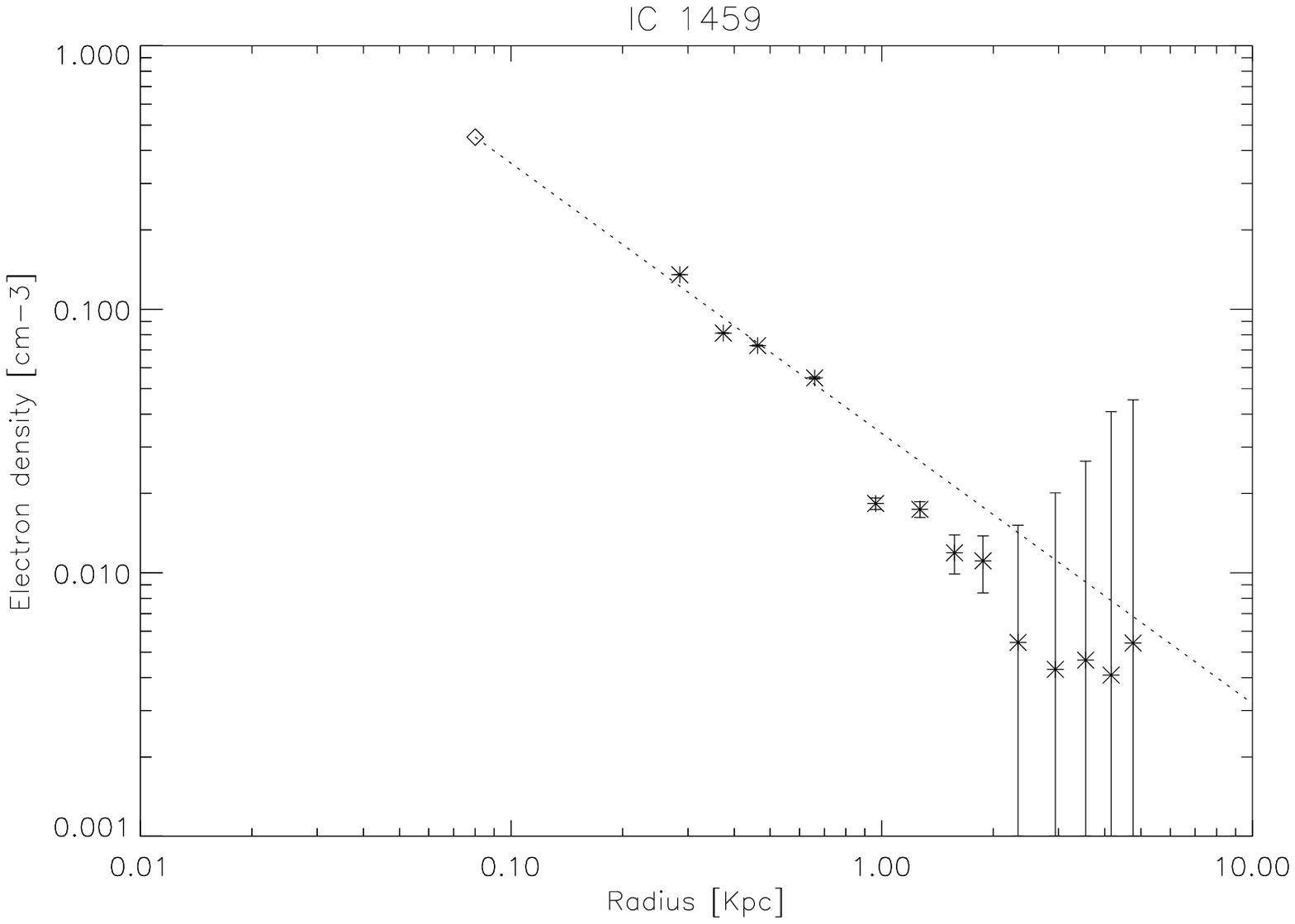,width=0.33\linewidth}
\psfig{figure=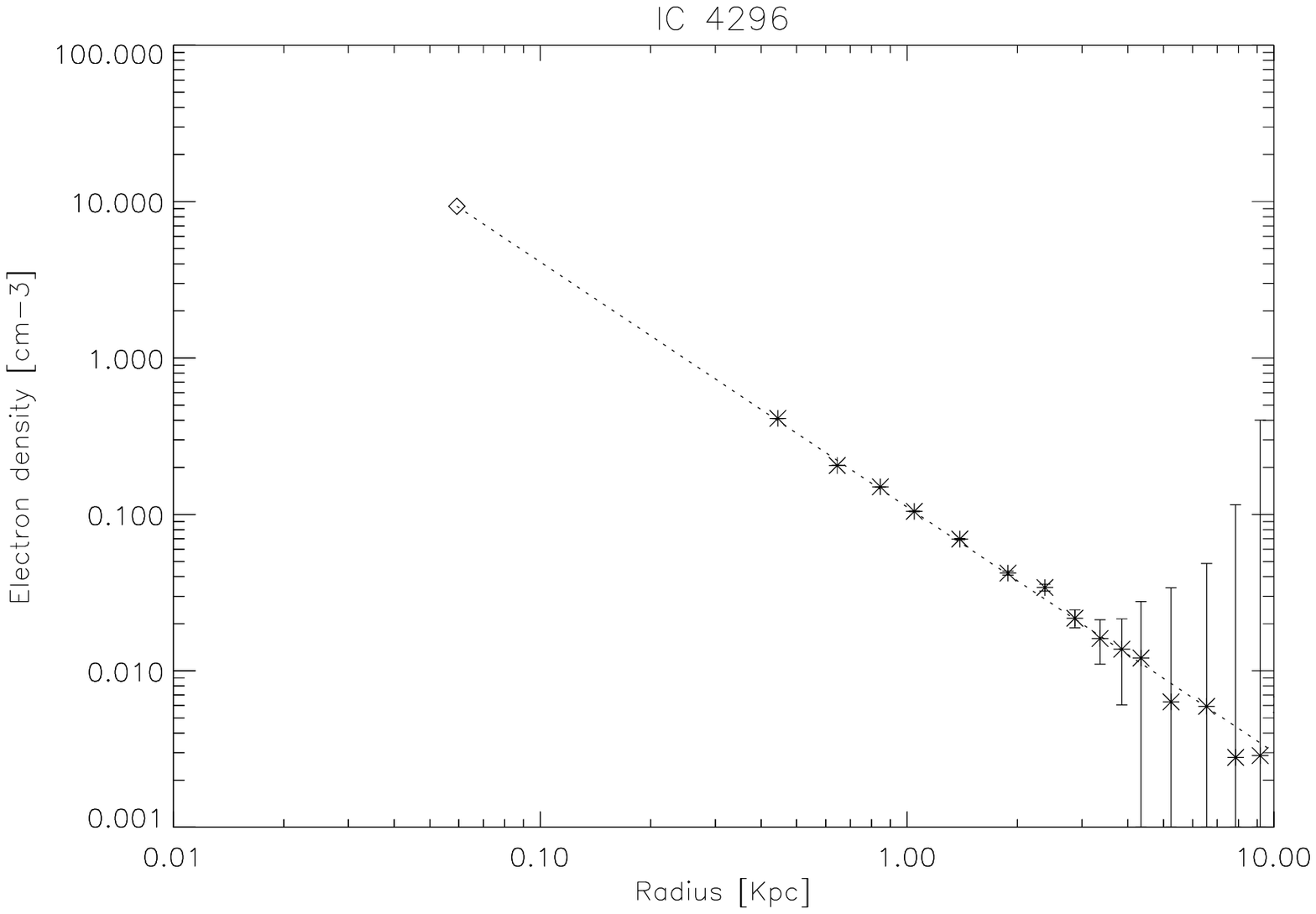,width=0.33\linewidth}}
\caption{The electron density profiles of the 30 objects, whose data quality
is sufficient to perform the deprojection of the brightness profile 
(see Sect. \ref{quality}). The solid lines represent the best-fit 
power-law relation to
the density profile.}
\label{neprof}
\end{figure*}

\end{document}